\documentclass[journal]{IEEEtran}

\usepackage{cite}
\usepackage{array}
\usepackage{color}
\usepackage{float}
\usepackage[cmex10]{amsmath}
\usepackage{nomencl}						
\usepackage[normalem]{ulem}			        
\usepackage{diagbox}						
\usepackage{slashbox}						
\usepackage{colortbl}						
\usepackage{multirow}						
\usepackage{tabularx}
\usepackage{placeins}
\usepackage{algorithm}
\usepackage{mathrsfs}
\usepackage{mathdots}
\usepackage{amssymb}
\usepackage{dblfloatfix}
\usepackage{amsthm}
\usepackage{arydshln}
\usepackage{color}
\usepackage[noend]{algpseudocode}
\usepackage{bm}
\usepackage{url}
\usepackage{threeparttable}
\usepackage[hidelinks]{hyperref}
\usepackage[T1]{fontenc}
\newcommand{\subparagraph}{}
\usepackage{fancyhdr} 
\usepackage{extarrows}
\usepackage{enumitem}
\usepackage{booktabs}

\usepackage{booktabs,siunitx,makecell}
\usepackage{pifont}
\usepackage{capt-of}

\newtheorem{theorem}{Theorem}
\newtheorem{definition}{Definition}

\newtheorem{lemma}{Lemma}

\ifCLASSINFOpdf
  \usepackage[pdftex]{graphicx}
	\graphicspath{{./figure/}}
  \DeclareGraphicsExtensions{.pdf,.jpeg,.png}
\else
  \usepackage[dvips]{graphicx}
  \graphicspath{{./figure/}}
  \DeclareGraphicsExtensions{.eps}
\fi

\ifCLASSOPTIONcompsoc
  \usepackage[caption=false,font=normalsize,labelfont=sf,textfont=sf]{subfig}
\else
  \usepackage[caption=false,font=footnotesize]{subfig}
\fi

\newcommand{\figref}[1]{\figurename~\ref{#1}}
\newcommand{\tabref}[1]{Table~\ref{#1}}

\ifCLASSOPTIONonecolumn

\else

\renewcommand{\arraystretch}{1.5}
\fi

\begin{document}
\setcounter{page}{1}
\vspace{-4\baselineskip}
\title{Quantitative Stability Region Estimation of Grid-Forming Inverters with Current Limiting via a State-Dependent Switched Energy Function}
\author{Yifan~Zhang, \IEEEmembership{Member, IEEE}, Peng~Huang, Yang~Wu, \IEEEmembership{Member, IEEE}

}
\IEEEaftertitletext{\vspace{-2.5\baselineskip}}
\ifCLASSOPTIONpeerreview
	\maketitle 
\else
	\maketitle
\fi
\thispagestyle{fancy}
\lhead{IEEE TRANSACTIONS ON POWER ELECTRONICS}
\rhead{\thepage}
\cfoot{}
\renewcommand{\headrulewidth}{0pt}
\pagestyle{fancy}

\begin{abstract}
With increasing renewable energy integration,
grid-forming (GFM) inverters are becoming key components
of future power systems. Compared with synchronous
generators (SGs), GFM inverters have limited overcurrent
capability and require current limiting. This can lead
to switching between normal and current-limited operating
modes under large disturbances, complicating synchronization
dynamics and posing a longstanding challenge to existing
transient-stability analysis methods. 
In this paper, a \emph{state-dependent
switched (SdS) energy function} theory is proposed for
transient-stability analysis of GFM inverters with current
limiting, whose dynamics can be abstracted as SdS nonlinear systems.
Regional energy functions are constructed for individual
operating modes and continuously matched at switching
interfaces, yielding a continuous piecewise energy function.
Hence, the derived conservative estimate of the stability
region (region of attraction, ROA) spans different
operating regions without prescribing subsequent current limiting activations and releases, a fundamental
distinction from conventional methods using signal-mode energy functions. 
Techniques for incorporating damping energy are extended to the proposed
framework to capture damping effects and further reduce the
conservatism of the ROA estimates. The methodology
applies to GFM inverters with different synchronization
control loops and current-limiting strategies. Experiments on a physical GFM inverter under various grid contingencies validate the proposed quantitative transient stability assessment.
\end{abstract}


\section{Introduction}
As renewable generation continues to develop, power systems are gradually transitioning from synchronous-generator (SG)-dominated to inverter-based-resource (IBR)-dominated systems. Among IBR technologies, grid-forming (GFM) inverters have shown great promise owing to their capability to provide frequency and voltage support \cite{Xiongfei_overview}. In this context, the transient synchronization stability of GFM inverters, i.e., their ability to maintain or regain synchronization following large disturbances such as voltage sags and phase jumps induced by grid contingencies, has attracted significant attention in recent years \cite{Xiongfei_overview}.

Transient-stability analysis of IBRs, particularly GFM inverters, has progressively advanced from qualitative characterization based on equilibrium existence, power--angle curves, and phase portraits \cite{Xiongfei_overview}, through equal-area-criterion (EAC)-based assessment \cite{chatterjee2025equal}, toward quantitative stability-region estimation via Lyapunov/energy-function-based direct methods \cite{shuai2018transient,fu2020large,lei2023transient,11540391}. The stability region, also referred to as the region of attraction (ROA), provides a direct criterion for transient-stability assessment without extensive time-domain simulations or repeated critical-clearing-time (CCT) searches \cite{chiang2011direct}. Compared with the EAC, the energy-function framework enables mathematically rigorous and conservative ROA estimation while accounting for damping \cite{shuai2018transient, lei2023transient}, revealing the quantitative relation between system parameters and ROA \cite{fu2020large}, and providing a unified stability criterion for different disturbances \cite{11540391}. It has therefore been widely applied and remains a promising approach for quantitative transient-stability assessment.

Current limiting, however, poses a major challenge to transient-stability analysis, particularly quantitative ROA estimation, for practical GFM inverters. Unlike SGs, which typically tolerate overcurrents of 5--7 pu \cite{wilson2018eu}, GFM inverters have a much lower overcurrent capability, typically around 1.5--2 pu due to the lower overcurrent capability of semiconductor power devices \cite{fan2022review}. Large grid disturbances can therefore readily drive a GFM inverter beyond its current capability and activate current-limiting control. Recent reviews of current-limiting strategies and the transient stability of GFM inverters
\cite{fan2022review,baeckeland2024overcurrent,baeckeland2025unified} have shown that current limiting introduces distinct large-signal dynamics and induces transitions between normal and current-limited operating modes, thereby substantially impacting transient stability. In addition, recent GFM grid-code in Great Britain and Europe \cite{NESO2026GridForming,ESO2023GBGFGuide,ENTSOE2025GFMPPM} explicitly encompass current-limit activation and release in transient-stability requirements. Accordingly, transient-stability analysis and ROA estimation should address the overall GFM dynamics spanning both normal and current-limited operating modes.

\begin{figure*}[!t]
    \centering
    \vspace{-6pt}
    \begin{minipage}[c]{0.6\linewidth}
        \centering
        \subfloat[GFM inverter system and synchronization control.]{
            \includegraphics[width=\linewidth]{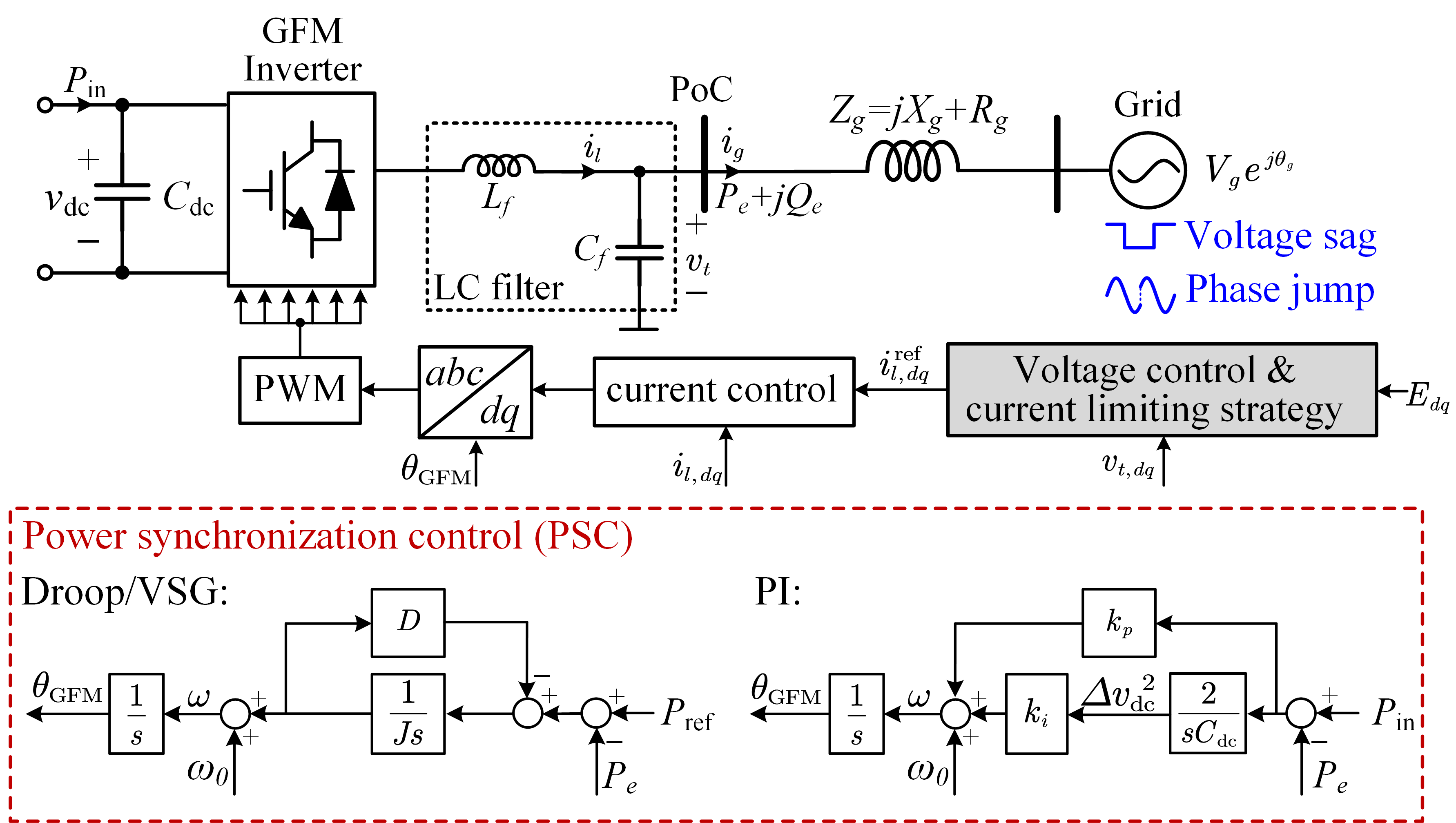}
        }
        \vspace{-3pt}
    \end{minipage}
    \hfill
    \begin{minipage}[c]{0.38\linewidth}
        \centering
        {\captionsetup[subfloat]{captionskip=1pt}
        \subfloat[Conventional circular current limiter.]{
            \includegraphics[width=0.73\linewidth]{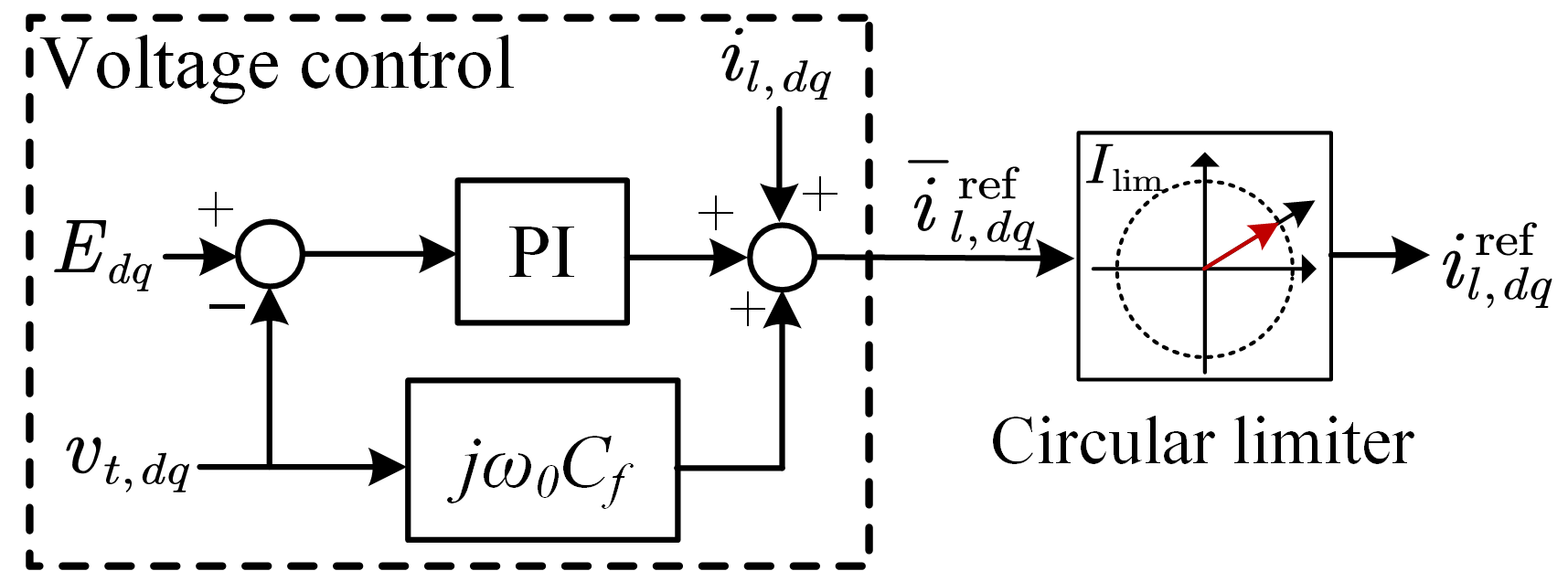}
        }}\\
        \vspace{-4pt}
        {\captionsetup[subfloat]{captionskip=0pt}
        \subfloat[threshold VI and current-limit-tracking (adaptive) VI.]{
            \includegraphics[width=0.95\linewidth]{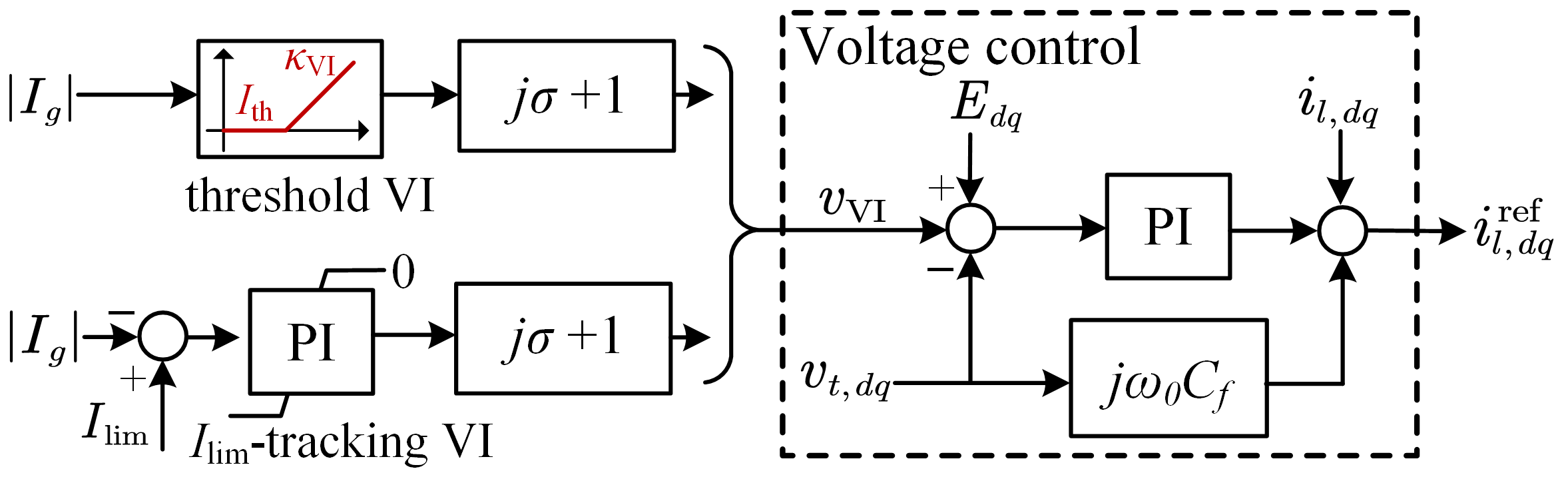}
        }}\\
        \vspace{-2pt}
        {\captionsetup[subfloat]{captionskip=0pt}
        \subfloat[EVI with circular current limiter.]{
            \includegraphics[width=0.79\linewidth]{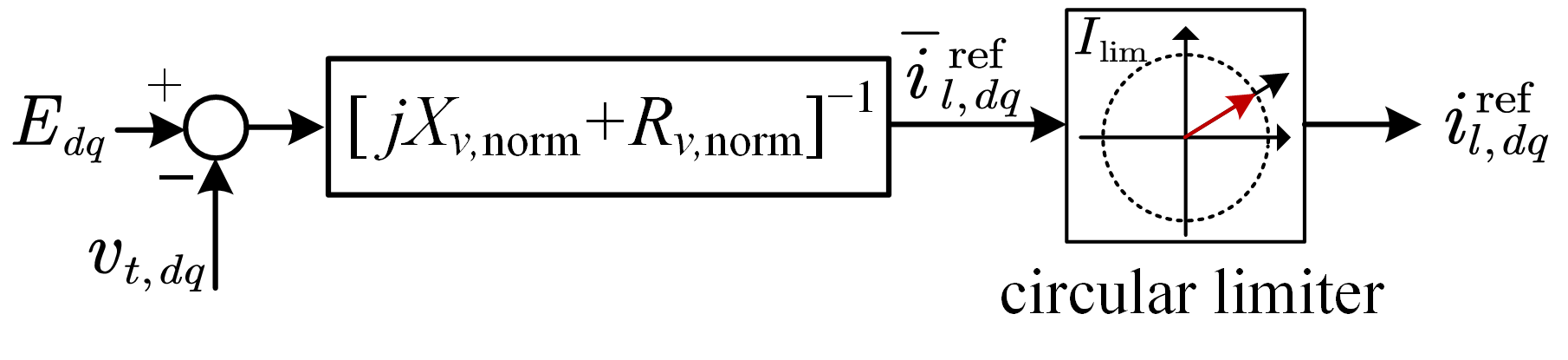}
        }}
        \vspace{-6pt}
    \end{minipage}
    \caption{GFM inverter system and representative current-limiting implementations.}
    \vspace{-8pt}
    \label{fig:Revisit_CL}
\end{figure*}
The transient-stability implications of specific current-limiting strategies have been investigated under different large disturbances
\cite{fan2022equivalent,zhan2024transient,rokrok2021transient,kkuni2024effects,8026165,jiang2026phase}, while comparative studies have examined different limiting strategies
\cite{lyu2025critical,baeckeland2025unified}.
Much of this work remains largely reliant on power--angle characteristics, using either first-order GFM models
\cite{fan2022equivalent,zhan2024transient,8026165}
or qualitative interpretation of power imbalance evolution along power-angle curves
\cite{fan2022equivalent,kkuni2024effects,jiang2026phase}.
EAC-based formulations in \cite{rokrok2021transient} can provide quantitative CCT estimates, but such estimate is specific to a prescribed disturbance and neglects damping effects.

Recent studies have begun to extend energy-function methods and corresponding ROA estimation to current-limited GFM inverters.
The studies in \cite{zhang2024estimating,liu2023dynamic} construct energy functions and estimate ROAs for the current-limited operating mode itself, without considering the subsequent transition back to normal operation.
Another line of work retains the normal-mode energy function and uses the current-limit boundary as an additional constraint, yielding conservative ROA estimates that exclude trajectories later entering current limitation \cite{fan2024transient,yang2025global,wang2025transient}.
\cite{liu2025state,luo2025estimation} consider prescribed transfers between fault-on and post-fault stages, for which stability is assessed by whether the transferred state lies within the ROA of the subsequent stage.
Both assume that current limiting is absent thereafter in the post-fault evolution---from fault clearance onward in \cite{liu2025state}, or following a prescribed brief transition interval in \cite{luo2025estimation}.
An overall ROA for the complete post-fault dynamics of a GFM inverter, irrespective of the subsequent pattern of current-limit activation and release, has yet to be developed.

This gap is also relevant to DC-voltage-controlled GFM inverters, another important GFM realization  \cite{Zhao2022VFM} that is also referred to as matching control \cite{8825565} or the frequency-following voltage-forming inverter \cite{Ai2024VFM}. 
It retains voltage-forming operation while enabling power tracking through DC-voltage regulation. 
Its transient stability has been investigated in \cite{Zhao2022VFM,luo2023design,9762041}, and energy-function-based ROA estimation has recently been developed in \cite{11540391}; however, current limiting has not been incorporated into the corresponding ROA analysis.

Under large disturbances, current limiting makes GFM inverter dynamics a nonlinear switched-system problem, motivating the use of switched-system stability theory \cite{liberzon2003switching}. 
For autonomous systems with state-dependent switching, suitably matched regional functions associated with individual operating modes can form a continuous Lyapunov function, enabling ROA estimation via the nonsmooth LaSalle invariance principle \cite{shevitz2002lyapunov}. Nevertheless, constructing such an energy function for GFM inverters with current limiting remains challenging. Integrating existing damping-energy estimates \cite{moon2000estimating,shuai2018transient,lei2023transient} into such switched dynamics presents a further difficulty.

To bridge the above gaps, this paper established \emph{state-dependent switched (SdS) energy function} for transient-stability analysis.
Specifically, regional energy functions are constructed for the normal and current-limited operating regions and continuously matched at their interfaces to form the SdS energy function.
The ROA is then estimated from the critical energy defined by the relevant unstable equilibrium point (UEP).
To further reduce ROA conservatism, damping-energy corrections are obtained analytically for classical GFM inverters and from the critical trajectory for DC-voltage-controlled GFM inverters.
Within the resulting ROA, convergence to the stable equilibrium point (SEP) is guaranteed regardless of how the current limiting subsequently activates and releases, a guarantee not provided by existing methods.
The framework is applicable to a class of current-limiting strategies and enables quantitative assessment of parameter effects and strategy comparison across disturbances.

\vspace{-2pt}
\section{Revisiting the Current Limiting Strategies and Their Large-Signal Switched Model}
The GFM inverter and its control structure are shown in
\figref{fig:Revisit_CL}(a). The grid is represented by the voltage
$V_g\angle\theta_g$ at the nominal angular frequency $\omega_0$ and is
connected to the inverter through $Z_g=R_g+jX_g$; $v_t$ and $P_e$ denote the voltage and output power at the point-of-connection (PoC), respectively. Throughout this paper, the inverter-to-grid power angle difference is defined as $\delta=\theta_{\mathrm{GFM}}-\theta_g$.

For a classical GFM inverter employing droop or VSG control, the dynamics can be expressed by the swing equation
\begin{equation}
\begin{cases}
    \dfrac{\mathrm{d}}{\mathrm{d}t}\delta =\omega,\\[2pt]
    J\dfrac{\mathrm{d}}{\mathrm{d}t}\omega
    =P_{\mathrm{ref}}-P_e(\delta)-D\omega,
\end{cases}
\label{GFMe}
\end{equation}
where $\omega$ is the inverter frequency deviation relative to the
grid, and $J$ and $D$ denote the virtual inertia and damping
coefficient, respectively.

A DC-voltage-controlled realization of power synchronization
control (PSC) is also illustrated in \figref{fig:Revisit_CL}(a).This structure admits an equivalent proportional--integral
(PI) form for power tracking \cite{11540391,9762041}.
The corresponding inverter dynamics are
\cite{11540391,9762041,Zhao2022VFM}
\begin{equation}
\begin{cases}
    \dfrac{\mathrm{d}}{\mathrm{d}t}\delta
    =k_p\!\left[P_{\mathrm{in}}-P_e(\delta)\right]
    +k_i\Delta v_{\mathrm{dc}}^2,\\[2pt]
    \dfrac{\mathrm{d}}{\mathrm{d}t}\Delta v_{\mathrm{dc}}^2
    =\dfrac{2}{C_{\mathrm{dc}}}
    \left[P_{\mathrm{in}}-P_e(\delta)\right],
\end{cases}
\label{VFMe}
\end{equation}
where
$\Delta v_{\mathrm{dc}}^2
=v_{\mathrm{dc}}^2-V_{\mathrm{dc,ref}}^2$ denotes the squared DC-voltage error, $P_{\text{in}}$ represents the DC-side input power, and $k_p$ and $k_i$ are the corresponding synchronization-control gains. For both realizations, current limiting modifies the synchronization dynamics through the output power $P_e(\delta)$.

Under normal operation, the output power applied to
both synchronization realizations in \eqref{GFMe} and \eqref{VFMe} is:
\begin{equation}
\hspace{-2mm}
\begin{aligned}
        P_{e,\mathrm{norm}}(\delta)=&\frac{X_{\Sigma ,\mathrm{norm}}EV_g\sin \delta}{\left| Z_{\Sigma ,\mathrm{norm}} \right|^2}+\frac{R_g\left( E^2-EV_g\cos \delta \right)}{\left| Z_{\Sigma ,\mathrm{norm}} \right|^2}\\&+\frac{R_{v,\mathrm{norm}}\left( EV_g\cos \delta -{V_g}^2 \right)}{\left| Z_{\Sigma ,\mathrm{norm}} \right|^2},
\end{aligned}
\label{Pe_norm}
\end{equation}
where
$Z_{v,\mathrm{norm}}
=R_{v,\mathrm{norm}}+jX_{v,\mathrm{norm}}$
denotes the constant virtual impedance under normal operation,
$Z_{\Sigma,\mathrm{norm}}
=Z_g+Z_{v,\mathrm{norm}}$
is the corresponding total impedance between the internal voltage source and the grid, with $X_{\Sigma,\mathrm{norm}}$ denoting its reactance.
The corresponding unconstrained current magnitude is
\[
|I_u(\delta)|
=
\frac{|Ee^{j\delta}-V_g|}
{|Z_{\Sigma,\mathrm{norm}}|}.
\]
Accordingly, the switching interface between the normal and current-limited modes is defined by
$|\delta|=\delta_{\mathrm{sw}}$, where the unconstrained current $I_u$ reaches the activation threshold $I_{\mathrm{sw}}$. This yields
\begin{equation}
\delta_{\mathrm{sw}}
=
\cos^{-1}\!\left(
\frac{E^2+V_g^2-I_{\mathrm{sw}}^2
|Z_{\Sigma,\mathrm{norm}}|^2}
{2EV_g}
\right).
\label{deltasw}
\end{equation}
The normal and current-limited operating regions are therefore denoted by
\[
\mathcal{R}_{\mathrm{norm}}
:=\{|\delta|<\delta_{\mathrm{sw}}\},
\qquad
\mathcal{R}_{\mathrm{lim}}
:=\{|\delta|\geq\delta_{\mathrm{sw}}\}.
\]

A broad range of current-limiting strategies for GFM inverters and their corresponding large-signal models for transient-stability analysis have been established and systematically reviewed in \cite{fan2022review,baeckeland2024overcurrent}; here, representative strategies and models are revisited to establish the basis for the subsequent SdS energy function construction. The corresponding control implementations are illustrated in \figref{fig:Revisit_CL}(b)--(d).
1) For the circular current limiter in \figref{fig:Revisit_CL}(b), its
current-limited inner-loop behavior can be represented by an equivalent variable virtual resistance
\cite{fan2022equivalent,lyu2025critical,jiang2026phase}.

\begin{table}[t!]
\centering
\caption{Parameterization of Current-Limiting Strategies}
\vspace{-4pt}
\label{tab:current_limit}
\setlength{\tabcolsep}{4pt}
{\footnotesize
\everymath{}
\everydisplay{}
\begin{tabular}{c|c|c|c}
\hline\hline
& $\delta_{\mathrm{sw}}$ in~\eqref{deltasw}
& $Z_{v,\mathrm{norm}}$
& $Z_{v,\mathrm{lim}}(\delta)$ \\
\hline
Circular limiter
& $I_{\mathrm{sw}}=I_{\lim}$
& $0$
& \eqref{Zv_VI}, $\phi_v=0$ \\
\hline
EVI
& $I_{\mathrm{sw}}=I_{\lim}$
& $Z_{v,\mathrm{norm}}$
& \eqref{Zv_VI}, $\phi_v\to\pi/2^{-}$ \\
\hline
Tracking VI
& $I_{\mathrm{sw}}=I_{\lim}$
& $0$
& \eqref{Zv_VI},  $\phi_v\to\pi/2^{-}$ 
\\
\hline
Threshold VI
& $I_{\mathrm{sw}}=I_{\mathrm{th}}$
& $0$
& \eqref{th_VI} \\
\hline\hline
\end{tabular}
}
\vspace{-6pt}
\end{table}

\begin{figure}[t!]
    \centering
    \vspace{-3pt}
    \includegraphics[width=0.4\textwidth]{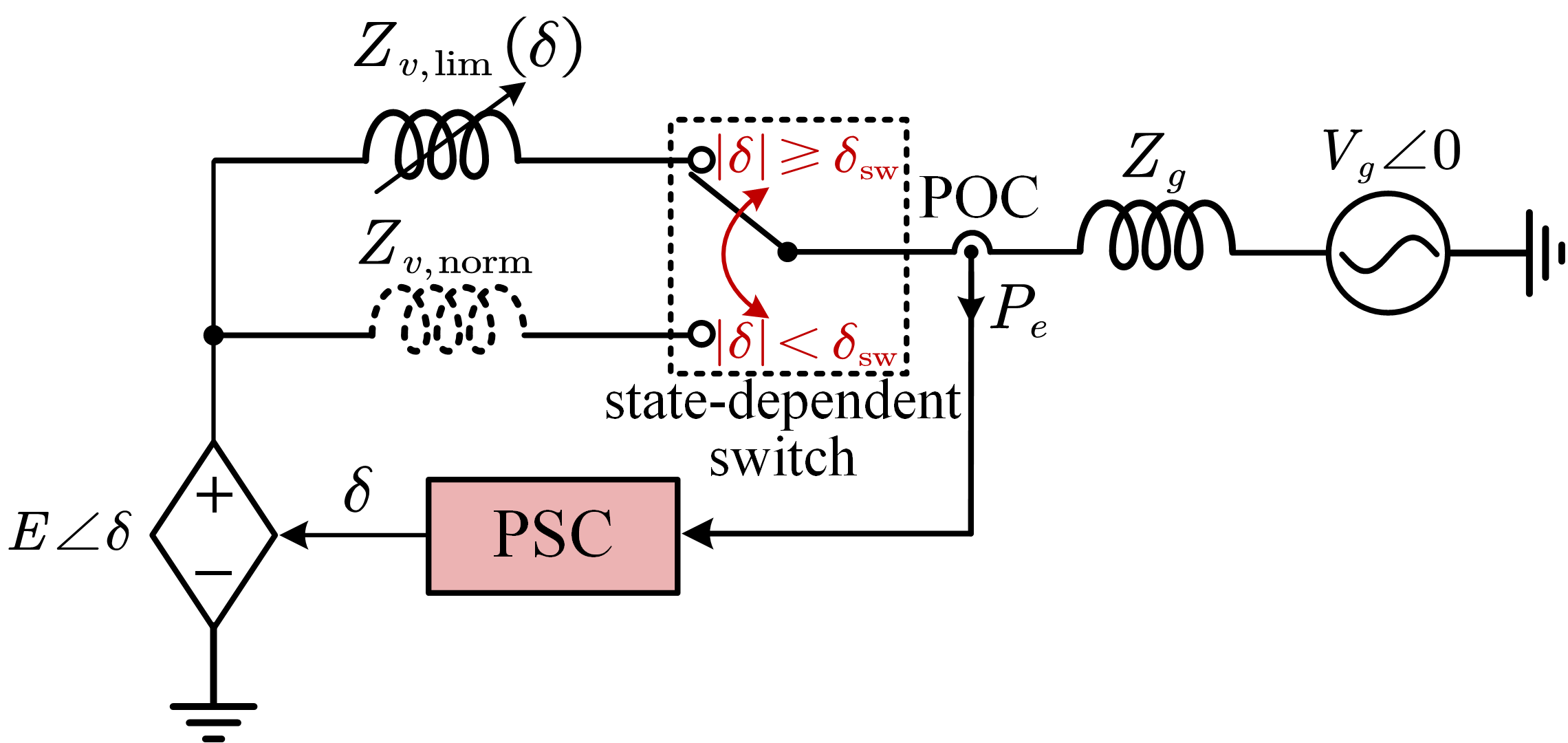}
    \vspace{-3pt}
    \caption{Unified state-dependent switched system representation for different current-limiting strategies. (Without loss of generality, the grid voltage angle is set to zero.)}
    \label{SDWS}	
    \vspace{-8pt}
\end{figure}

2) The threshold VI, illustrated by the upper implementation in
\figref{fig:Revisit_CL}(c), activates an additional virtual impedance once the current exceeds $I_{\mathrm{th}}$
\cite{Wang2025TSEM,Qoria2020CriticalClearingTime},
\begin{equation}
\begin{aligned}
    X_{v,\lim}=&\sigma R_{v,\lim},\,\\
    R_{v,\lim}(\delta)=&
    \left\{
    \begin{array}{@{}l@{\;}l}
    0, & |I_g|<I_{\mathrm{th}}\\
    \kappa_{\mathrm{VI}}\left(|I_g|-I_{\mathrm{th}}\right),
    & |I_g|\ge I_{\mathrm{th}}
    \end{array}
    \right.
\end{aligned}
\label{th_VI}
\end{equation}
where $\sigma=X_{v,\mathrm{lim}}/R_{v,\mathrm{lim}}$ is the virtual-impedance $X/R$ ratio. Since $I_g(\delta)$ is a function of $\delta$, the resulting virtual impedance is itself
$\delta$-dependent. Because threshold VI does not enforce a hard current limit, the activation threshold $I_{\mathrm{th}}$ is set below the prescribed current limit $I_{\mathrm{lim}}$ in practical designs \cite{fan2022review}.

3) The lower implementation in \figref{fig:Revisit_CL}(c) instead uses a current-limit-tracking loop to adapt the virtual-impedance magnitude and enforce $|I_g|=I_{\lim}$ \cite{nasr2023controlling,rosso2021implementation}. 

Another hard-limiting realization illustrated in \figref{fig:Revisit_CL}(d) is the embedded-virtual-impedance (EVI) implementation \cite{zhan2024transient}. It is also well-known as the virtual-admittance implementation \cite{fan2022review,kkuni2024effects}, where the voltage error is directly converted into the current reference through the virtual admittance before the circular limiter. Despite their different implementations, the current-limit-tracking VI and EVI share the same large-signal equivalent: the magnitude of a virtual impedance is adapted while its impedance angle is preserved so that
\begin{equation}
\frac{|Ee^{j\delta}-V_g|}
{|Z_{v,\mathrm{lim}}+Z_g|} = I_{\lim}.
\label{I_hard}
\end{equation}
Accordingly, the equivalent virtual impedance during current
limiting can be written compactly as
\begin{equation}
Z_{v,\mathrm{lim}}(\delta)
=
\lambda(\delta)e^{j\phi_v},
\label{Zv_VI}
\end{equation}
where
$
\lambda(\delta)
=
-a+
\sqrt{
a^2+\Delta(\delta)/I_{\lim}^2
-|Z_g|^2
},
$
$\Delta(\delta)=E^2+V_g^2-2EV_g\cos\delta$,
$a=R_g\cos\phi_v+X_g\sin\phi_v$, and
$\phi_v=\tan^{-1}\sigma$.
For $\phi_v=0$, \eqref{Zv_VI} reduces to a variable resistance, which corresponds to the equivalent circuit model of the circular limiter. Thus, the circular limiter, current-limit-tracking VI, and EVI share
the same current-limited impedance form. The EVI differs from the tracking VI mainly in normal operation, where a fixed virtual impedance $Z_{v,\mathrm{norm}}$ is retained. For clarity, the corresponding impedance forms under normal and
current-limited operation are summarized in \tabref{tab:current_limit}.

\begin{figure}[t!]
    \centering
    \vspace{-3pt}
    \includegraphics[width=0.46\textwidth]{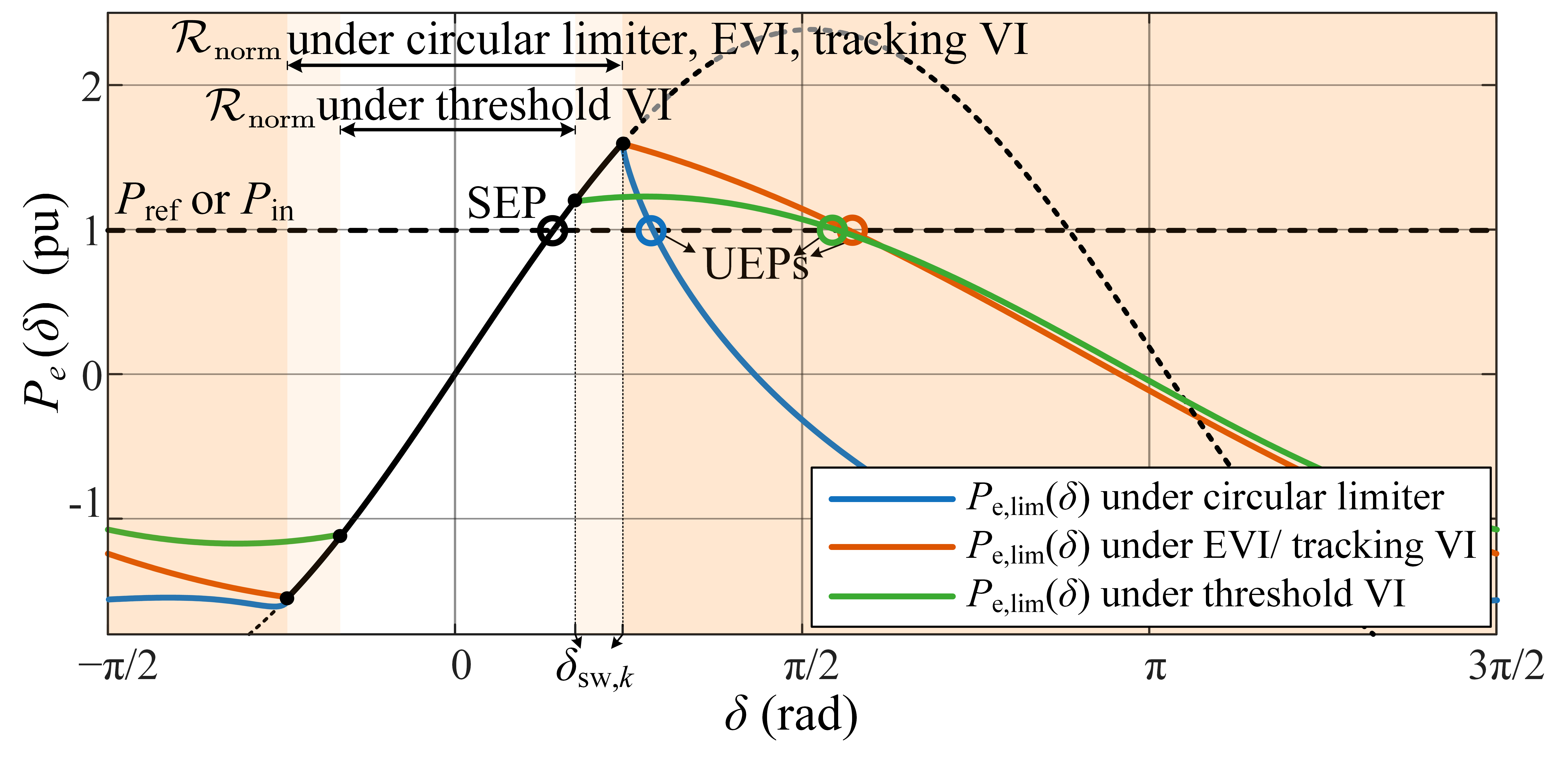}
    \vspace{-9pt}
    \caption{Piecewise $P_{\mathrm{e}}$ curves under different current-limiting
    strategies. The black solid, black dotted, and colored curves denote
    $P_{\mathrm{e,norm}}$, $P_{\mathrm{e}}$ without current limiting, and
    $P_{\mathrm{e,lim}}$, respectively. The open circles mark the corresponding
    system UEPs, while the two orange shades indicate the strategy-dependent current-limited operating regions $\mathcal{R}_{\mathrm{lim}}$.}
    \vspace{-3pt}
    \label{Pe_curve}	
    \vspace{-5pt}
\end{figure}

Under current-limited operation, the PoC active power for the 
strategies discussed above can be expressed as
\begin{equation}
\hspace{-2mm}
\begin{aligned}
P_{e,\lim}(\delta)
={}&
\frac{
R_{\Sigma,\lim}V_g(E\cos\delta-V_g)
+
X_{\Sigma,\lim}EV_g\sin\delta
}{
R_{\Sigma,\lim}^2+X_{\Sigma,\lim}^2
}\\
&+R_g|I_g(\delta)|^2,
\end{aligned}
\label{Pe_lim}
\end{equation}
where
$R_{\Sigma,\lim}=R_g+R_{v,\lim}$ and
$X_{\Sigma,\lim}=X_g+X_{v,\lim}$. For hard current-limiting, $|I_g(\delta)|=I_{\lim}$.

Current-limit activation and release render the post-fault large-signal model a state-dependent switched system, as illustrated in \figref{SDWS}. Accordingly, the output power $P_e$
becomes the state-dependent piecewise function
\begin{equation}
P_e(\delta)=
\begin{cases}
P_{e,\mathrm{norm}}(\delta),
& \delta\in\mathcal R_{\mathrm{norm}},\\
P_{e,\lim}(\delta),
& \delta\in\mathcal R_{\mathrm{lim}}.
\end{cases}
\label{Pe_switched}
\end{equation}
which, together with \eqref{GFMe} or \eqref{VFMe}, constitutes the
post-fault large-signal model for transient-stability analysis. The resulting $P_e$ for different current-limit strategies is illustrated in \figref{Pe_curve}.

Other strategies include, for example, priority-based limiters \cite{rokrok2021transient,8026165}, which exhibit current-forming behavior in the current-limited mode, with the GFM inverter modeled as a PSC-controlled current source. The corresponding $P_e$ is likewise a piecewise function, yielding a state-dependent switched model. Hence, the SdS energy function construction developed in the next section is not restricted to a particular current-limiting type. However, from a practical standpoint, such current-forming operation does not retain the natural voltage-source response of a GFM inverter and is therefore not aligned with recent grid-code requirements \cite{NESO2026GridForming,ESO2023GBGFGuide,ENTSOE2025GFMPPM}. Accordingly, to keep the methodological development focused, the following analysis considers the 
current-limiting strategies in \figref{fig:Revisit_CL}(b)--(d), with particular emphasis on the circular limiter and EVI as hard current-limiting realizations.


\section{Proposed State-Dependent Switched Energy Function and ROA Estimation}
\subsection{Theoretical basis of SdS energy function}
In an autonomous state-dependent switched system, the current state $\mathbf{x}$ uniquely determines both the switching decision and the active operating mode. Such a system can be expressed as \cite{liberzon2003switching}:
\begin{equation}
\dot{\mathbf{x}}
=
f_{\sigma(\mathbf{x})}(\mathbf{x}),
\qquad
\sigma(\mathbf{x})\in\mathcal{P},
\label{eq:sdss}
\end{equation}
where $\mathcal{P}:=\{1,2,\ldots,m\}$ denotes the set of operating modes. The state space is partitioned into corresponding operating regions $\mathcal{R}_{p}$, $p\in\mathcal{P}$. When $\mathbf{x}\in\mathcal{R}_{p}$, mode $p$ is active, i.e., $\sigma(\mathbf{x})=p$, and the dynamics are governed by corresponding regional vector
field $f_{p}(\mathbf{x})$, that is 
\begin{equation}
\dot{\mathbf{x}}
=
f_{p}(\mathbf{x}),
\qquad
\mathbf{x}\in\mathcal{R}_{p}.
\label{eq:regional_dynamics}
\end{equation}
For the present application, it suffices to consider the common case where system trajectory crosses switching interfaces directly.


\textit{Premise (Strict Crossing):}
The switching interface\footnote{$h_{pq}$ is a scalar function that equals zero on the switching interface and its nonzero gradient $\nabla h_{pq}$ gives the normal direction.} between two adjacent operating
regions is denoted by
\begin{equation}
\Sigma_{pq}
:=
\overline{\mathcal R_p}
\cap
\overline{\mathcal R_q}
=
\left\{
\mathbf x:
h_{pq}(\mathbf x)=0
\right\}.
\end{equation}
The adjacent vector fields satisfy
\begin{equation}
\left[
\nabla h_{pq}(\mathbf x)^{\rm T}
f_p(\mathbf x)
\right]
\left[
\nabla h_{pq}(\mathbf x)^{\rm T}
f_q(\mathbf x)
\right]
>0,
\quad
\mathbf x\in\Sigma_{pq}.
\end{equation}
This condition means that the normal components of the two
adjacent vector fields have the same nonzero sign at the
switching interface. Hence, the trajectory passes through the interface directly without remaining on it, and no Filippov sliding motion occurs \cite{liberzon2003switching}.

Building on Lyapunov stability theory for switched systems \cite{shevitz2002lyapunov, wu2001lyapunov, bacciotti2004nonsmooth}, we define the SdS energy function as follows.

\begin{definition}[SdS energy function]
\label{def:sdsef}
A continuous function
$\mathcal{W}:\mathbb{R}^{n}\rightarrow\mathbb{R}$
is called a state-dependent switched (SdS) energy function
for \eqref{eq:sdss} if it can be expressed as
\begin{equation}
\mathcal{W}(\mathbf{x})
=
\sum_{p\in\mathcal{P}}
\chi_{p}(\mathbf{x})W_{p}(\mathbf{x}),
\label{eq:sdsef_composite}
\end{equation}
where $\chi_p(\mathbf{x})$ equals $1$ for
$\mathbf{x}\in\mathcal{R}_p$ and $0$ otherwise,
and the regional functions satisfy the following conditions.

\emph{1) Negative semi-definite derivative:}
For each operating region $\mathcal{R}_{p}$, the corresponding
regional energy function $W_p(\mathbf{x})$ is continuously
differentiable and satisfies
\begin{equation}
\dot W_p(\mathbf{x})
=
\nabla W_p(\mathbf{x})^{\mathrm T}
f_p(\mathbf{x})
\le0,
\qquad
\mathbf{x}\in\mathcal R_p.
\label{eq:regional_energy_derivative}
\end{equation}

\emph{2) Continuity across switching interfaces:}
For every pair of adjacent operating regions, the corresponding
regional energy functions satisfy
\begin{equation}
W_p(\mathbf{x})
=
W_q(\mathbf{x}),
\qquad
\mathbf x\in\Sigma_{pq}.
\end{equation}
\end{definition}

\begin{figure}[t!]
\centering
\vspace{-3pt}
{\captionsetup[subfloat]{captionskip=0pt}
\subfloat[Geometric interpretation of the SdS energy function and the corresponding ROA estimate.]{\includegraphics[width=0.35\textwidth]{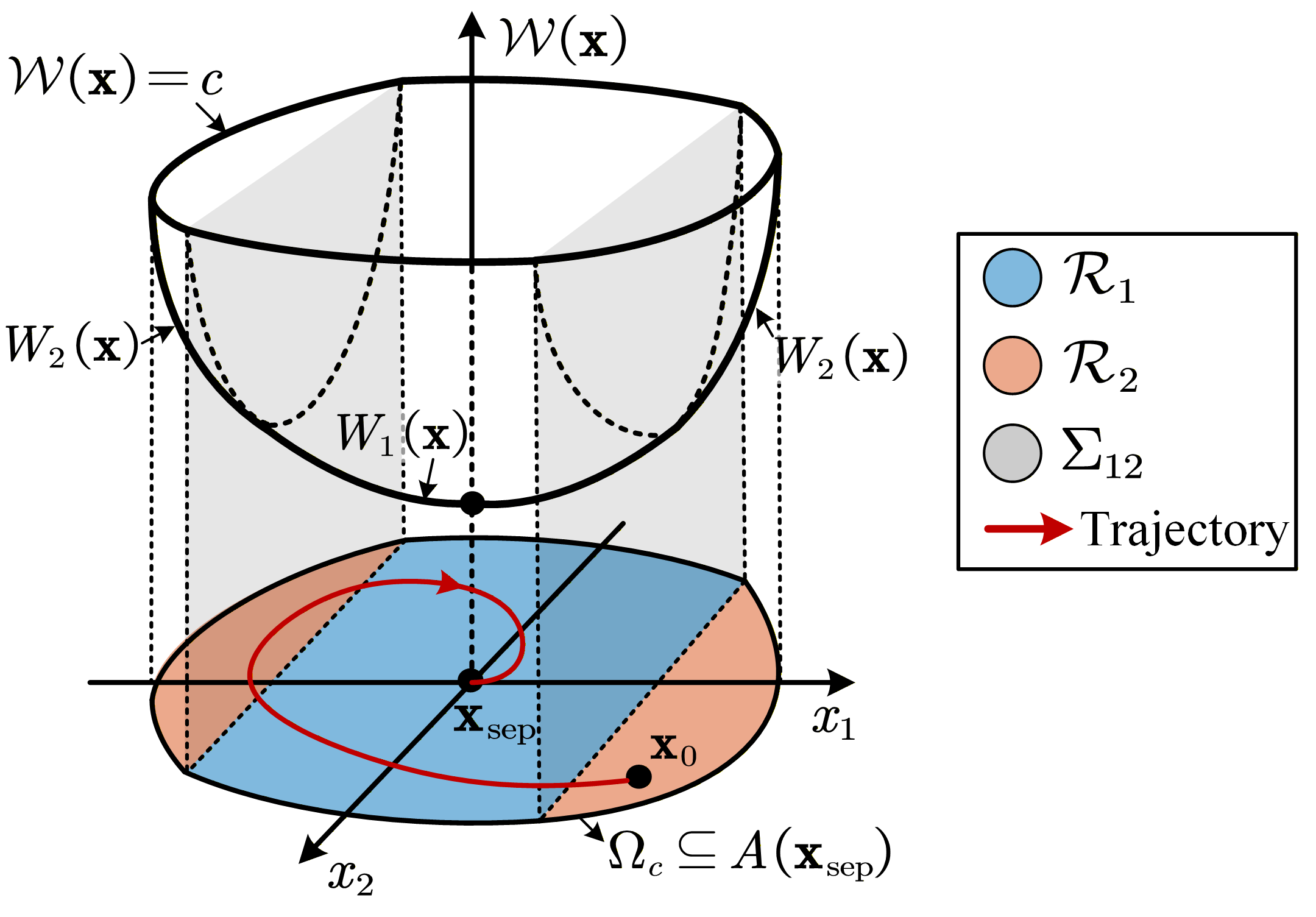}}
}\\
\vspace{-3pt}
{\captionsetup[subfloat]{captionskip=-1pt}
\subfloat[Evolution of the SdS energy function under state-dependent switching.]{\includegraphics[width=0.31\textwidth]{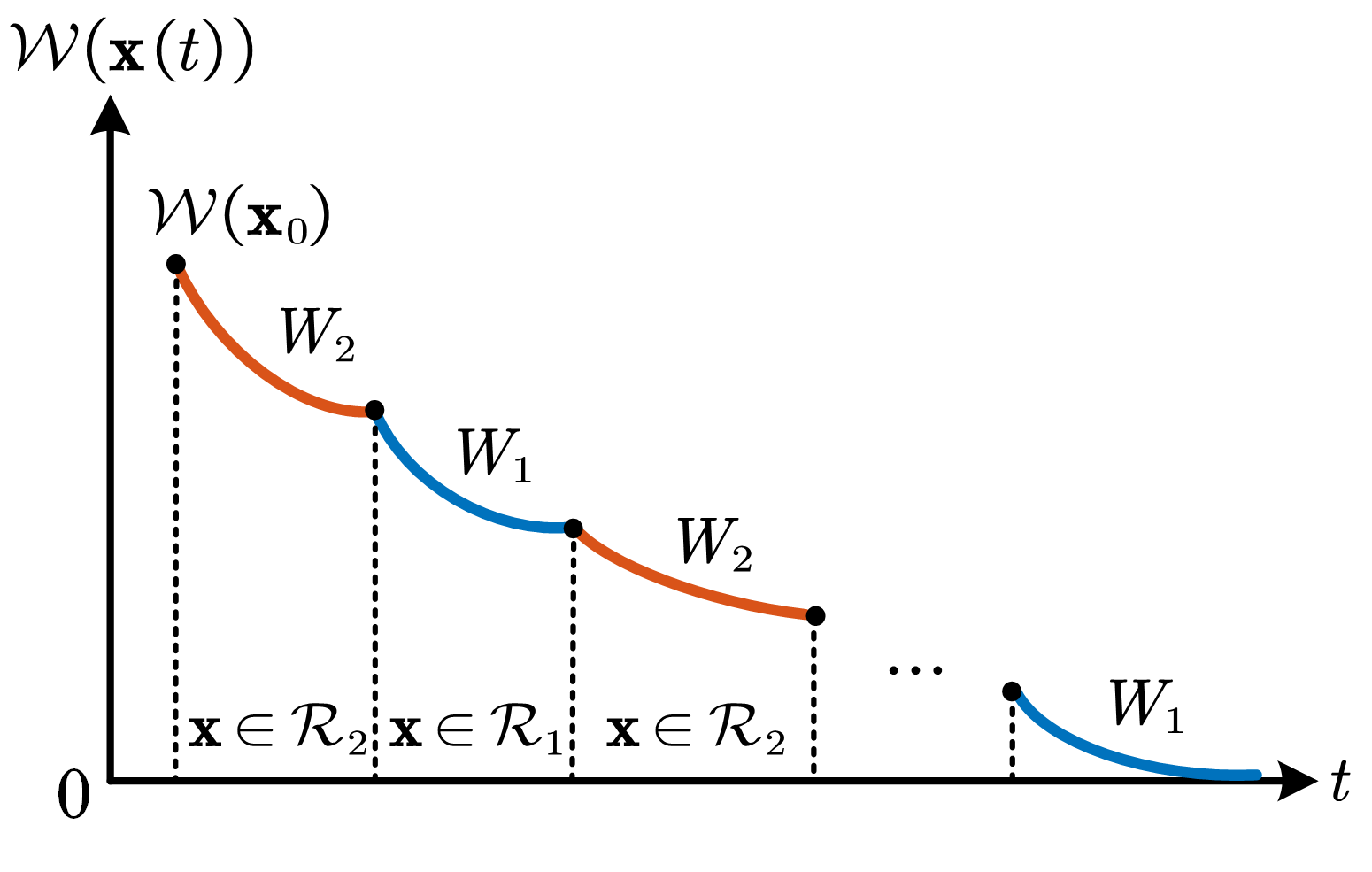}}
}
\vspace{-2pt}
\caption{Illustration of ROA estimation using the proposed SdS energy function.}
\label{SdSEF}
\vspace{-8pt}
\end{figure} 

Compared with the conventional energy function
\cite{chiang2011direct}, the proposed SdS energy function requires not only
that the energy decreases within each operating region, but
also that adjacent regional energy functions are continuous
across their common switching interface. Consequently,
$\mathcal W(\mathbf x)$ is a continuous and piecewise-$C^{1}$
function. Physically, switching only changes the analytical
expression used to evaluate the system energy, without
introducing any instantaneous energy jump.


LaSalle's invariance principle underpins ROA estimation using conventional energy functions \cite{nonlinearbook_Khalil} and has been extended to switched systems \cite{shevitz2002lyapunov}. Under the strict-crossing premise, its application to the proposed SdS energy function yields the following result.

\begin{theorem}[LaSalle principle for SdS energy function]
\label{thm:sdsef_lasalle}
Consider system \eqref{eq:sdss} under the strict-crossing
premise, and let $\mathcal{W}$ be an SdSEF satisfying
Definition~\ref{def:sdsef}. Let $\mathbf{x}_{\rm sep}$ denote
the SEP. For a compact and connected
sublevel set
\begin{equation}
\Omega_c
:=
\left\{
\mathbf{x}\in\mathbb{R}^{n}:
\mathcal{W}(\mathbf{x})\leq c
\right\},
\label{eq:sdsef_sublevel_set}
\end{equation}
define the zero-derivative set as
\begin{equation}
\mathcal{Z}_{c}
:=
\bigcup_{p\in\mathcal{P}}
\left\{
\mathbf{x}\in
\Omega_c\cap\overline{\mathcal{R}}_{p}:
\dot W_p(\mathbf{x})=0
\right\}.
\label{eq:sdsef_zero_derivative_set}
\end{equation}
If $\{\mathbf{x}_{\rm sep}\}$ is the largest invariant subset
of $\mathcal{Z}_{c}$, then every state $\mathbf{x}_0$ starting in stability domain $\Omega_c$ converges to the SEP, i.e.,
\begin{equation}
\lim_{t\rightarrow\infty}\mathbf{x}(t)
=
\mathbf{x}_{\rm sep},
\qquad
\forall\,\mathbf{x}_0\in\Omega_c.
\label{eq:sdsef_convergence}
\end{equation}
\end{theorem}
\begin{proof}
See Appendix~\ref{app:sdsef_lasalle}.
\end{proof}

In Theorem~\ref{thm:sdsef_lasalle}, $\Omega_c$ provides an estimate of the ROA of the SEP, $A(\mathbf{x}_{\rm sep})$, as illustrated in \figref{SdSEF}(a). 
Following the UEP-based approach in conventional
energy-function \cite{chiang2011direct},
the energy of the relevant UEP is selected as the critical energy for ROA estimation, and
$\Omega_c
=
\left\{
\mathbf{x}:
\mathcal{W}(\mathbf{x})
\le
\mathcal{W}(\mathbf{x}_{\rm uep})
\right\}$. 

The proposed SdS energy function is a special case of the multiple-Lyapunov-function approach \cite{liberzon2003switching}, with regional functions joined into a single continuous energy function, thereby enabling direct ROA estimation using its sublevel sets. This distinguishes it from what is termed an SdS energy function in \cite{liu2020state}, which permits energy jumps at switching. 
As illustrated in \figref{SdSEF}(b), the SdS energy function is non-increasing within each operating mode and remains continuous as trajectories cross switching interfaces under the strict-crossing premise. Regardless of how the system switches between operating modes, every trajectory starting within the estimated ROA converges to the SEP.



\subsection{SdS energy function under current limiting strategies}
\begin{figure}[t!]
    \centering
    \vspace{-3pt}
    \includegraphics[width=0.45\textwidth]{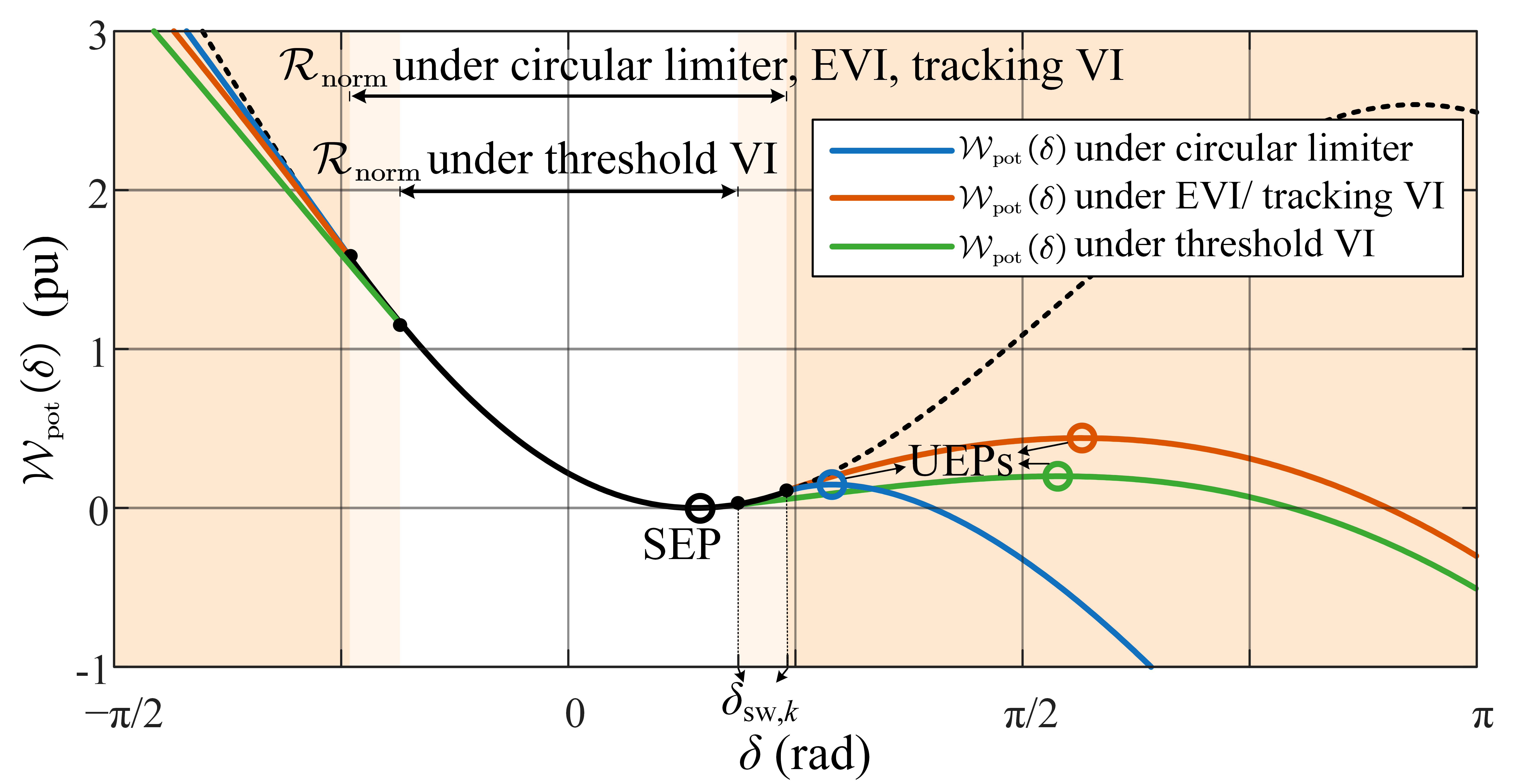}
    \vspace{-9pt}
    \caption{Potential-energy curves $\mathcal{W}_{\mathrm{pot}}(\delta)$. The color scheme is the same as in \figref{Pe_curve}. Colors denote different current-limiting strategies, and matching open circles mark the corresponding system UEPs. The black solid curve is shared by all strategies in the normal operating region, whereas the black dashed curve represents the conventional potential-energy function without current limiting.}
    \label{Wpot_curve}	
    \vspace{-10pt}
\end{figure}

To construct SdS energy function, regional energy functions are constructed in the normal operating region $\mathcal{R}_{\mathrm{norm}}$ and the current-limited operating region $\mathcal{R}_{\mathrm{lim}}$ separately, and matched at their interfaces $\delta_{\text{sw}}$ to satisfy the continuity condition in Definition~\ref{def:sdsef}. The SdS energy functions for two grid-forming strategies are constructed below.

1) For a classical GFM inverter, the potential energy in
$\mathcal{R}_{\mathrm{norm}}$ retains the conventional form\footnote{With
the parameters in \tabref{TableA}, all three strategies
have the same total impedance in normal operation and
hence share the same SEP and normal-mode dynamics.}.
Using \eqref{Pe_norm} yields
\begingroup
\small
\begin{equation}
\begin{aligned}
	\mathcal{W} _{\mathrm{pot}}(\delta )=&\int{\left[ P_{e,\mathrm{norm}}(\delta )-P_{\mathrm{ref}} \right] \mathrm{d}\delta}\\
	=&\left( \frac{R_gE^2-R_{v,\mathrm{norm}}V_{g}^{2}}{|Z_{\Sigma ,\mathrm{norm}}|^2}-P_{\mathrm{ref}} \right) \delta \\
    &-EV_g\frac{X_{\Sigma ,\mathrm{norm}}\cos \delta -(R_{v,\mathrm{norm}}-R_g)\sin \delta}{|Z_{\Sigma ,\mathrm{norm}}|^2},\\
\end{aligned}
\label{eq:Wpot_norm}
\end{equation}
\endgroup
In $\mathcal{R}_{\mathrm{lim}}$, continuity at
$\delta=\pm\delta_{\mathrm{sw}}$ is enforced by matching
the potential energy to its normal-mode boundary values:
\begingroup
\small
\begin{equation}
\begin{aligned}
&\mathcal{W}_{\mathrm{pot}}(\delta)=
\\[-2pt]
&\quad
\left\{
\begin{array}{@{}l@{\;}l@{}}
\displaystyle
\mathcal{W}_{\mathrm{pot}}(\delta_{\mathrm{sw}})
+\int_{\delta_{\mathrm{sw}}}^{\delta}
\left[P_{e,\lim}(\delta)-P_{\mathrm{ref}}\right]\mathrm{d}\delta,
& \delta\geq\delta_{\mathrm{sw}},
\\[6pt]
\displaystyle
\mathcal{W}_{\mathrm{pot}}(-\delta_{\mathrm{sw}})
+\int_{-\delta_{\mathrm{sw}}}^{\delta}
\left[P_{e,\lim}(\delta)-P_{\mathrm{ref}}\right]\mathrm{d}\delta,
& \delta\leq-\delta_{\mathrm{sw}}.
\end{array}
\right.
\end{aligned}
\label{eq:Wpot_lim}
\end{equation}
\endgroup
Together, \eqref{eq:Wpot_norm} and \eqref{eq:Wpot_lim}
define a piecewise potential-energy function that is
continuous across the switching interfaces. 
Although the analytical representation of \eqref{eq:Wpot_lim}
may involve standard elliptic integrals, these can be evaluated efficiently using numerical routines or precomputed lookup tables. 
The potential-energy curves for the three current-limiting
strategies represented by different colors are shown in different colors in \figref{Wpot_curve}.

\begin{figure}[t!]
\centering
\vspace{-3pt}
{\captionsetup[subfloat]{captionskip=-1pt}
\subfloat[Classical GFM.]{\includegraphics[width=0.45\textwidth]{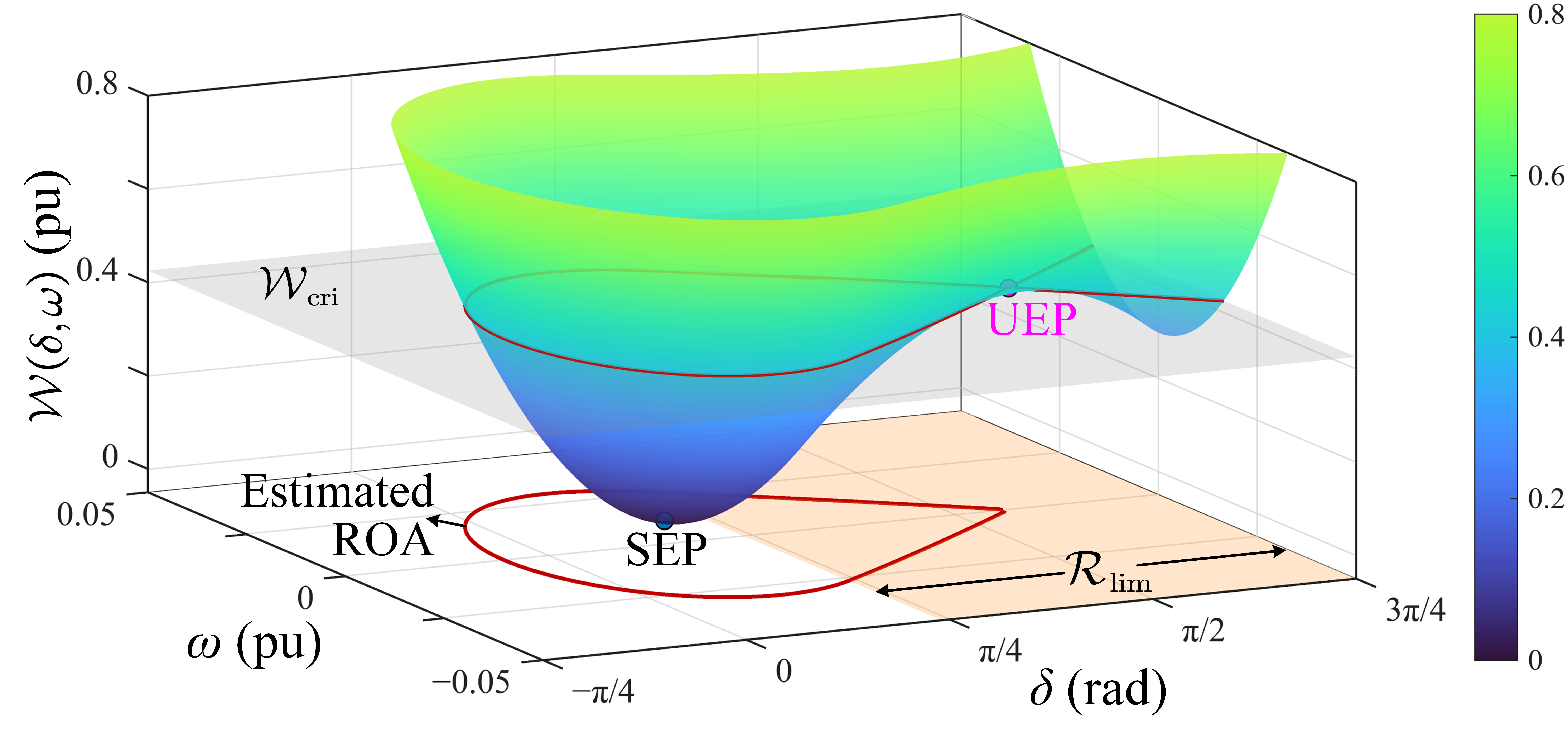}}
}\\
\vspace{-3pt}
{\captionsetup[subfloat]{captionskip=-1pt}
\makebox[\linewidth][c]{%
    \hspace*{1mm}%
    \subfloat[DC-voltage-controlled GFM.]{\includegraphics[width=0.45\textwidth]{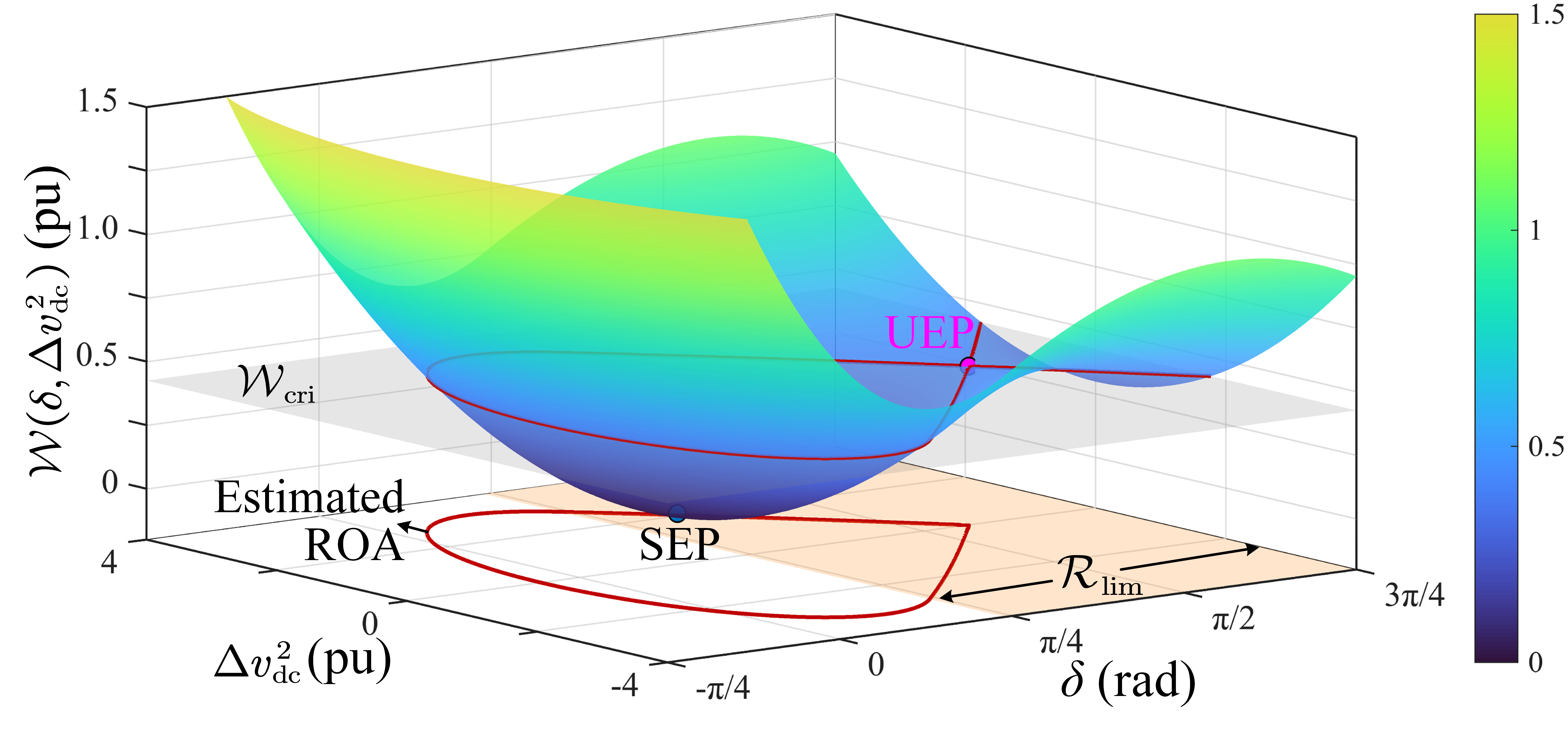}}%
}
}
\vspace{-6pt}
\caption{SdS energy functions and corresponding ROA estimates, using the EVI current-limiting strategy as an illustrative example.}
\label{ROA3d}
\vspace{-8pt}
\end{figure} 

The kinetic energy of the classical GFM inverter is
\begin{equation}
\mathcal{W}_{\mathrm{kin}}(\omega)
=
\frac{1}{2}J\omega^2,
\label{eq:Wkin_GFM}
\end{equation}
which is continuous throughout the phase plane.
The resulting SdS energy function is therefore
\begin{equation}
\mathcal{W}(\delta,\omega)
=
\mathcal{W}_{\mathrm{kin}}(\omega)
+
\mathcal{W}_{\mathrm{pot}}(\delta).
\label{eq:W_GFM}
\end{equation}
In both operating regions, its derivative satisfies
\begin{equation}
\dot{\mathcal{W}}
=
-D\omega^2
\leq 0,
\label{eq:Wdot_GFM}
\end{equation}
fulfilling the remaining condition in Definition~\ref{def:sdsef}.

Therefore, the relevant boundary UEP determines the critical energy
$\mathcal{W}_{\mathrm{cri}}
=\mathcal{W}(\mathbf{x}_{\rm uep})$.
The connected component of
$\{\mathbf{x}:\mathcal{W}(\mathbf{x})
<\mathcal{W}_{\mathrm{cri}}\}$
containing the normal-operation SEP provides an ROA estimate.
Taking EVI as an example, \figref{ROA3d}(a) shows the
continuous energy surface with the SEP at the bottom of
the energy well. The projected red critical-energy contour encloses the
estimated ROA, which spans both $\mathcal{R}_{\mathrm{norm}}$
and $\mathcal{R}_{\mathrm{lim}}$ (shaded in orange).

2) For the DC-voltage-controlled GFM inverter, the
PI-controller energy formulation in \cite{11540391}
gives the same potential-energy construction in
\eqref{eq:Wpot_norm} and \eqref{eq:Wpot_lim}, with
$P_{\mathrm{ref}}$ replaced by $P_{\mathrm{in}}$.
Thus, the curves in \figref{Wpot_curve} also apply.
The kinetic-energy term becomes
\begin{equation}
\mathcal{W}_{\mathrm{kin}}(\Delta v_{\mathrm{dc}}^2)
=
\frac{C_{\mathrm{dc}}k_i}{4}
\left(\Delta v_{\mathrm{dc}}^2\right)^2,
\label{eq:Wkin_VFM}
\end{equation}
which remains continuous across the switching interfaces.
The corresponding SdS energy function is
\begin{equation}
\mathcal{W}(\delta,\Delta v_{\mathrm{dc}}^2)
=
\mathcal{W}_{\mathrm{kin}}(\Delta v_{\mathrm{dc}}^2)
+
\mathcal{W}_{\mathrm{pot}}(\delta).
\label{eq:W_VFM}
\end{equation}
Using \eqref{VFMe}, its derivative is nonpositive in both
operating regions:
\begin{equation}
\dot{\mathcal{W}}
=
\begin{cases}
-k_p\left[P_{\mathrm{in}}
-P_{e,\mathrm{norm}}(\delta)\right]^2,
& \delta\in\mathcal{R}_{\mathrm{norm}},
\\[3pt]
-k_p\left[P_{\mathrm{in}}
-P_{e,\lim}(\delta)\right]^2,
& \delta\in\mathcal{R}_{\mathrm{lim}}.
\end{cases}
\label{eq:Wdot_VFM}
\end{equation}
As illustrated in \figref{ROA3d}(b) for EVI current-limiting strategy, the projected
red contour encloses the estimated ROA in the
$(\delta,\Delta v_{\mathrm{dc}}^2)$ plane, extending across
both normal and current-limited operating regions.

The analysis shows that the proposed method requires no prior knowledge of the
current-limit activation and release sequence.
Another key advantage is that the energy function in
\eqref{eq:W_GFM} or \eqref{eq:W_VFM} is analytical and explicitly characterizes
the ROA estimate, establishing a direct quantitative link
to system parameters.
The energy is calculated directly from the state,
avoiding time-consuming trajectory integration and
iterative computation of the stability boundary.

\begin{figure}[t!]
\centering
\vspace{-3pt}
{\captionsetup[subfloat]{captionskip=-1pt}
\subfloat[Circular limiter.]{\includegraphics[width=0.44\textwidth]{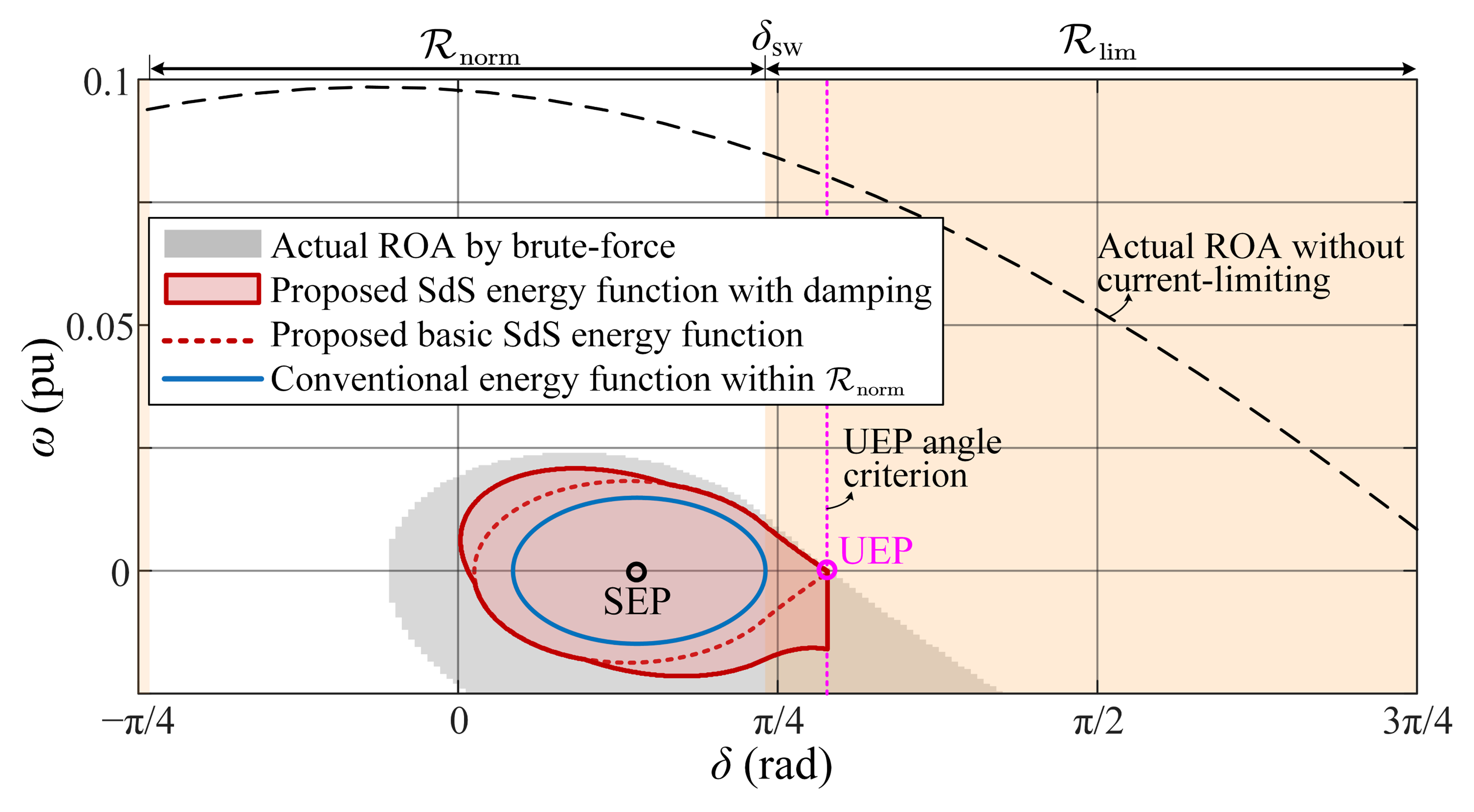}}
}\\
\vspace{-3pt}
{\captionsetup[subfloat]{captionskip=-1pt}
\makebox[\linewidth][c]{%
    \subfloat[EVI.]{\includegraphics[width=0.44\textwidth]{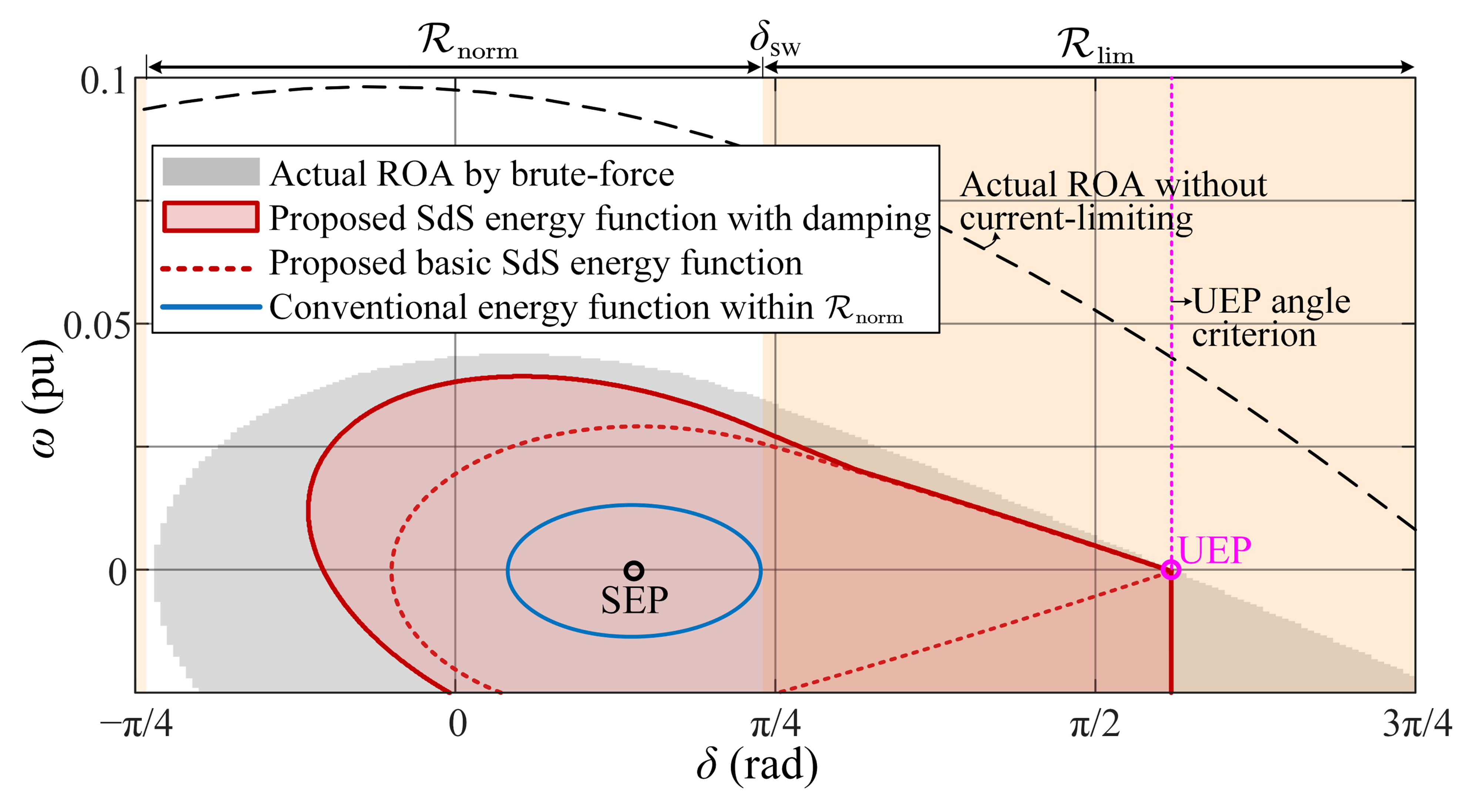}}%
}
}
\vspace{-3pt}
{\captionsetup[subfloat]{captionskip=-1pt}
\makebox[\linewidth][c]{%
    \hspace*{-1.2mm}%
    \subfloat[Threshold VI.]{\includegraphics[width=0.44\textwidth]{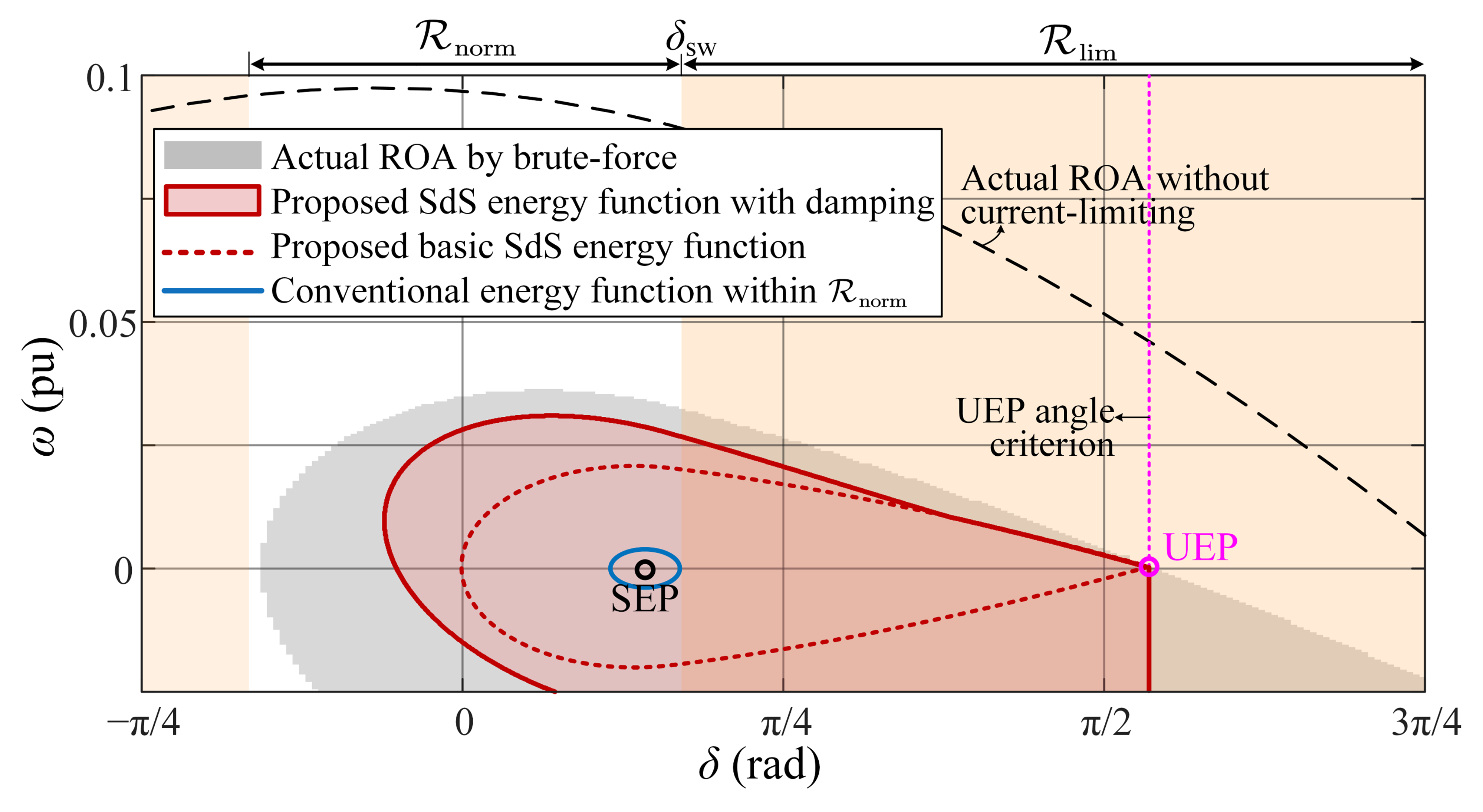}}%
}
}
\vspace{-6pt}
\caption{ROA estimation for classical GFM inverters under different current-limiting strategies: effect of damping-energy incorporation and comparison with existing methods.}
\label{GFMROA}
\vspace{-8pt}
\end{figure}

\begin{figure*}[t!]
\centering
\vspace{-5pt}
{\captionsetup[subfloat]{captionskip=0pt}
\subfloat[Cirular limiter in $\mathcal{R} _{\lim}$.]{\includegraphics[width=0.3\textwidth]{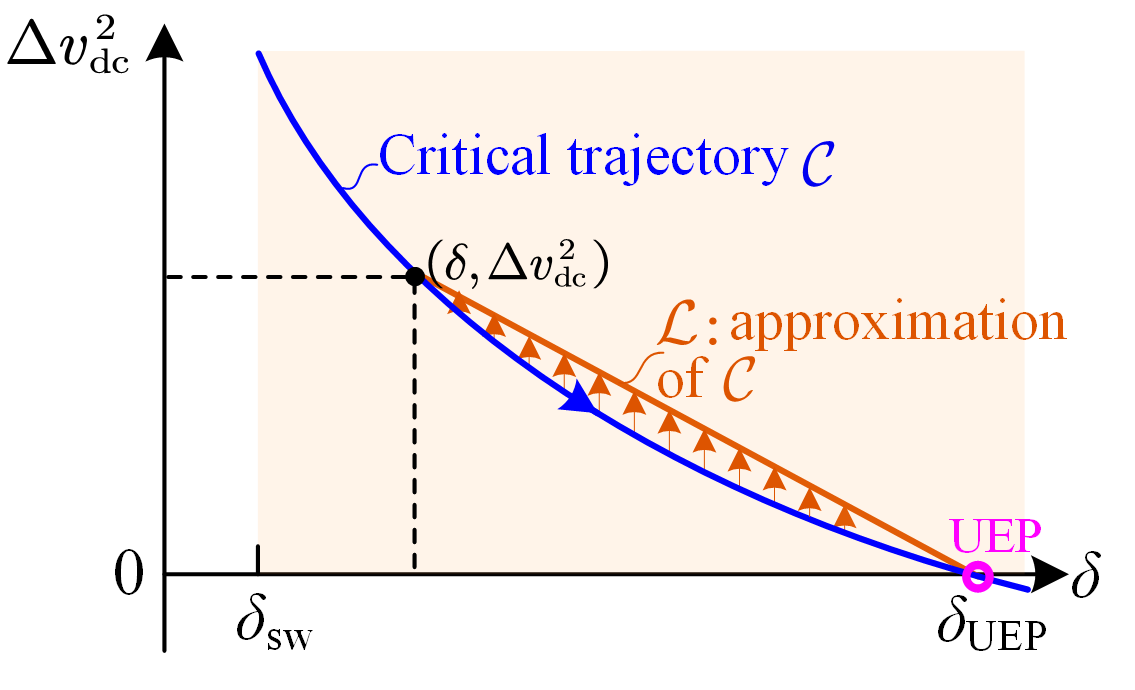}}}
\hfil
{\captionsetup[subfloat]{captionskip=0pt}
\subfloat[EVI strategy in $\mathcal{R} _{\lim}$.]{\includegraphics[width=0.3\textwidth]{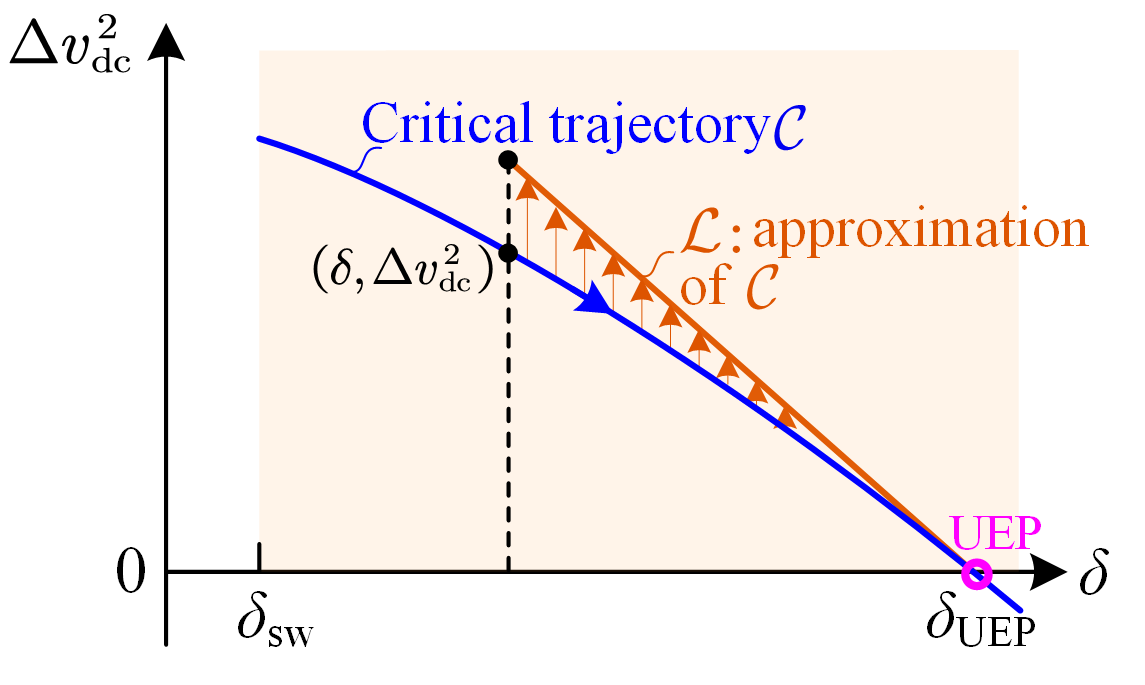}}}
\hfil
{\captionsetup[subfloat]{captionskip=0pt}
\subfloat[all strategies in $\mathcal{R} _{\mathrm{norm}}$.]{\includegraphics[width=0.32\textwidth]{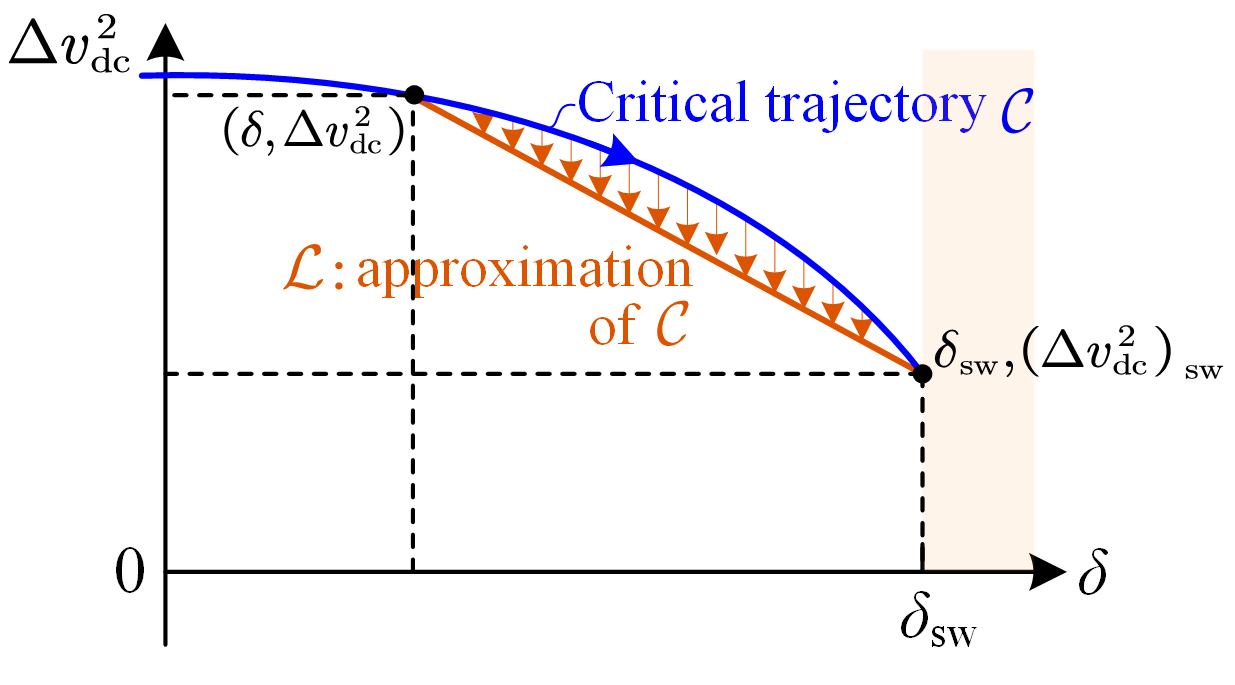}}}
\vspace{-3pt}
\caption{CTA-based damping-energy estimation. Blue curves
and orange lines denote the critical trajectory $\mathcal C$
and its linear approximation $\mathcal L$, respectively.}
\vspace{-10pt} 
\label{CTA}
\end{figure*}

\begin{figure*}[t!]
\centering
\vspace{-5pt}
{\captionsetup[subfloat]{captionskip=0pt}
\subfloat[Cirular limiter.]{\includegraphics[width=0.32\textwidth]{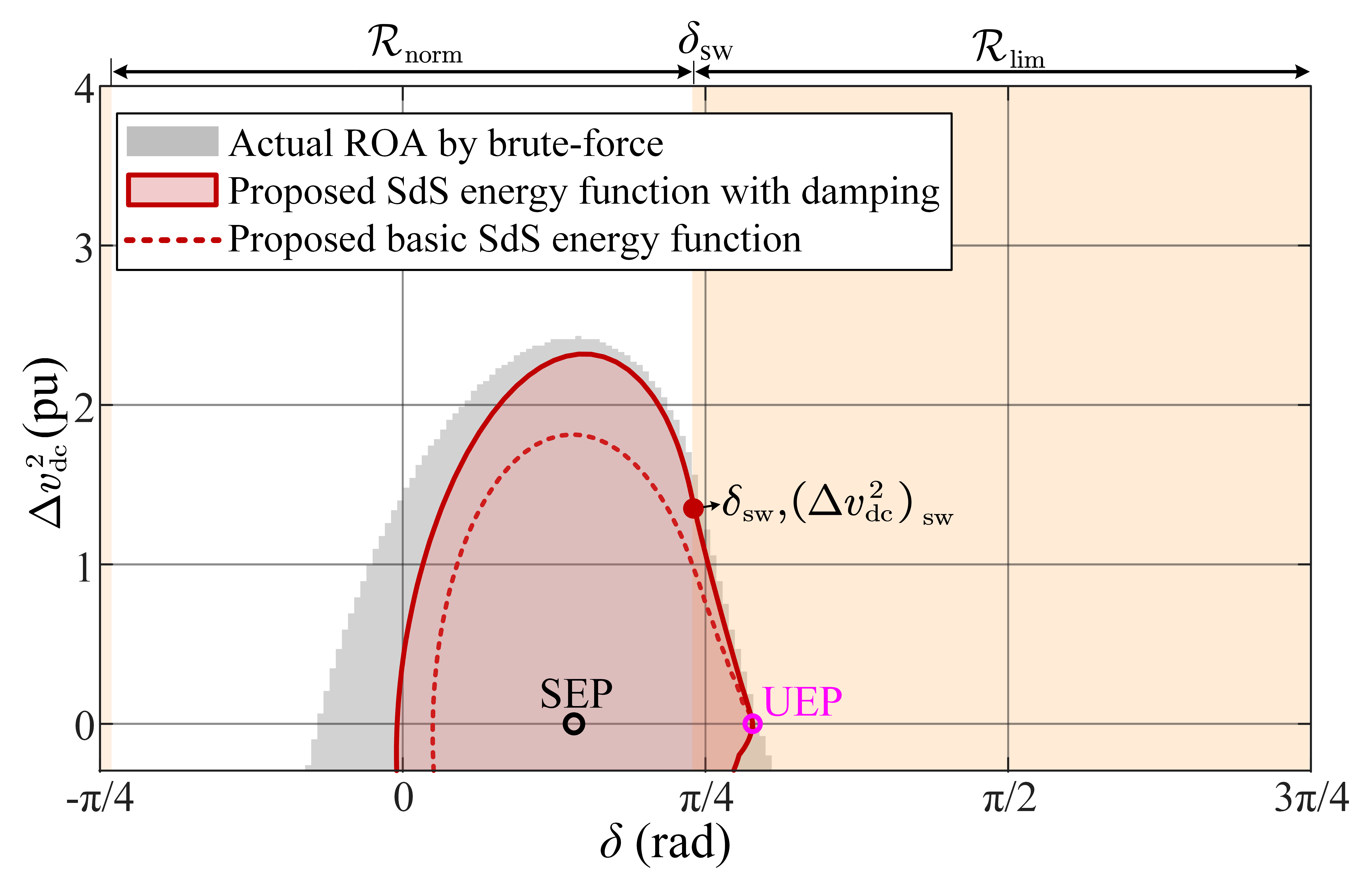}}}
{\captionsetup[subfloat]{captionskip=0pt}
\subfloat[EVI strategy .]{\includegraphics[width=0.32\textwidth]{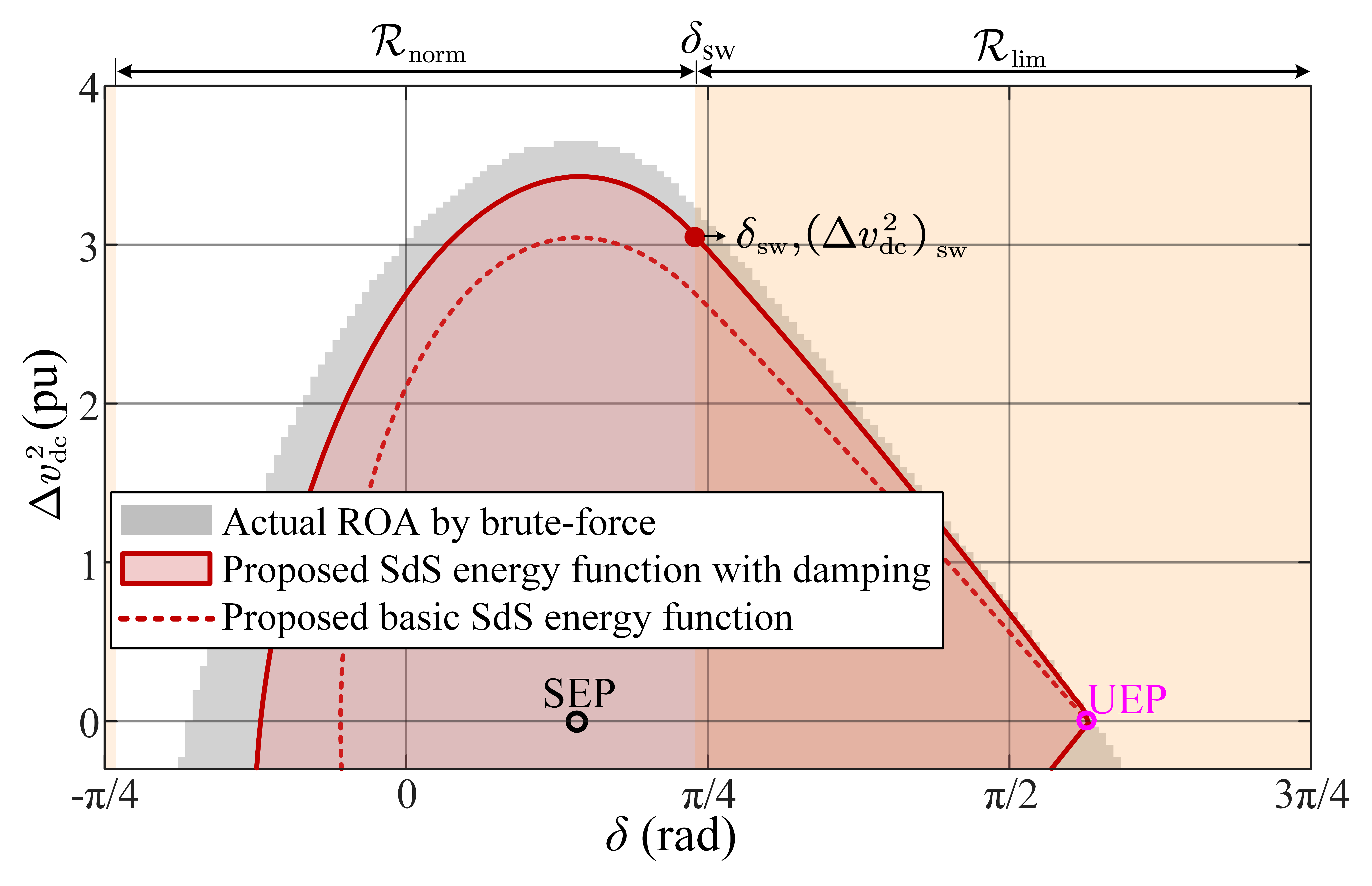}}}
{\captionsetup[subfloat]{captionskip=0pt}
\subfloat[Threshold VI strategy.]{\includegraphics[width=0.32\textwidth]{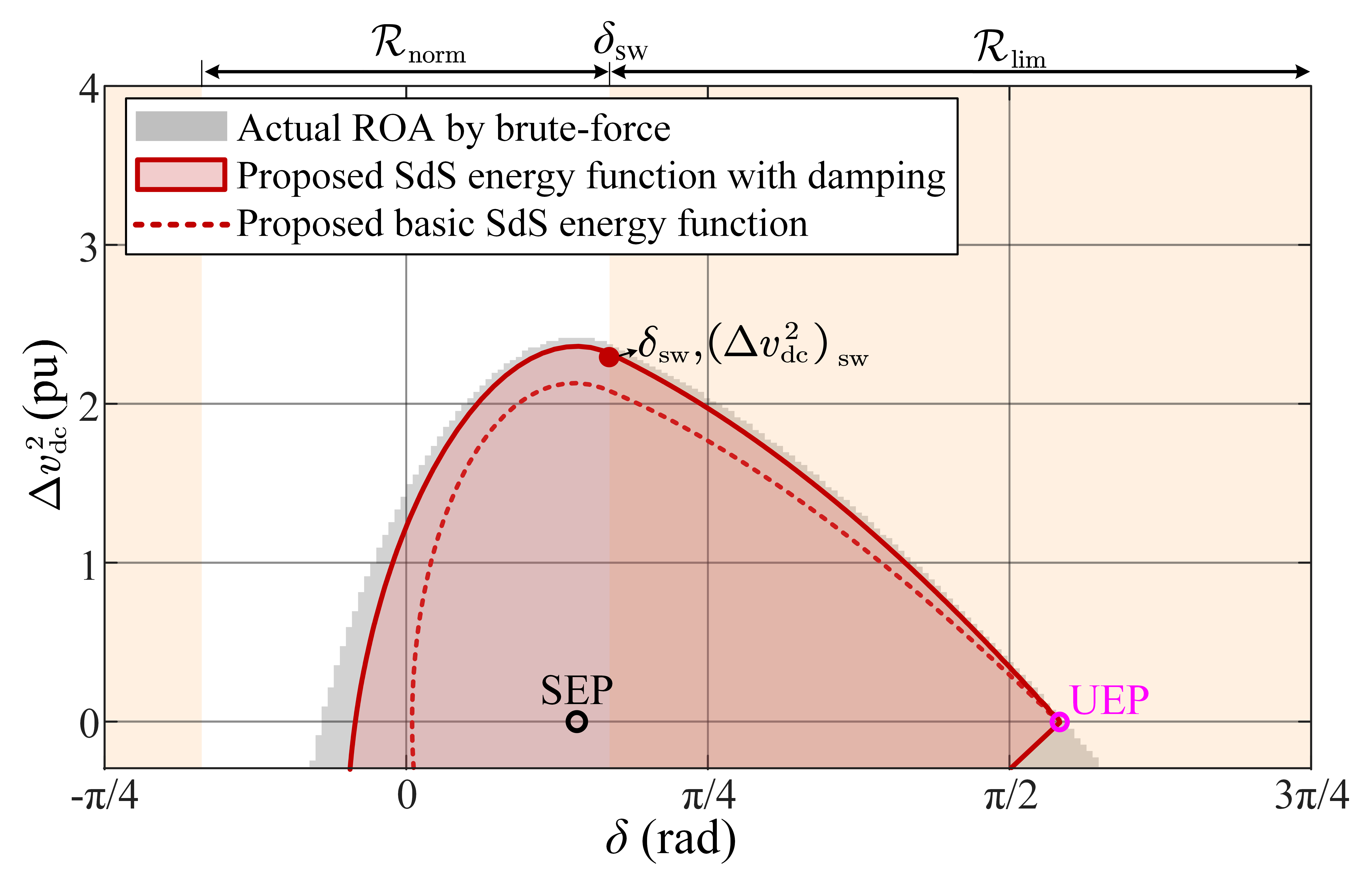}}}
\vspace{-3pt}
\caption{ROA estimation using the basic SdS energy function
with and without CTA-based damping-energy incorporation.}
\vspace{-10pt} 
\label{VFMROA}
\end{figure*}

\subsection{Refinement through damping-energy incorporation}
The proposed SdS energy function is further refined by incorporating damping energy to obtain less conservative ROA estimates, given the large damping in GFM inverters.
\subsubsection{Estimation for classical GFM inverters} \mbox{}\\ \hspace*{\parindent}
Damping-energy estimation was developed for power systems
in \cite{moon2000estimating} and subsequently adopted in
GFM transient-stability studies, e.g.,
\cite{shuai2018transient}. Following this
construction, the damping contribution is estimated as
\begin{equation}
\mathcal{W}_{\mathrm{d},\lambda}(\delta,\omega)
=
\lambda D\omega(\delta-\delta_{\mathrm{SEP}})
+\frac{\lambda D^2}{2J}(\delta-\delta_{\mathrm{SEP}})^2,
\label{eq:Wd_lambda}
\end{equation}
where $\delta_{\mathrm{SEP}}$ denotes the power angle at the SEP and
$\lambda\in[0,1]$ is an estimation parameter. Incorporating
\eqref{eq:Wd_lambda} into the preceding construction gives
\begin{equation}
\mathcal{W}_{\lambda}(\delta,\omega)
=
\mathcal{W}(\delta,\omega)
+\mathcal{W}_{\mathrm{d},\lambda}(\delta,\omega),
\label{eq:W_lambda}
\end{equation}
with $\lambda=0$ recovering the basic SdS energy function. The damping term $\mathcal{W}_{\mathrm{d},\lambda}$
is continuous across the switching interfaces as required by Definition~\ref{def:sdsef}.
Using \eqref{GFMe}, the regional time derivative of
$\mathcal{W}_{\lambda}$ is
\begin{equation}
\begin{aligned}
\dot{\mathcal{W}}_{\lambda}
={}&-(1-\lambda)D\omega^2
+\frac{\lambda D}{J}(\delta-\delta_{\mathrm{SEP}})\\
&\times
\begin{cases}
P_{\mathrm{ref}}-P_{e,\mathrm{norm}}(\delta),
& \delta\in\mathcal{R}_{\mathrm{norm}},\\
P_{\mathrm{ref}}-P_{e,\lim}(\delta),
& \delta\in\mathcal{R}_{\lim}.
\end{cases}
\end{aligned}
\label{eq:Wdot_lambda}
\end{equation}
As for a stable system, the range of $\delta$ is $[\delta_{\mathrm{UEP}}-2\pi,\delta_{\mathrm{UEP}}]$, where
$(\delta-\delta_{\mathrm{SEP}})[P_{\mathrm{ref}}-P_{e,\mathrm{norm}}(\delta)]\leq0$
and
$(\delta-\delta_{\mathrm{SEP}})[P_{\mathrm{ref}}-P_{e,\lim}(\delta)]\leq0$
hold in $\mathcal{R}_{\mathrm{norm}}$ and
$\mathcal{R}_{\lim}$, respectively, as illustrated in
\figref{Pe_curve}.
The second term in \eqref{eq:Wdot_lambda} is therefore
nonpositive, giving $\dot{\mathcal{W}}_{\lambda}\leq0$. The ROA estimates for different $\lambda$ values are
combined by taking their union, with the critical energy
taken as $\mathcal{W}_{\lambda}(\delta_{\mathrm{UEP}},0)$
as described above.

The SdS energy function incorporating damping yields the
ROA estimates shown by the red solid boundaries and enclosed
shading in \figref{GFMROA}.
Compared with the basic SdS energy estimates
(red dashed curves), incorporating damping enlarges the
estimated regions and reduces conservatism.
The gray regions represent the actual ROA obtained by brute-force
grid search, and the proposed estimates lie within actual ROA in all three cases.

The comparison in \figref{GFMROA} also shows that retaining
current limiting substantially reduces the post-fault ROA
relative to that predicted when current limiting is
neglected (black dashed curves).
This means that restoration of the grid voltage alone
does not restore the ROA predicted by an
unrestricted normal-mode model. 

With identical normal-mode dynamics, EVI current-limiting strategy yields a larger ROA than the circular limiter, demonstrating the benefit of
its predominantly inductive equivalent impedance over
the variable resistive equivalent of the circular limiter.
For threshold VI strategy, $I_{\mathrm{th}}<I_{\mathrm{lim}}$
leads to earlier impedance activation and a narrower
$\mathcal{R}_{\mathrm{norm}}$.
This earlier intervention reduces utilization of the
available current capability, giving a smaller ROA than
EVI despite the same virtual-impedance $X/R$ ratio $\sigma$.

\figref{GFMROA} also compares existing transient-stability
assessment methods that account for current limiting.
Using the conventional energy function constructed from
the normal-mode dynamics \cite{fan2024transient,yang2025global,wang2025transient}
yields much smaller ROA estimates (blue curves), which
must be confined to $\mathcal{R}_{\mathrm{norm}}$ and
cannot account for current-limited operation or
switching between modes.
The UEP-angle criterion
\cite{fan2022equivalent,zhan2024transient,8026165}
(magenta dashed lines) is nonconservative when applied
to the GFM model containing inertia: kinetic energy can carry a
trajectory beyond $\delta_{\mathrm{UEP}}$ even when its
initial angle lies below it.

\subsubsection{CTA-based estimation for DC-voltage-controlled GFM inverters}
\mbox{}\\ \hspace*{\parindent}
For the DC-voltage-controlled GFM inverter, equivalent damping is generated by $k_p$ instead of $D$ in classic GFM \cite{11540391}.
The energy dissipated along the critical trajectory
$\mathcal C$ from $(\delta,\Delta v_{\mathrm{dc}}^2)$ approaching UEP is
\begin{equation}
E_d(\delta,\Delta v_{\mathrm{dc}}^2)=
\frac{k_pC_{\mathrm{dc}}}{2}
\int_{\mathcal C}
[P_{\mathrm{in}}-P_e(\delta)]
\,\mathrm{d}(\Delta v_{\mathrm{dc}}^2),
\label{eq:damping_energy}
\end{equation}
By energy balance, the stability boundary satisfies
\begin{equation}
\mathcal W(\delta,y)-E_d(\delta,y)
=\mathcal W_{\mathrm{cri}},
\qquad
\mathcal W_{\mathrm{cri}}
=\mathcal W(\delta_{\mathrm{UEP}},0).
\label{eq:critical_energy_balance}
\end{equation}
Following \cite{Qu2024,lei2023transient}, critical trajectory approximation (CTA) is adopted to further refine the basic ROA estimate. The underlying idea is to use geometric bounds on $\mathcal C$ to represent the path-dependent integral by an analytical expression that depends only on the state and is therefore path-independent.

Integration by parts gives
\begin{equation}
\begin{aligned}
E_d(\delta,\Delta v_{\mathrm{dc}}^2)&=
\frac{k_pC_{\mathrm{dc}}}{2}
\bigl\{
-[P_{\mathrm{in}}-P_e(\delta)]\Delta v_{\mathrm{dc}}^2
+\mathcal B(\delta)
\bigr\},\\
\mathcal B(\delta):&=
\int_{\delta}^{\delta_{\mathrm{UEP}}}
(\Delta v_{\mathrm{dc}}^2)(\gamma)
P_e'(\gamma)\,\mathrm{d}\gamma,
\end{aligned}
\label{eq:damping_energy_ibp}
\end{equation}
where only $\mathcal B$ is path-dependent, and a lower bound on $\mathcal B$ gives a conservative damping-energy estimate.
For the circular limiter and EVI strategies, which are
the focus here, $P_{e,\lim}'(\delta)<0$ along the critical
trajectory in $\mathcal R_{\lim}$, so upper bounds on
$(\Delta v_{\mathrm{dc}}^2)_{\mathcal C}$ are used to
approximate $\mathcal B$, as developed below.

\begin{lemma}
\label{lem:cta_curvature}
For \eqref{VFMe}, consider the critical trajectory approaching
the UEP. Over $\delta_{\mathrm{sw}}<\delta<\delta_{\mathrm{UEP}}$,
this trajectory is convex (concave) if $P_{e,\lim}(\delta)$
is convex (concave) throughout the same interval.
\end{lemma}

\begin{proof}
See Appendix~\ref{app:cta_curvature}.
\end{proof}

For the circular limiter, the convexity of
$P_{e,\lim}(\delta)$ implies a convex critical trajectory
by Lemma~\ref{lem:cta_curvature}.
The secant $\mathcal L$ joining
$(\delta,\Delta v_{\mathrm{dc}}^2)\in\mathcal C$
to the UEP therefore lies above $\mathcal C$,
as shown in \figref{CTA}(a).
The integral term in \eqref{eq:damping_energy_ibp}
is bounded by
\begingroup
\edef\savedbaselineskip{\the\baselineskip} 
\small
\setlength{\baselineskip}{\savedbaselineskip} 
\begin{equation}
\begin{aligned}
\mathcal B_{\lim}(\delta)
\geq&
\widehat{\mathcal B}_{\lim}
(\delta,\Delta v_{\mathrm{dc}}^2)\\
&=
\frac{\Delta v_{\mathrm{dc}}^2}
{\delta_{\mathrm{UEP}}-\delta}
\int_{\delta}^{\delta_{\mathrm{UEP}}}
(\delta_{\mathrm{UEP}}-\gamma)
P_{e,\lim}'(\gamma)\,\mathrm{d}\gamma.
\end{aligned}
\label{eq:B_sec_definition}
\end{equation}
\endgroup
For the EVI strategy, the critical trajectory is concave and lies
below its tangent at the UEP, as shown in \figref{CTA}(b).
Thus,
\begingroup
\edef\savedbaselineskip{\the\baselineskip} 
\small
\setlength{\baselineskip}{\savedbaselineskip} 
\begin{equation}
\begin{aligned}
\mathcal B_{\lim}(\delta)
\geq&
\widehat{\mathcal B}_{\lim}
(\delta,\Delta v_{\mathrm{dc}}^2)\\
&=
-m_{\mathrm{UEP}}
\int_{\delta}^{\delta_{\mathrm{UEP}}}
(\delta_{\mathrm{UEP}}-\gamma)
P_{e,\lim}'(\gamma)\,\mathrm{d}\gamma,
\end{aligned}
\label{eq:I_tangent}
\end{equation}
\endgroup
where
$m_{\mathrm{UEP}}
=-2P_{e,\lim}'(\delta_{\mathrm{UEP}})/
(C_{\mathrm{dc}}\lambda_s)<0$
is the slope of the critical trajectory at the UEP,
and $\lambda_s$ is the negative eigenvalue of
\eqref{VFMe} linearized at the UEP. In $\mathcal R_{\lim}$, substituting
$\widehat{\mathcal B}_{\lim}$ for $\mathcal B$ in
\eqref{eq:damping_energy_ibp} gives the damping-energy
estimate $\widehat E_d$. According to \eqref{eq:critical_energy_balance},
the estimated ROA boundary point at the switching
interface is obtained as
$(\delta_{\mathrm{sw}},
(\Delta v_{\mathrm{dc}}^2)_{\mathrm{sw}})$,
with the corresponding damping-energy estimate
denoted by $\widehat E_{d,\mathrm{sw}}$.

In normal operating region $\mathcal R_{\mathrm{norm}}$, the critical trajectory
is concave where $P_{e,\mathrm{norm}}'(\delta)>0$.
The secant joining $(\delta,\Delta v_{\mathrm{dc}}^2)$
to $(\delta_{\mathrm{sw}},
(\Delta v_{\mathrm{dc}}^2)_{\mathrm{sw}})$ provides the required lower bound\cite{lei2023transient}, as illustrated in \figref{CTA}(c). The damping-energy estimate in $\mathcal R_{\mathrm{norm}}$
then simplifies to
\begin{equation}
\begin{aligned}
\widehat E_d(\delta,\Delta v_{\mathrm{dc}}^2)
=&\widehat E_{d,\mathrm{sw}}
+\frac{k_pC_{\mathrm{dc}}}{2}
\frac{
(\Delta v_{\mathrm{dc}}^2)_{\mathrm{sw}}
-\Delta v_{\mathrm{dc}}^2
}{
\delta_{\mathrm{sw}}-\delta
}\\
&\times
\int_{\delta}^{\delta_{\mathrm{sw}}}
[P_{\mathrm{in}}-P_{e,\mathrm{norm}}(\gamma)]
\,\mathrm{d}\gamma.
\end{aligned}
\label{eq:Ed_norm_cta}
\end{equation}
Accordingly, the damping-energy estimates in
$\mathcal R_{\mathrm{norm}}$ and $\mathcal R_{\lim}$
join continuously at this switching point.
The resulting $\widehat E_d$ depends only on the system state.
The corresponding energy function is
\begin{equation}
\mathcal W_{\mathrm{CTA}}(\delta,\Delta v_{\mathrm{dc}}^2)
=\mathcal W(\delta,\Delta v_{\mathrm{dc}}^2)-\widehat E_d(\delta,\Delta v_{\mathrm{dc}}^2).
\label{eq:W_CTA}
\end{equation}

\figref{VFMROA} shows the resulting ROA estimates.
The basic SdS energy function already covers much
of the actual ROA (gray regions) obtained by brute-force
grid search, while incorporating damping estimated by CTA further reduces conservatism.
No overestimation is observed in any of the three cases shown.

The preceding development establishes the proposed
SdS energy-function methodology and its associated ROA estimate. It is worth noting that the ROA is an inherent property of the post-fault system dynamics which is independent of the disturbance or fault type. If the energy of the fault-clearing state is lower than the critical energy, the subsequent trajectory is predicted
to remain within the estimated ROA and converge to the SEP.

\subsection{Discussion of Parameter Effects on the Estimated ROA}
The proposed SdS energy function directly relates
system parameters to the estimated ROA, enabling
effective assessment of their influence on transient stability.

\begin{figure}[t!]
\centering
\vspace{-3pt}
{\captionsetup[subfloat]{captionskip=-1pt}
\subfloat[Virtual inertia $J$.]{%
\includegraphics[width=0.24\textwidth]{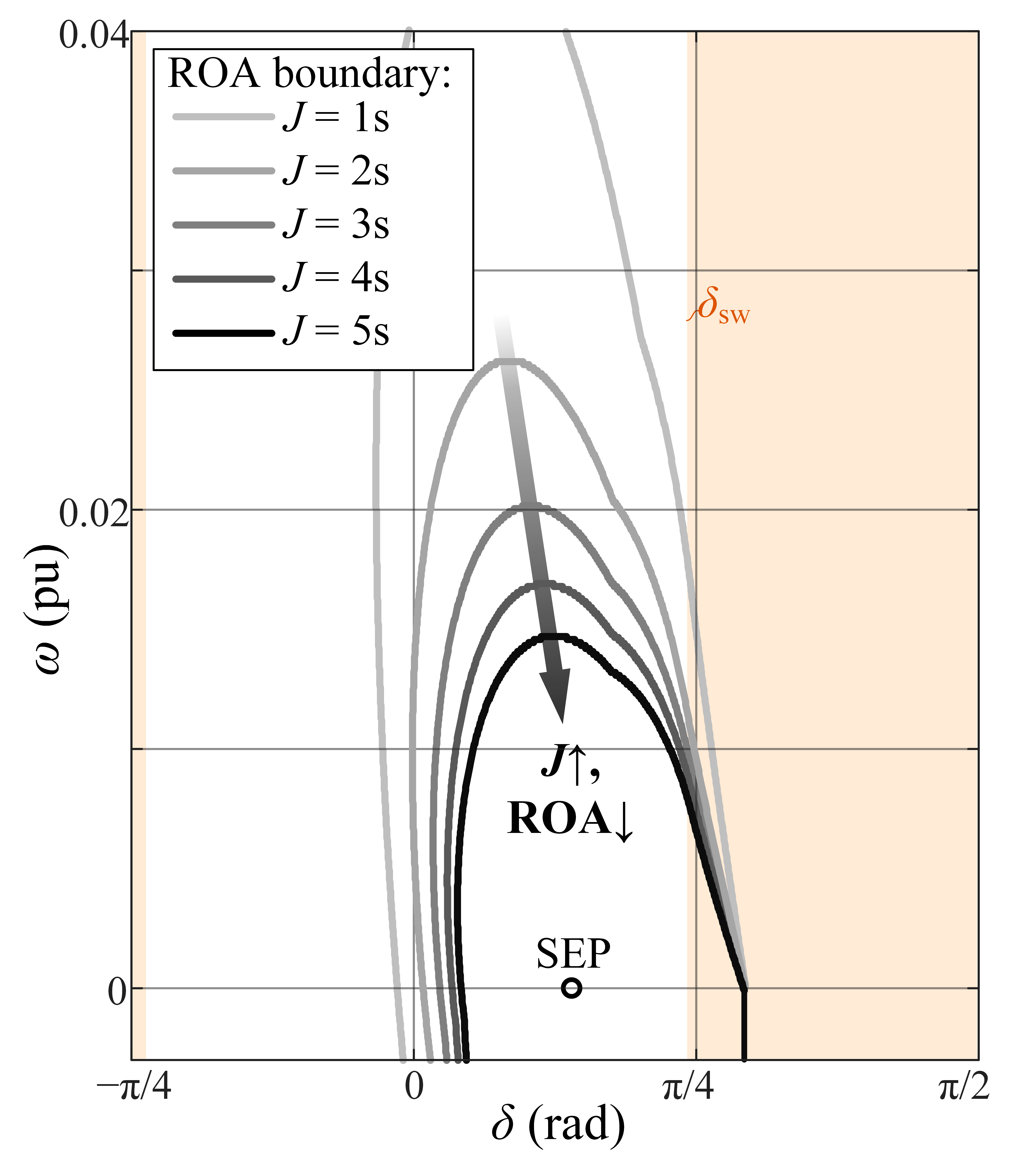}}
}
{\captionsetup[subfloat]{captionskip=-1pt}
\subfloat[Integral gain $k_i$.]{%
\includegraphics[width=0.24\textwidth]{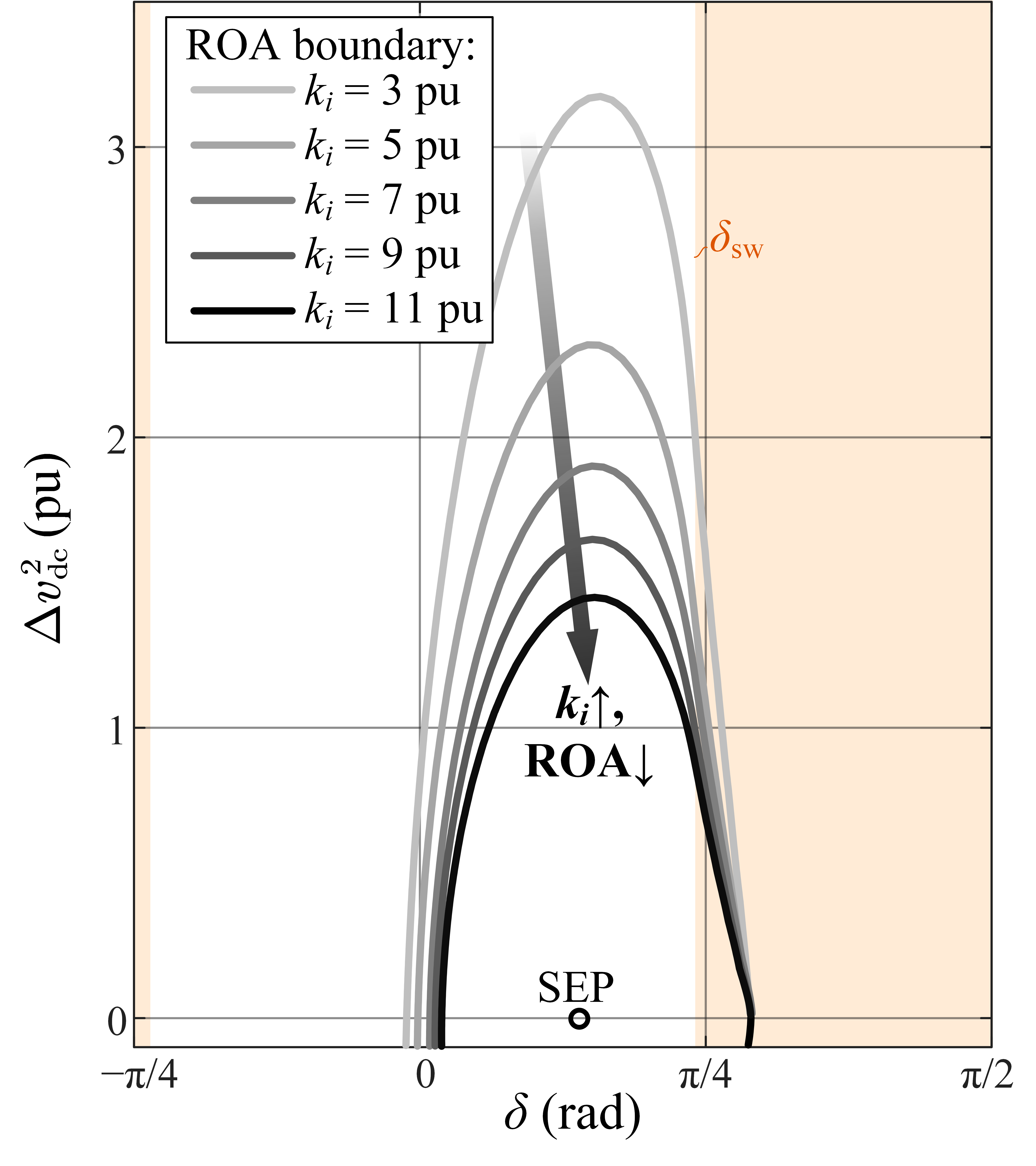}}%
}
\vspace{-3pt}
\caption{Effects of $J$ and $k_i$ on the estimated ROAs
of classical and DC-voltage-controlled GFM inverters,
respectively, using the circular limiter as an
illustrative example.}
\label{change1}
\vspace{-6pt}
\end{figure}

\begin{figure}[t!]
\centering
\vspace{-3pt}
{\captionsetup[subfloat]{captionskip=-1pt}
\subfloat[Damping coefficient $D$.]{%
\includegraphics[width=0.24\textwidth]{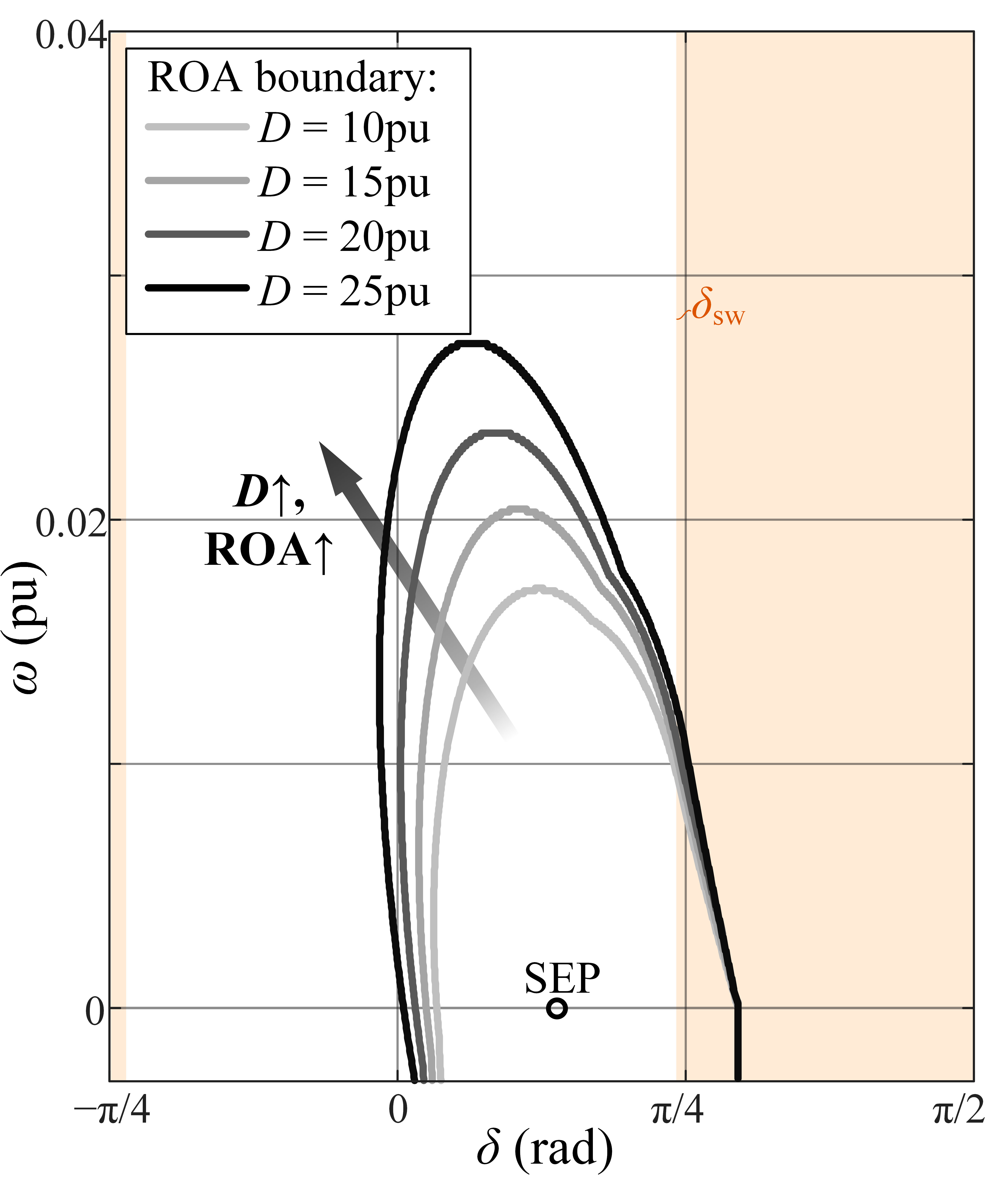}}
}
{\captionsetup[subfloat]{captionskip=-1pt}
\subfloat[Proportional gain $k_p$.]{%
\includegraphics[width=0.24\textwidth]{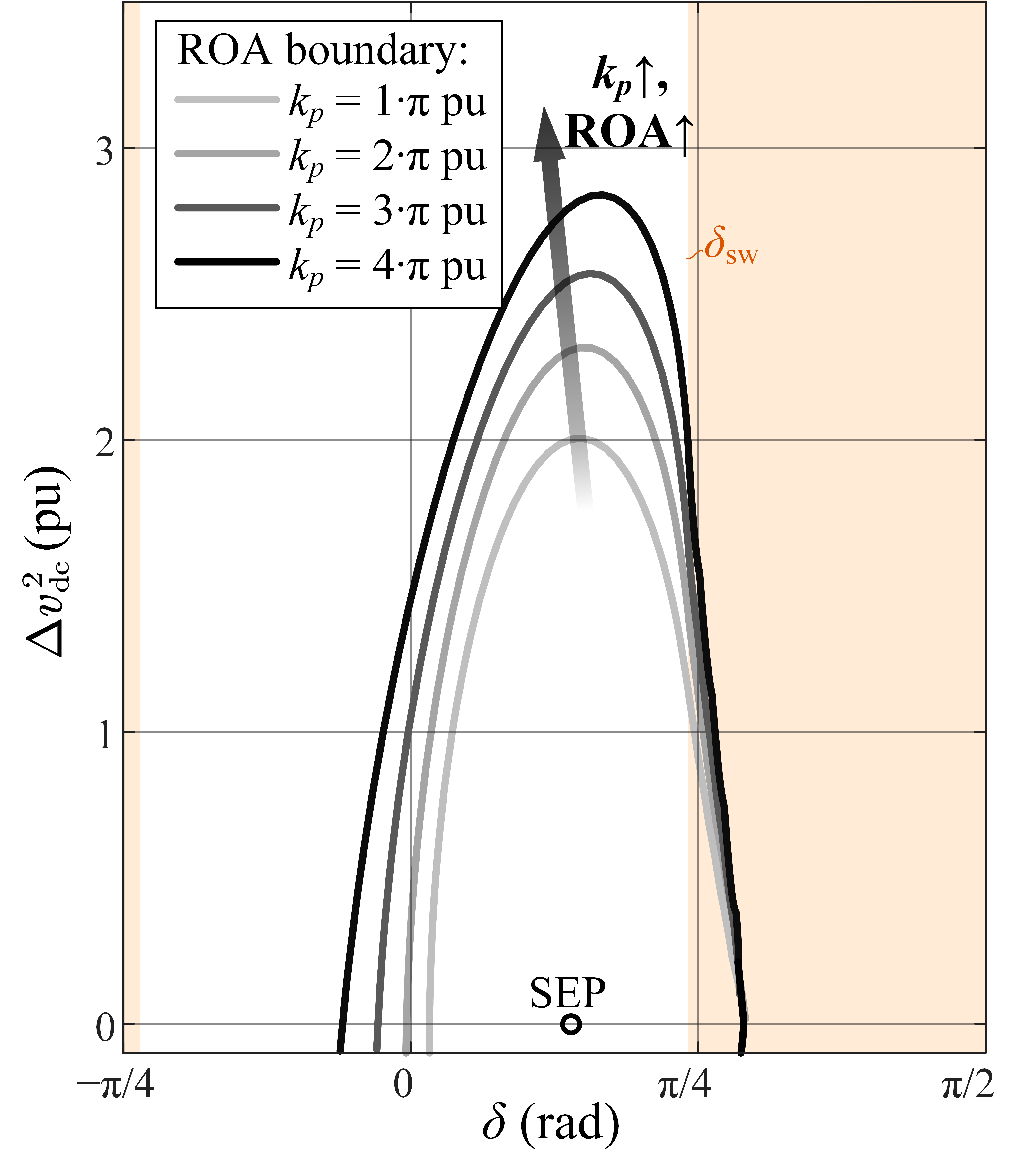}}%
}
\vspace{-4pt}
\caption{Effects of $D$ and $k_p$ on the estimated ROAs
of classical and DC-voltage-controlled GFM inverters,
respectively, using the circular limiter as an
illustrative example.}
\label{change2}
\vspace{-6pt}
\end{figure}

\begin{figure}[t!]
\centering
\vspace{-4pt}
{\captionsetup[subfloat]{captionskip=-1pt}
\subfloat[Circular current limiter.]{%
\includegraphics[width=0.24\textwidth]{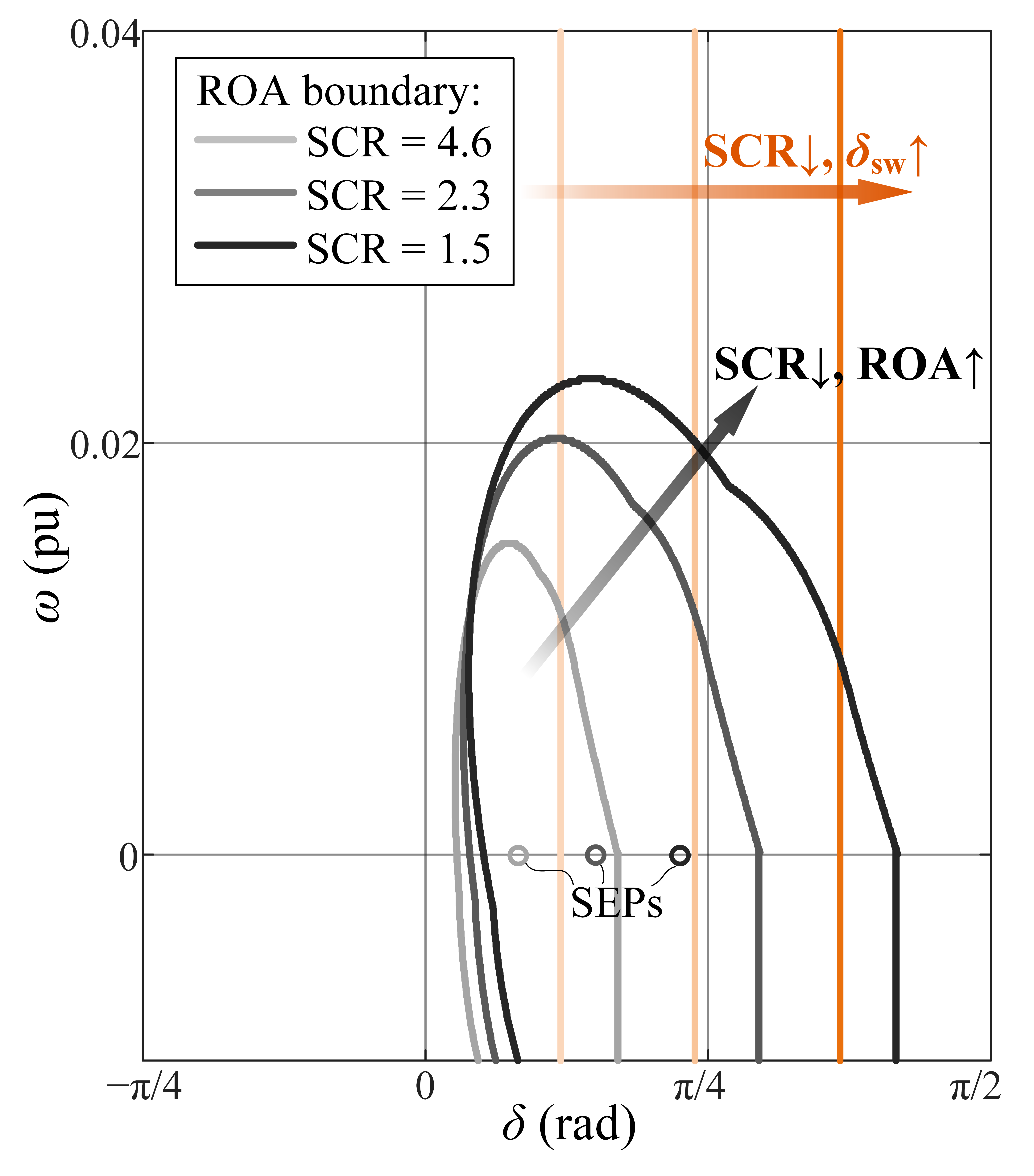}}
}
{\captionsetup[subfloat]{captionskip=-1pt}
\subfloat[EVI current-limiting strategy.]{%
\includegraphics[width=0.241\textwidth]{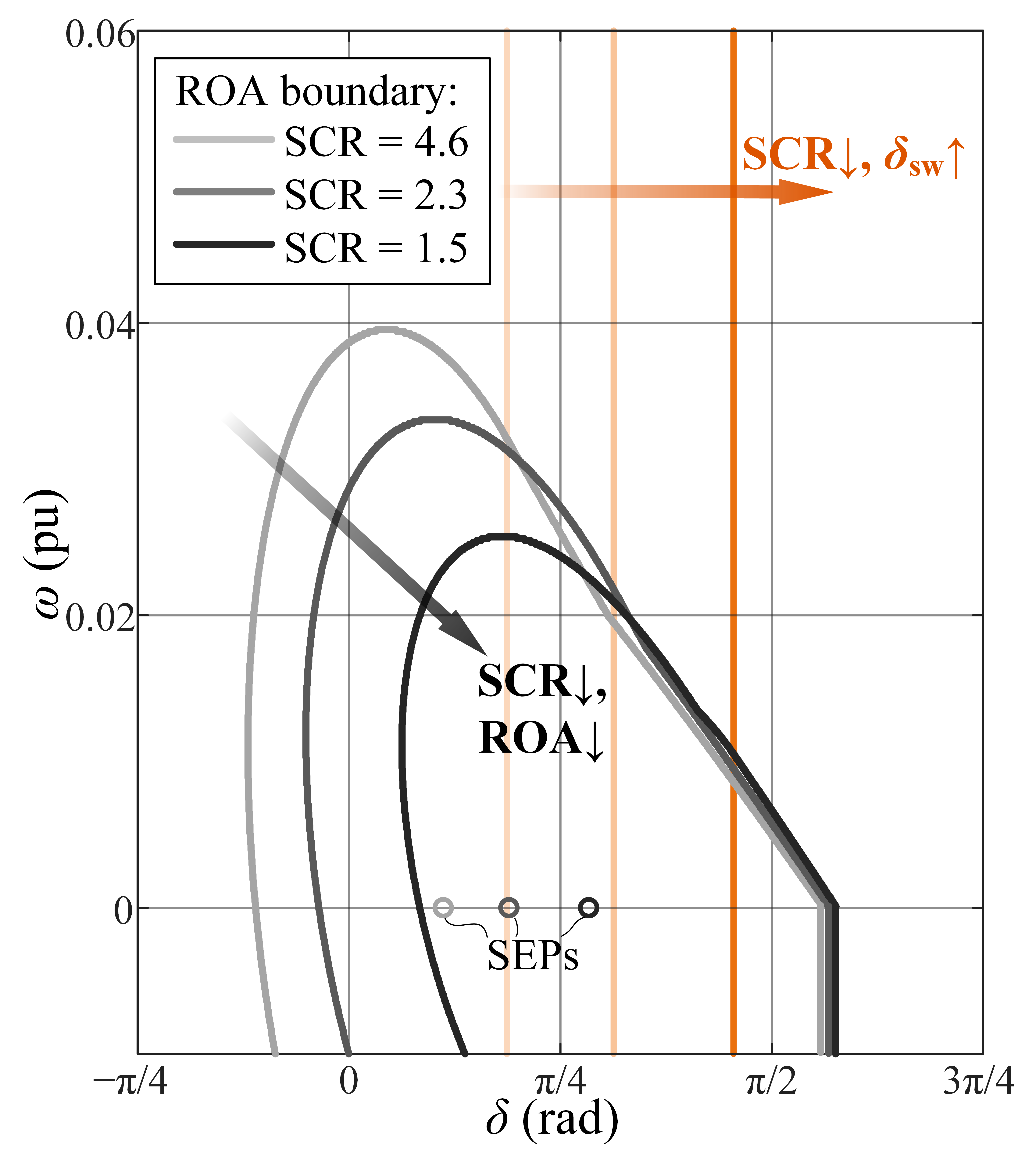}}%
}
\vspace{-4pt}
\caption{Effects of grid strength on the estimated ROA
under different current-limiting strategies, using
a classical GFM inverter as an illustrative example.}
\label{change3}
\vspace{-6pt}
\end{figure}

\subsubsection{Influence of Virtual Inertia $J$ and Integral Gain $k_i$}
\mbox{}\\ \hspace*{\parindent}
\figref{change1} shows the effects of virtual inertia
$J$ and integral gain $k_i$, with darker curves
representing larger parameter values.
Reducing $J$ enlarges the estimated ROA of the
classical GFM inverter.
Under a voltage sag, however, smaller inertia may
produce greater fault-on excursions in angle and
frequency, so a larger ROA does not necessarily imply
a longer CCT. By contrast, smaller inertia does improve
phase-jump tolerance.
For the DC-voltage-controlled GFM inverter, reducing
$k_i$ likewise enlarges the estimated ROA.
Nevertheless, $J$ and $k_i$ remain subject to RoCoF and
power-tracking requirements, respectively.

\subsubsection{Influence of Damping Coefficient $D$ and Proportional Gain $k_p$}
\mbox{}\\ \hspace*{\parindent}
\figref{change2} shows the effects of $D$ and $k_p$,
with darker curves corresponding to larger values.
Since the proposed SdS energy function incorporates
damping energy, it can effectively capture the effects
of $D$ and $k_p$ on the ROA.
Increasing $D$ enlarges the estimated ROA of the
classical GFM inverter, and increasing $k_p$ has
the same effect for the DC-voltage-controlled GFM inverter.
In practice, $D$ is limited by power-regulation
capability, and $k_p$ by small-signal stability of
the whole inverter control system.

\subsubsection{Influence of Grid Strength}
\mbox{}\\ \hspace*{\parindent}
\figref{change3} compares the ROA estimates under
different short-circuit ratios\footnote{The
grid-impedance $X/R$ ratio is held constant as SCR varies.} (SCRs)
taking the classical GFM inverter as an example.
Darker shades indicate lower SCRs for the ROA
boundaries, SEP markers, and switching-interface
lines at $\delta=\delta_{\mathrm{sw}}$.

For the circular limiter in \figref{change3}(a),
reducing SCR increases $\delta_{\mathrm{sw}}$ and
widens the normal operating region
$\mathcal R_{\mathrm{norm}}$, because the larger
grid impedance restricts current for a given power angle.
The overall ROA estimate also grows in size as the grid
weakens, with its right boundary extending markedly and the SEP shifting rightward.
This contrasts with the larger ROA predicted for
a stronger grid when current limiting is neglected.

For the EVI strategy in \figref{change3}(b),
a stronger grid likewise narrows the normal operating
region. For example, at $\mathrm{SCR}=4.6$, the SEP
lies very close to the switching interface, so
current limiting can be activated by a smaller
power-angle excursion from the SEP.  Nevertheless, the overall ROA estimate increases
with SCR, so grid weakening reduces the attraction
region, opposite to the circular-limiter case. Using a conventional energy function based on normal-mode dynamics \cite{fan2024transient}
would instead yield the contradictory conclusion that a smaller ROA estimate exists as SCR increases. This is because the critical energy is limited by
the boundary of $\mathcal R_{\mathrm{norm}}$,
thereby excluding potentially stable trajectories
that traverse $\mathcal R_{\lim}$.

\begin{figure}[b!]
    \centering
    \vspace{-8pt}
    \includegraphics[width=0.482\textwidth]{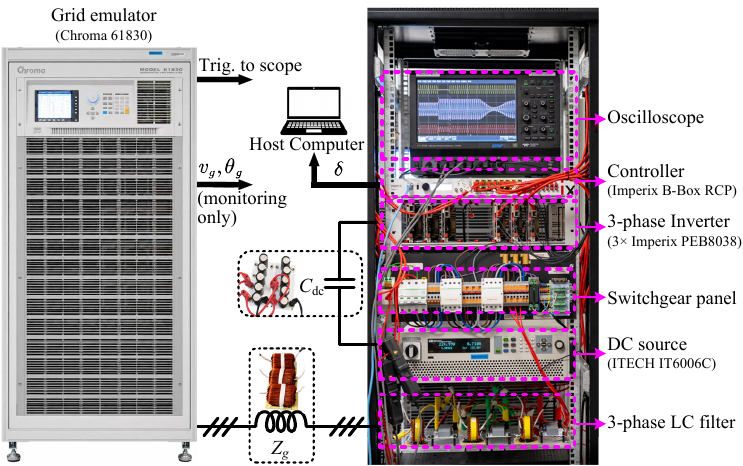}
    \vspace{-9pt}
    \caption{Experimental setup of the grid-connected GFM inverter system.}
    \label{Platform}
    \vspace{-6pt}
\end{figure}

\section{Experimental Validation and Application\\Case Studies}
Experiments on a practical GFM inverter under different
grid contingencies demonstrate the application of the
proposed SdS energy function and validate its
transient-stability predictions.
The experimental setup in \figref{Platform} consists
of a Chroma grid emulator, a three-phase power converter,
and an ITECH programmable DC power supply.
The grid emulator applies the disturbances, and the GFM
inverter control is implemented on an Imperix B-Box RCP.
Detailed system parameters and experimental settings
are listed in \tabref{TableA}.
All experimental results are presented with fault
inception aligned at $t=0$.

\begin{figure}[t!]
    \centering
    \vspace{-8pt}
        \subfloat[Time-domain responses for $t_{\mathrm{fault}}=78~\mathrm{ms}$ (stable).]
        {%
            \includegraphics[width=0.35\textwidth]
            {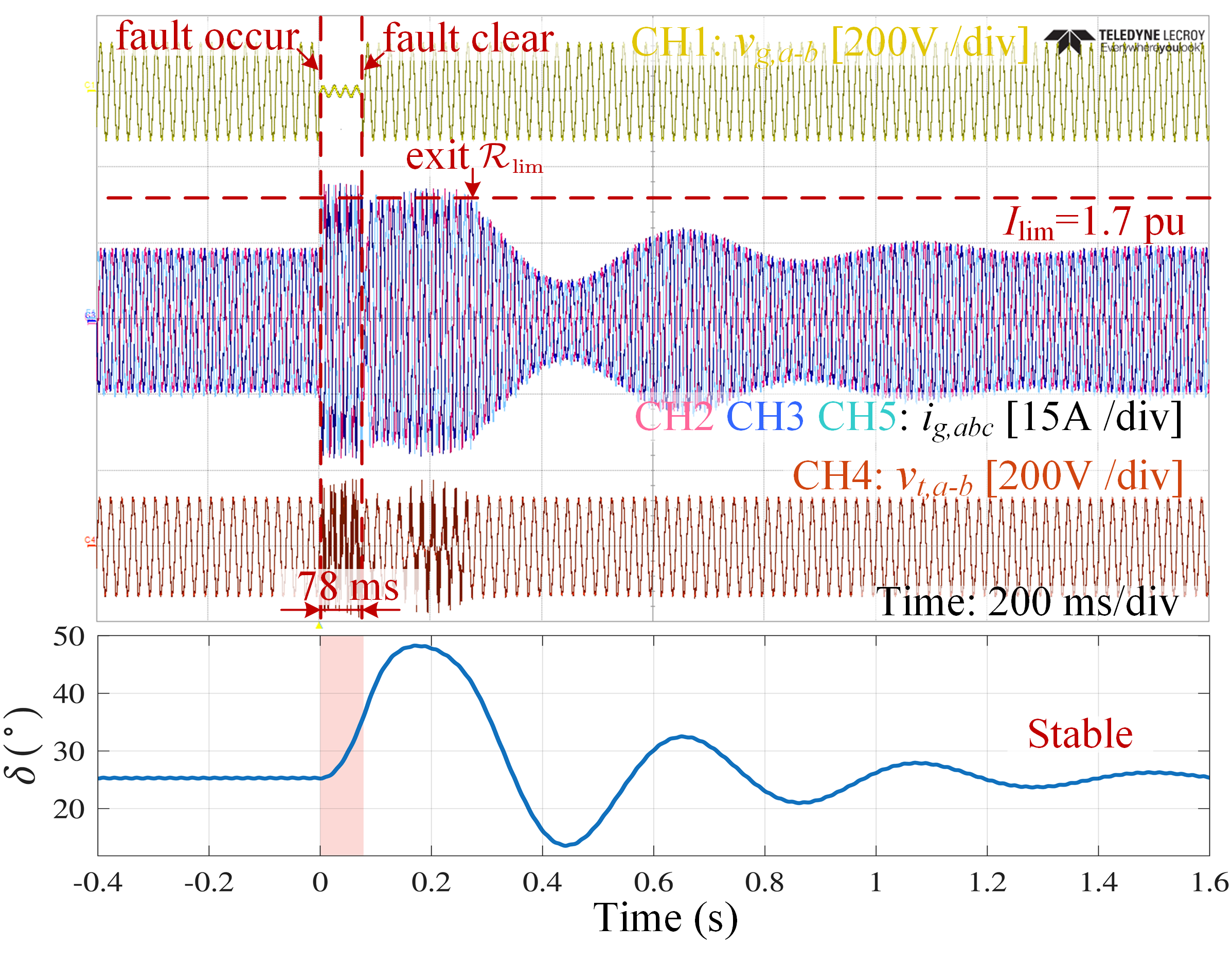}%
        }\\
        \vspace{-4pt}
        \subfloat[Time-domain responses for $t_{\mathrm{fault}}=81~\mathrm{ms}$ (unstable).]
        {%
            \includegraphics[width=0.35\textwidth]
            {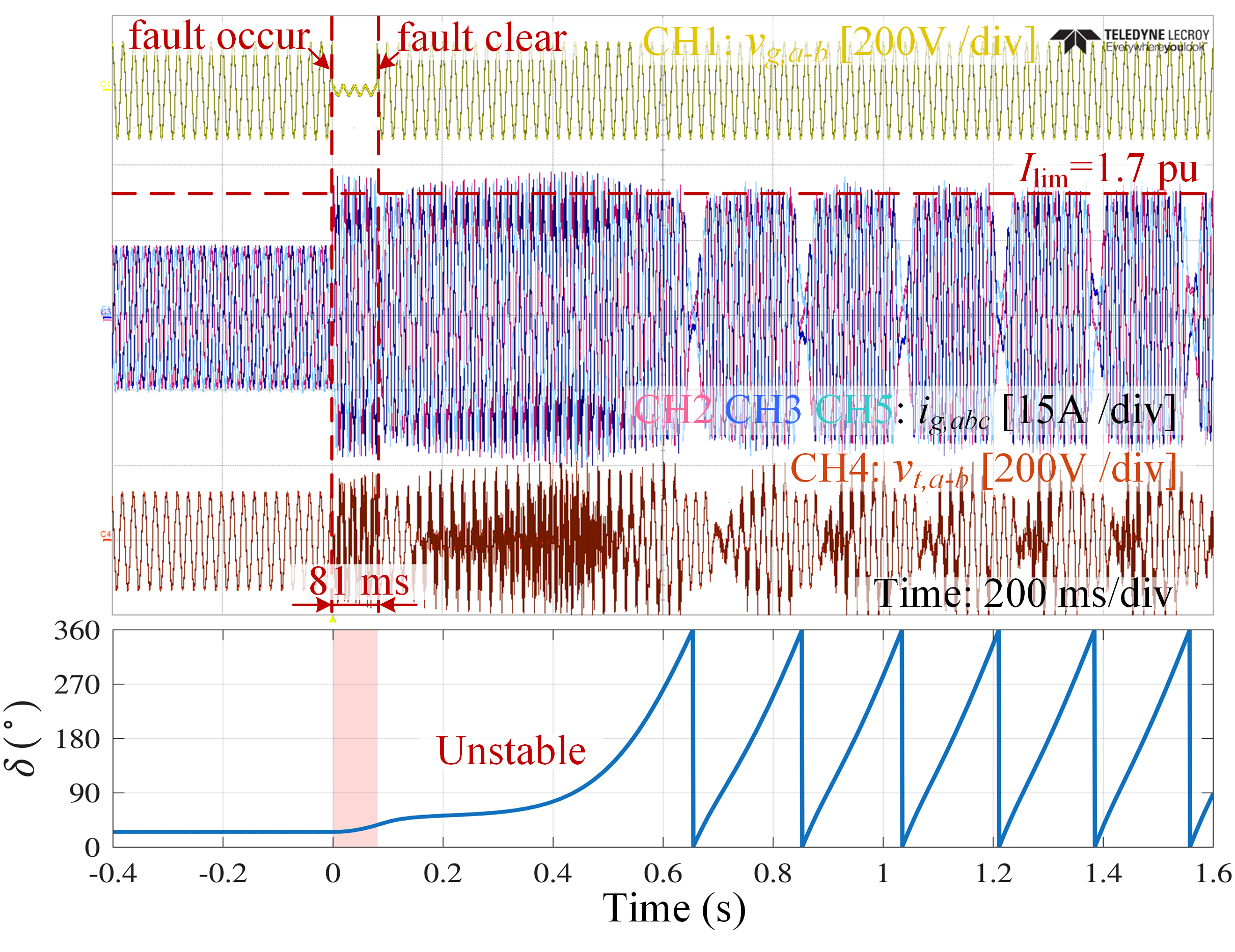}%
        }\\
        \vspace{-4pt}
        \subfloat[SdS energy function $\mathcal W_{\lambda}$ for $t_{\mathrm{fault}}=78$ and $81~\mathrm{ms}$ (CCT $=\text{80}~\mathrm{ms}$).]{%
            \includegraphics[width=0.35\textwidth]
            {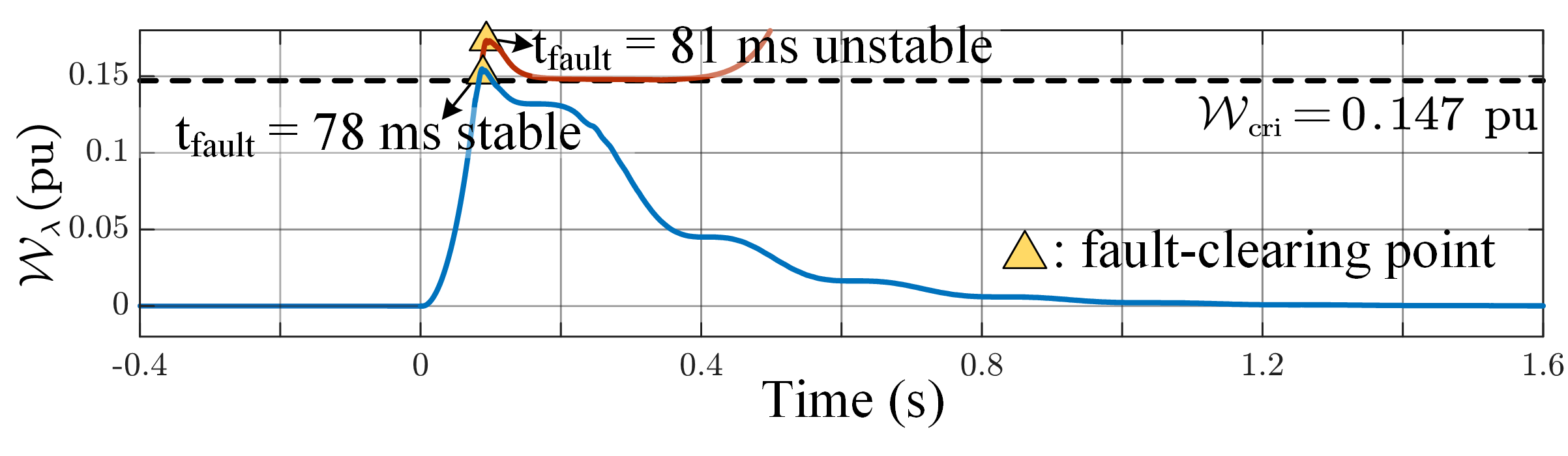}%
        }
        \caption{Experimental results of a classical GFM inverter with a circular limiter under a voltage sag fault.}
        \label{GFMsag1}
    \vspace{-8pt}
\end{figure}

\begin{figure*}[t!]
    \centering
    {\captionsetup[subfloat]{captionskip=0pt}

    \begin{minipage}[t]{0.32\textwidth}
        \centering
        \vspace{0pt}
        \subfloat[Time-domain responses for $t_{\mathrm{fault}}=118~\mathrm{ms}$ (stable).]{%
            \includegraphics[width=\linewidth]
            {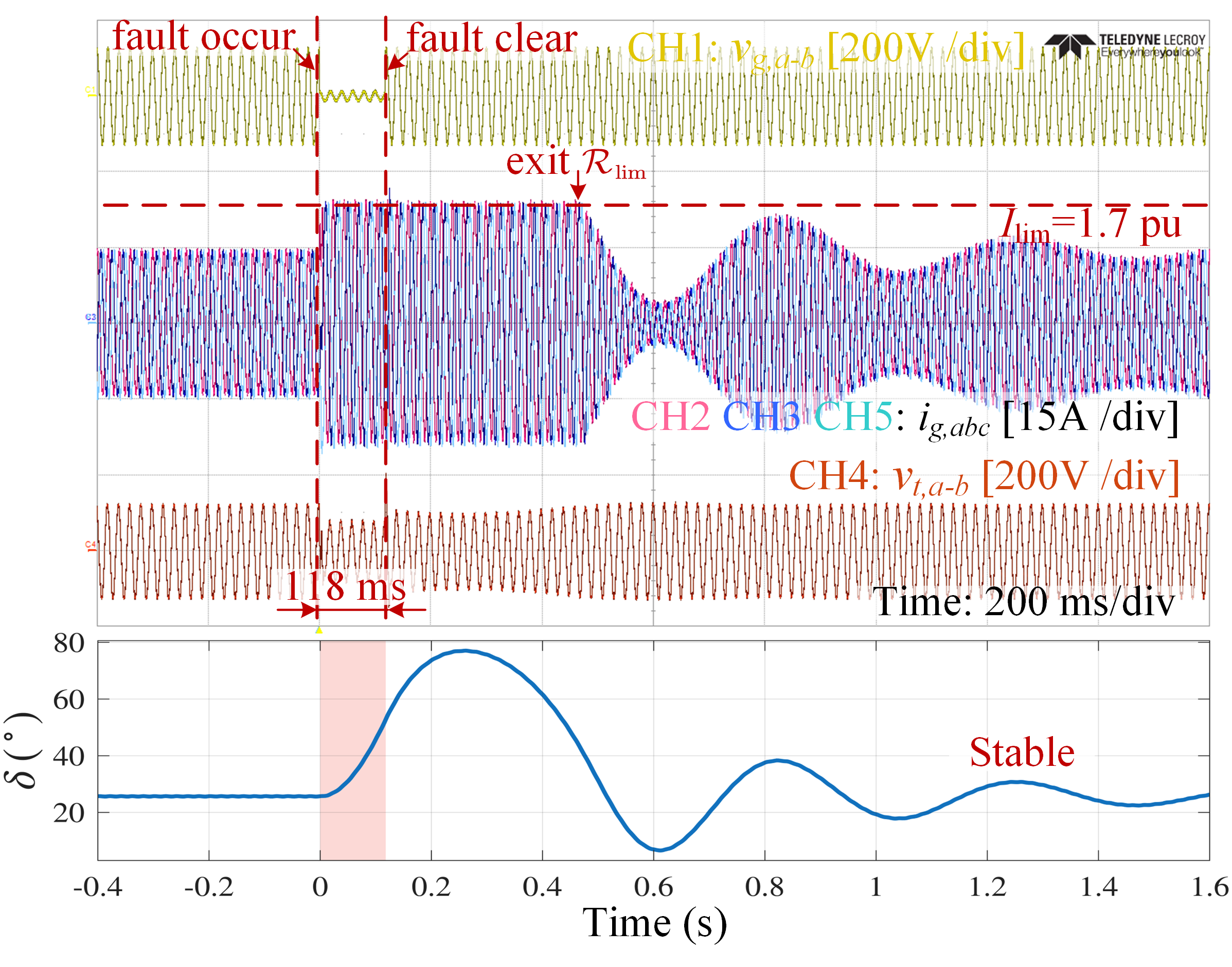}%
        }\\
        \vspace{-4pt}
        \subfloat[Time-domain responses for $t_{\mathrm{fault}}=137~\mathrm{ms}$ (unstable).]{%
            \includegraphics[width=\linewidth]
            {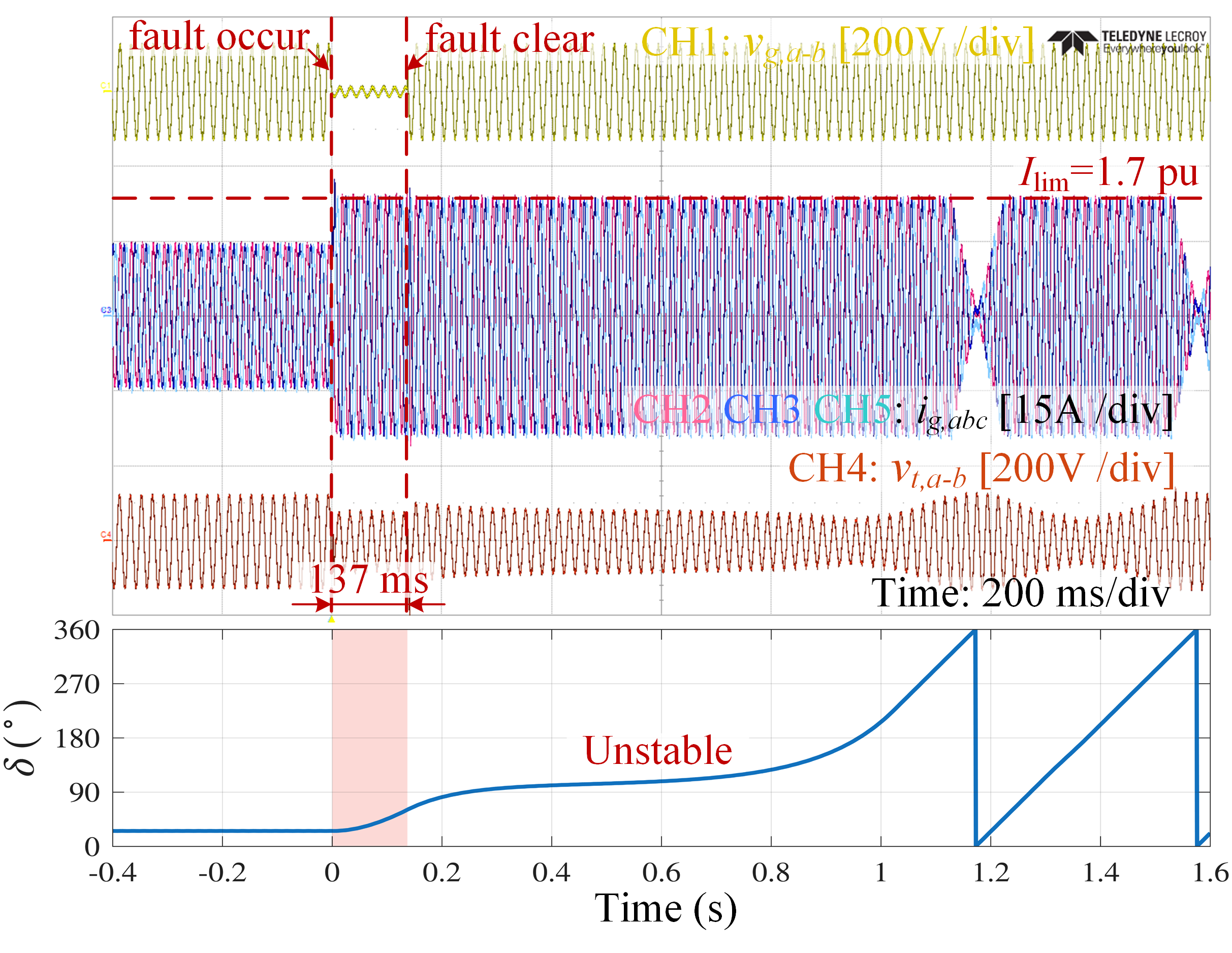}%
        }\\
        \vspace{-4pt}
        \subfloat[SdS energy function $\mathcal W_{\lambda}$ for $t_{\mathrm{fault}}=118$ and $137~\mathrm{ms}$ (CCT $=\text{136}~\mathrm{ms}$).]{%
            \includegraphics[width=\linewidth]
            {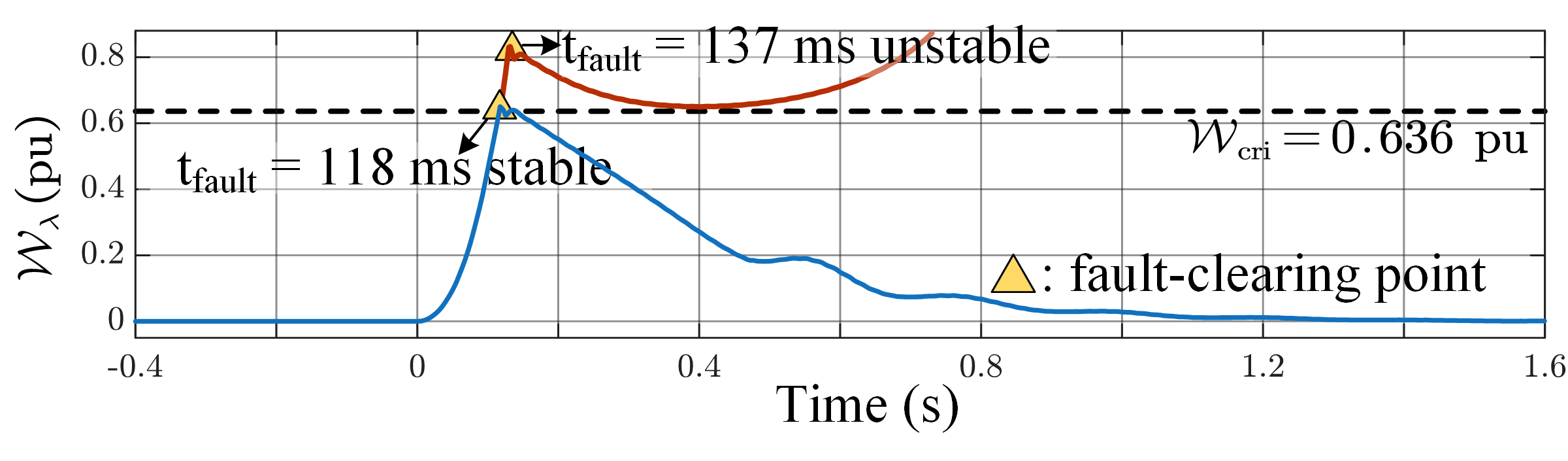}%
        }
        \vspace{-2pt}
        \caption{Experimental results of a classical GFM inverter with the EVI current-limiting strategy under a voltage sag fault.}
        \label{GFMsag2}
    \end{minipage}%
    \hfill
    \begin{minipage}[t]{0.32\textwidth}
        \centering
        \vspace{0pt}
        \subfloat[Time-domain responses for $t_{\mathrm{fault}}=98~\mathrm{ms}$ (stable).]{%
            \includegraphics[width=\linewidth]
            {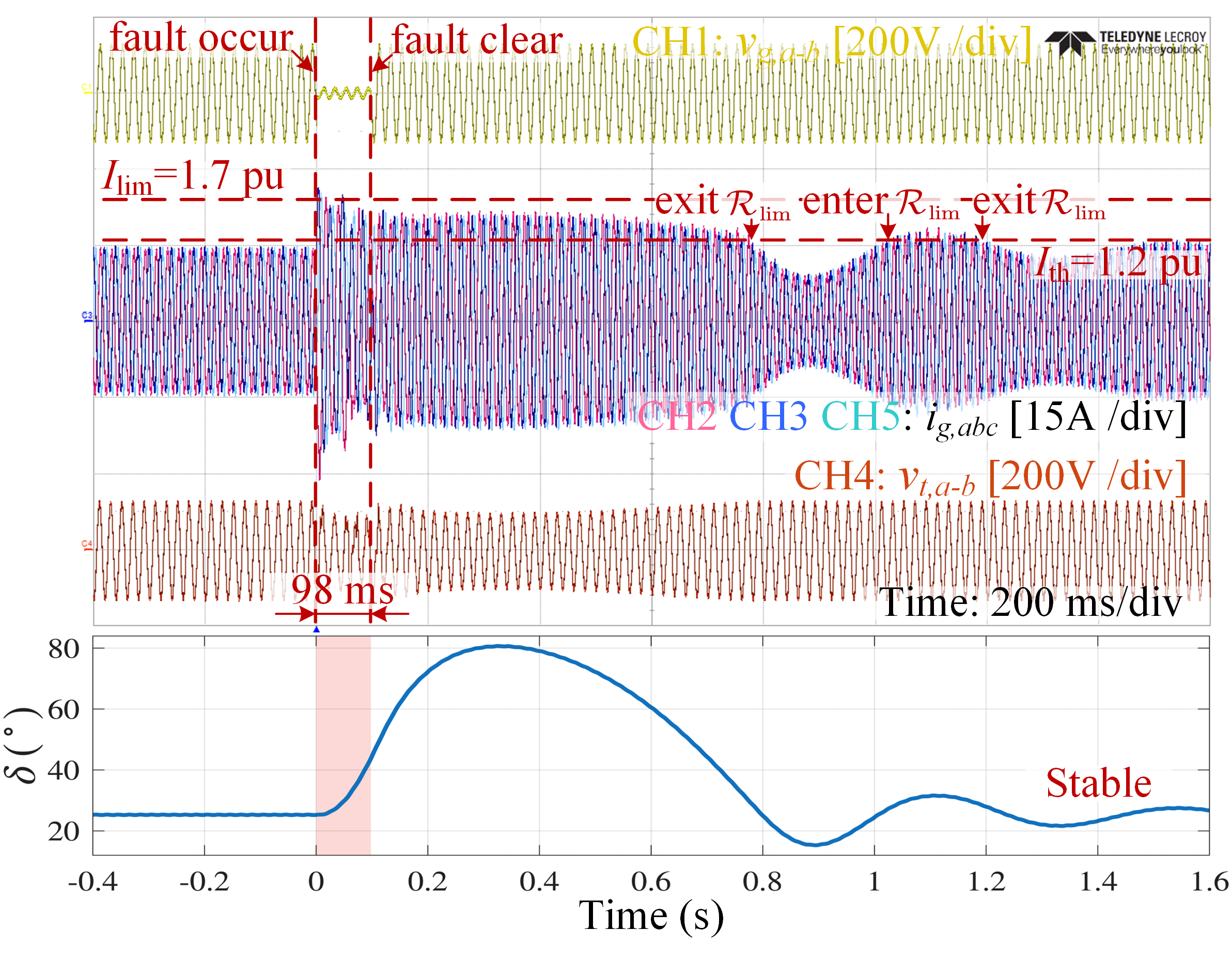}%
        }\\
        \vspace{-4pt}
        \subfloat[Time-domain responses for $t_{\mathrm{fault}}=111~\mathrm{ms}$ (unstable).]{%
            \includegraphics[width=\linewidth]
            {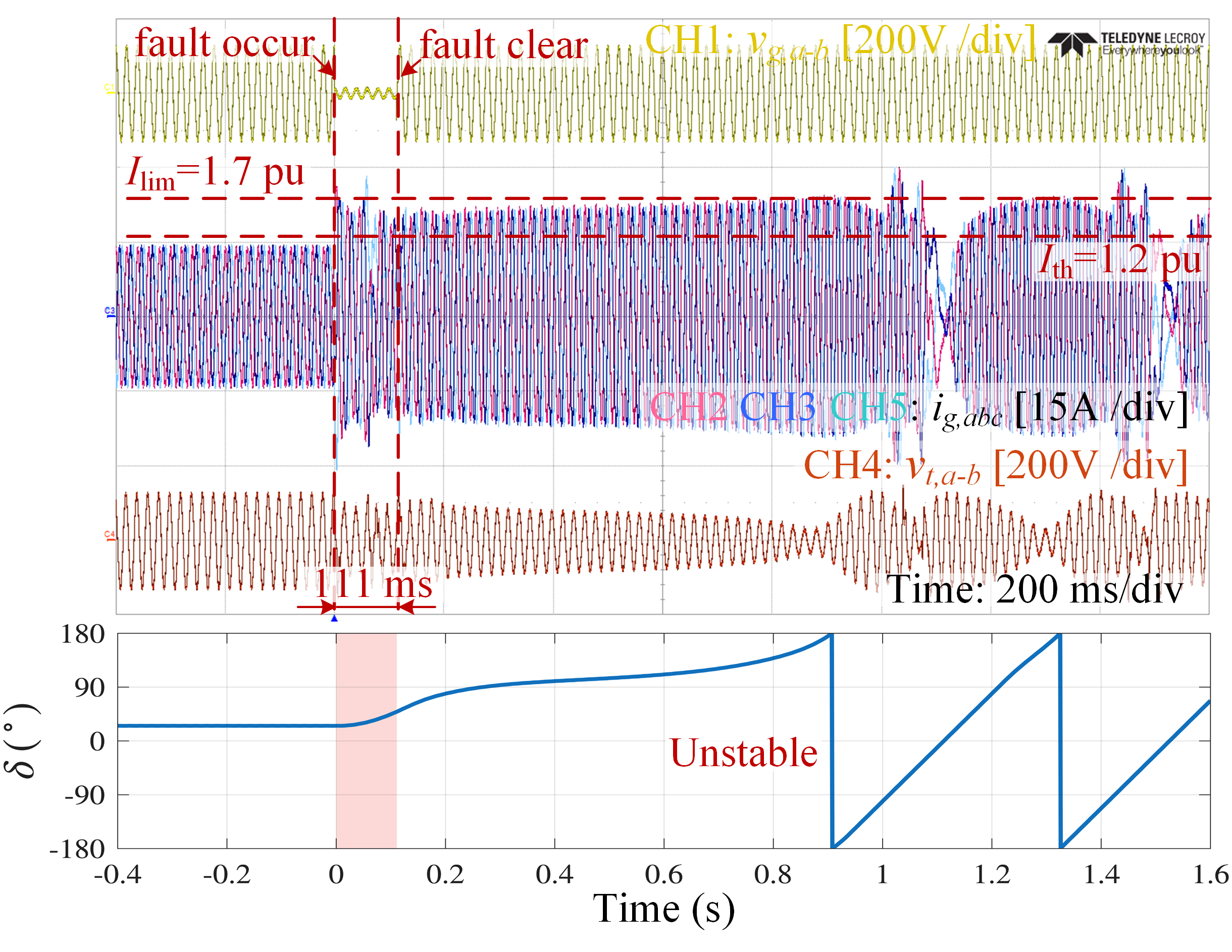}%
        }\\
        \vspace{-4pt}
        \subfloat[Value of SdS energy function $\mathcal W_{\lambda}$ for $t_{\mathrm{fault}}=98$ and $111~\mathrm{ms}$ (CCT $=\text{110}~\mathrm{ms}$).]{%
            \includegraphics[width=\linewidth]
            {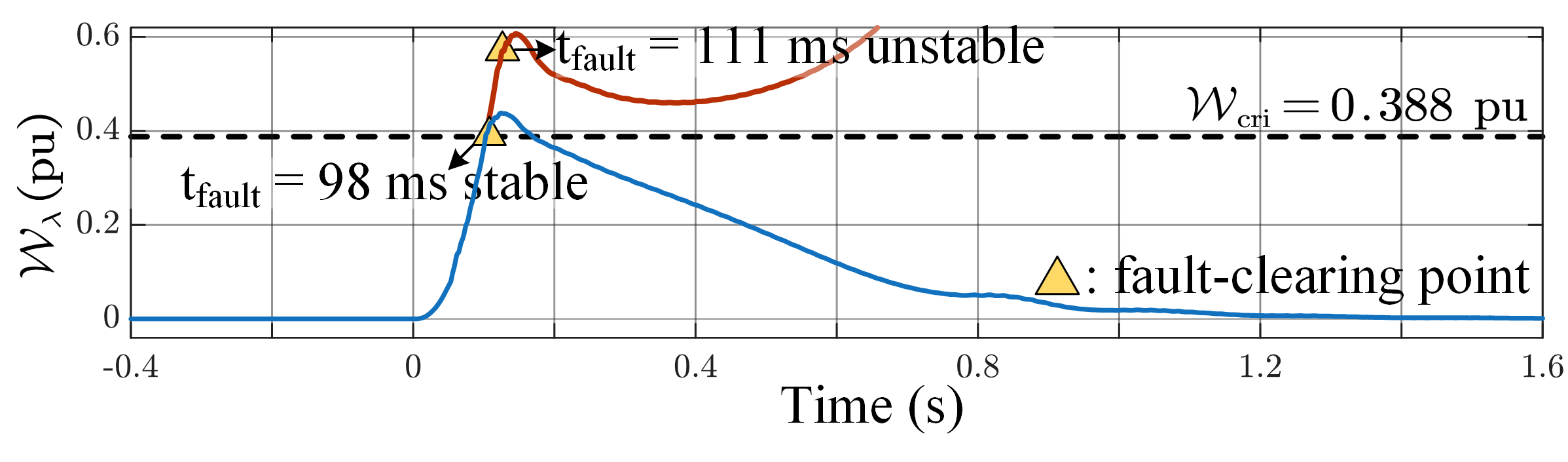}%
        }
        \vspace{-2pt}
        \caption{Experimental results of a classical GFM inverter with the threshold VI current-limiting strategy under a voltage sag fault.}
        \label{GFMsag3}
    \end{minipage}%
    \hfill
    \begin{minipage}[t]{0.32\textwidth}
        \centering
        \vspace{0pt}
        \subfloat[Time-domain responses for phase jump value $\Delta\theta_g=44^\circ$ (stable).]
        {%
            \includegraphics[width=\linewidth]
            {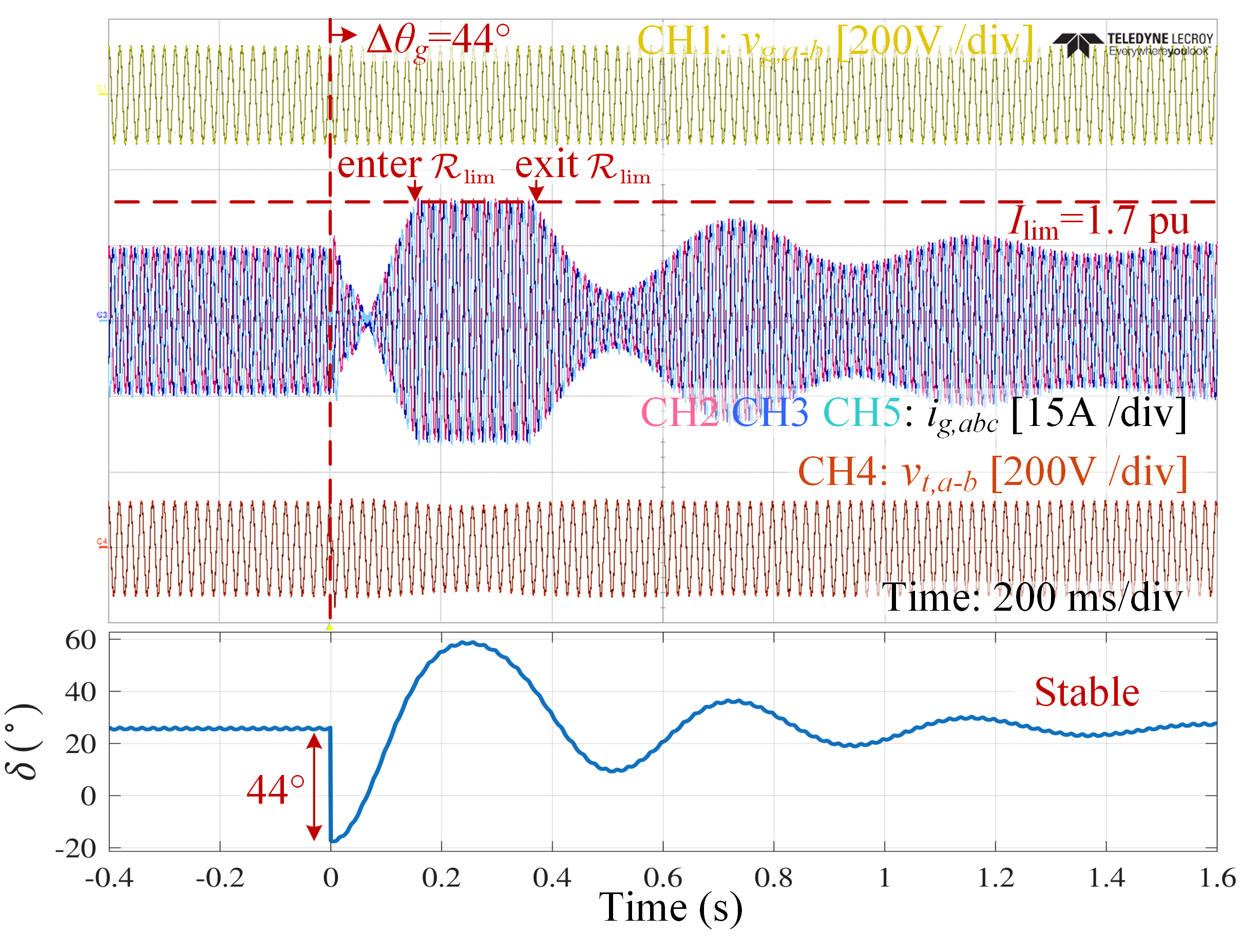}%
        }\\
        \vspace{-4pt}
        \subfloat[Time-domain responses for phase jump value $\Delta\theta_g=60^\circ$ (unstable).]
        {%
            \includegraphics[width=\linewidth]
            {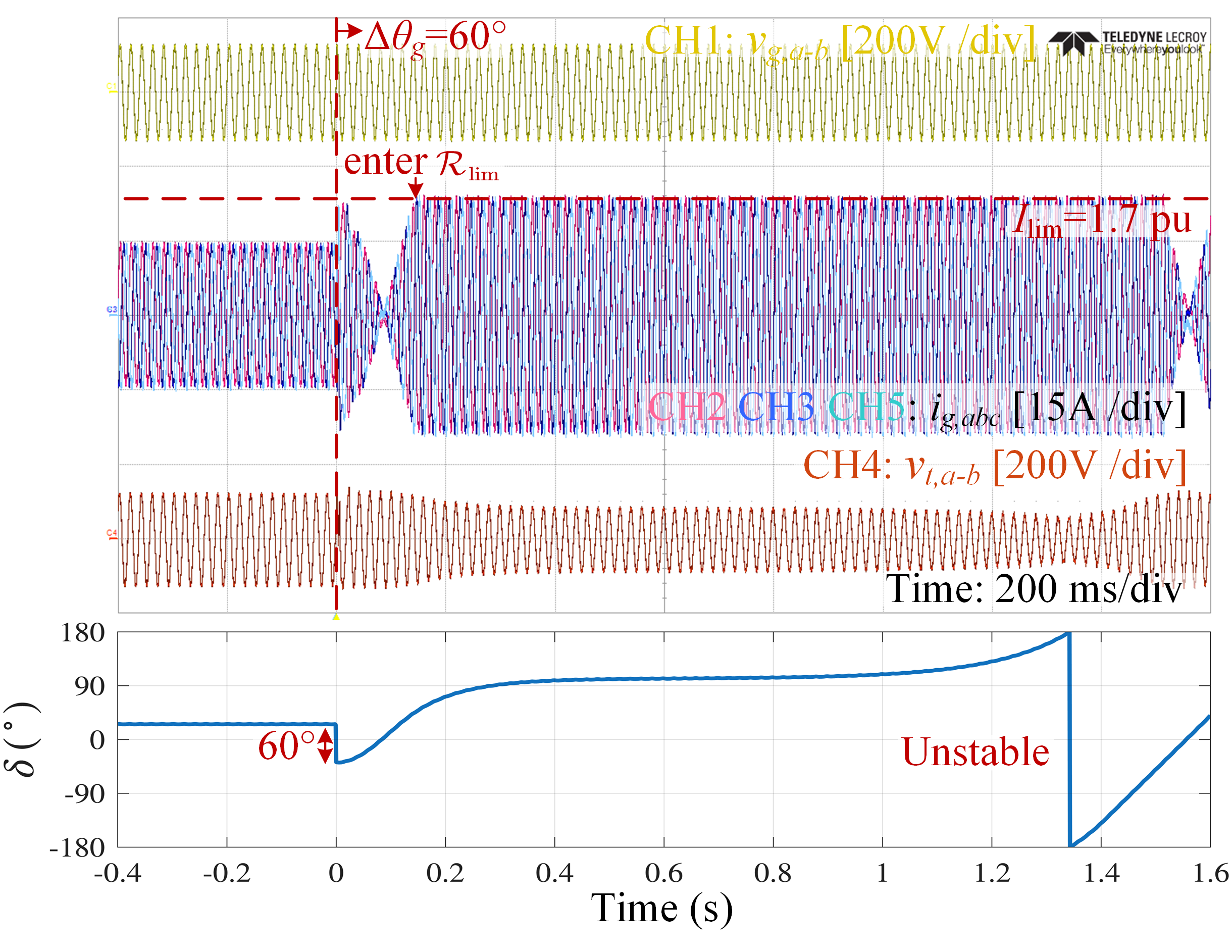}%
        }\\
        \subfloat[SdS energy function $\mathcal W_{\lambda}$ for $\Delta\theta_g=44^\circ$, $60^\circ$, $-74^\circ$, and $-76^\circ$.]{%
            \includegraphics[width=\linewidth]
            {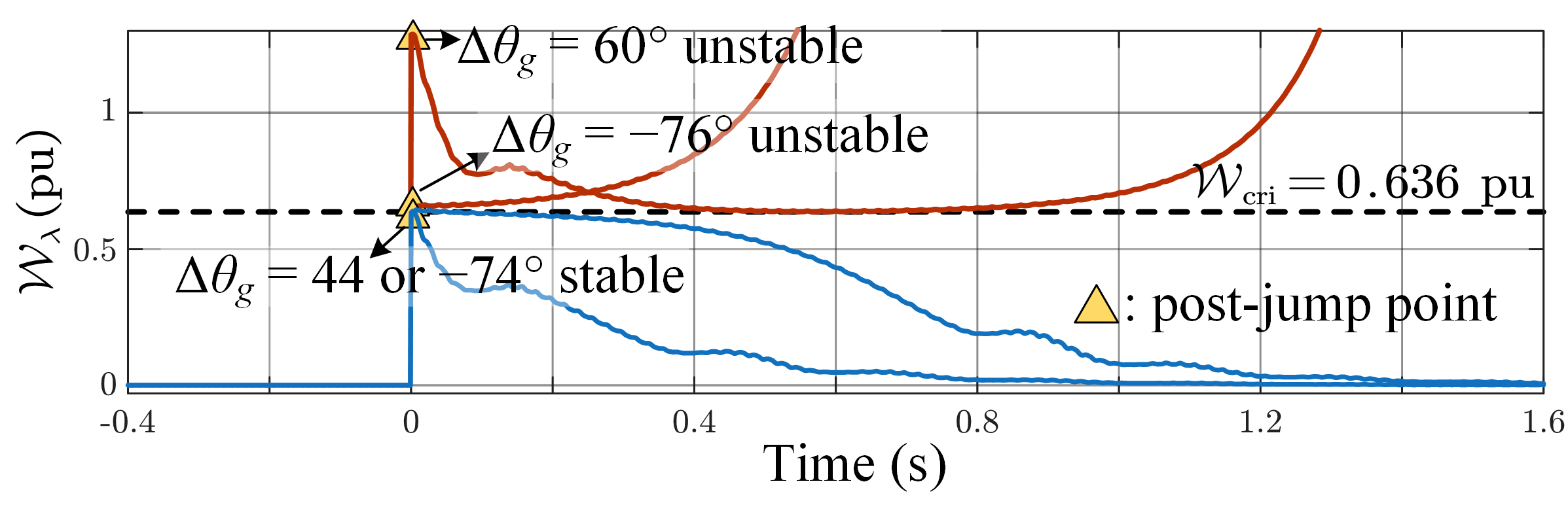}%
        }
        \vspace{-2pt}
        \caption{Experimental results of a classical GFM inverter with the EVI current-limiting strategy under phase jumps.}
        \label{GFMpj}
        \vspace{-6pt}
    \end{minipage}
    }
\end{figure*}

\subsection{Classical GFM Inverter}
\subsubsection{Voltage Sag Fault}
\mbox{}\\ \hspace*{\parindent}
The experimental results for the classical GFM inverter
with a circular current limiter under a voltage sag
fault are shown in \figref{GFMsag1}.
The fault duration is set to
$t_{\mathrm{fault}}=78~\mathrm{ms}$, corresponding to
the CCT predicted by the proposed SdS energy function.
The time-domain response in \figref{GFMsag1}(a)
remains stable as predicted, with the current effectively
limited during the fault.
After fault clearance,
the inverter continue operating in current-limited mode $\mathcal R_{\lim}$
until it returns to the normal operating mode at around $0.24~\mathrm{s}$.
The corresponding blue energy trace in
\figref{GFMsag1}(c) lies near
$\mathcal W_{\mathrm{cri}}$ at fault clearance, with
a small discrepancy consistent with experimental
uncertainties. It subsequently decays through both
current-limited and normal operation toward the SEP
energy, shown as zero.

For the EVI current-limiting strategy, the predicted
CCT increases to $118~\mathrm{ms}$.
Setting $t_{\mathrm{fault}}$ to this value gives
the stable time-domain response in \figref{GFMsag2}(a).
The inverter limits its current during the fault and
continues operating under current limiting after
fault clearance, returning to the normal operating
mode only at around $0.47~\mathrm{s}$.
This again demonstrates that grid-voltage restoration
does not imply release of current limiting,
while continued current-limited operation does not
imply transient instability.
The corresponding energy trace in \figref{GFMsag2}(c)
lies near $\mathcal W_{\mathrm{cri}}$ at fault clearance
and subsequently decays to the SEP energy, consistent
with the predicted convergence to the SEP.

For the threshold VI current-limiting strategy, the
predicted CCT of $98~\mathrm{ms}$ also yields a stable
time-domain response in \figref{GFMsag3}(a).
After fault clearance, the inverter remains in the
current-limited operating region $\mathcal R_{\lim}$
until about $0.78~\mathrm{s}$.
Current limiting is reactivated at about
$1.02~\mathrm{s}$ and released again at about
$1.2~\mathrm{s}$.
Despite this repeated current-limit activation and
release, the energy in \figref{GFMsag3}(c) decays
from near $\mathcal W_{\mathrm{cri}}$ at fault
clearance toward the origin.
This behavior further illustrates a key advantage of the
proposed SdS energy function: convergence from the
estimated ROA is guaranteed without prescribing the
sequence, number, or timing of current-limit activations
and releases, or the time spent in either operating mode.

The experimentally measured CCTs for the circular
limiter, EVI, and threshold VI are $80$, $136$, and
$110~\mathrm{ms}$, respectively, confirming the
conservatism of the theoretical predictions.
Increasing $t_{\mathrm{fault}}$ to $81$, $137$, and
$111~\mathrm{ms}$, respectively, causes loss of
synchronization in all three cases, as shown by
the time-domain responses in \figref{GFMsag1}(b),
\figref{GFMsag2}(b), and \figref{GFMsag3}(b).
In each case, the corresponding red energy trace
in panel (c) exceeds $\mathcal W_{\mathrm{cri}}$
by a certain margin at fault clearance and fails
to return to the origin. For comparison, using the conventional energy function restricted to the normal operating
mode \cite{fan2024transient,yang2025global,wang2025transient}
yields much smaller CCT estimates of only $66$, $56$, and
$20~\mathrm{ms}$, respectively.

Both the circular limiter and the EVI current-limiting
strategy effectively constrain the current amplitude
to $I_{\lim}$ in the stable and unstable tests.
For the threshold VI strategy,
$I_{\mathrm{th}}<I_{\lim}$ causes earlier
virtual-impedance activation and reduces utilization
of the available current capability relative to EVI.
Its virtual-impedance response is also constrained by
the voltage-control-loop bandwidth, and brief current overshoots
above $I_{\lim}$ are visible in both
\figref{GFMsag3}(a) and (b).
The EVI strategy therefore provides a longer CCT and
more effective current limitation in these experiments.

\begin{figure}[t!]
    \centering
    \includegraphics[width=0.36\textwidth]{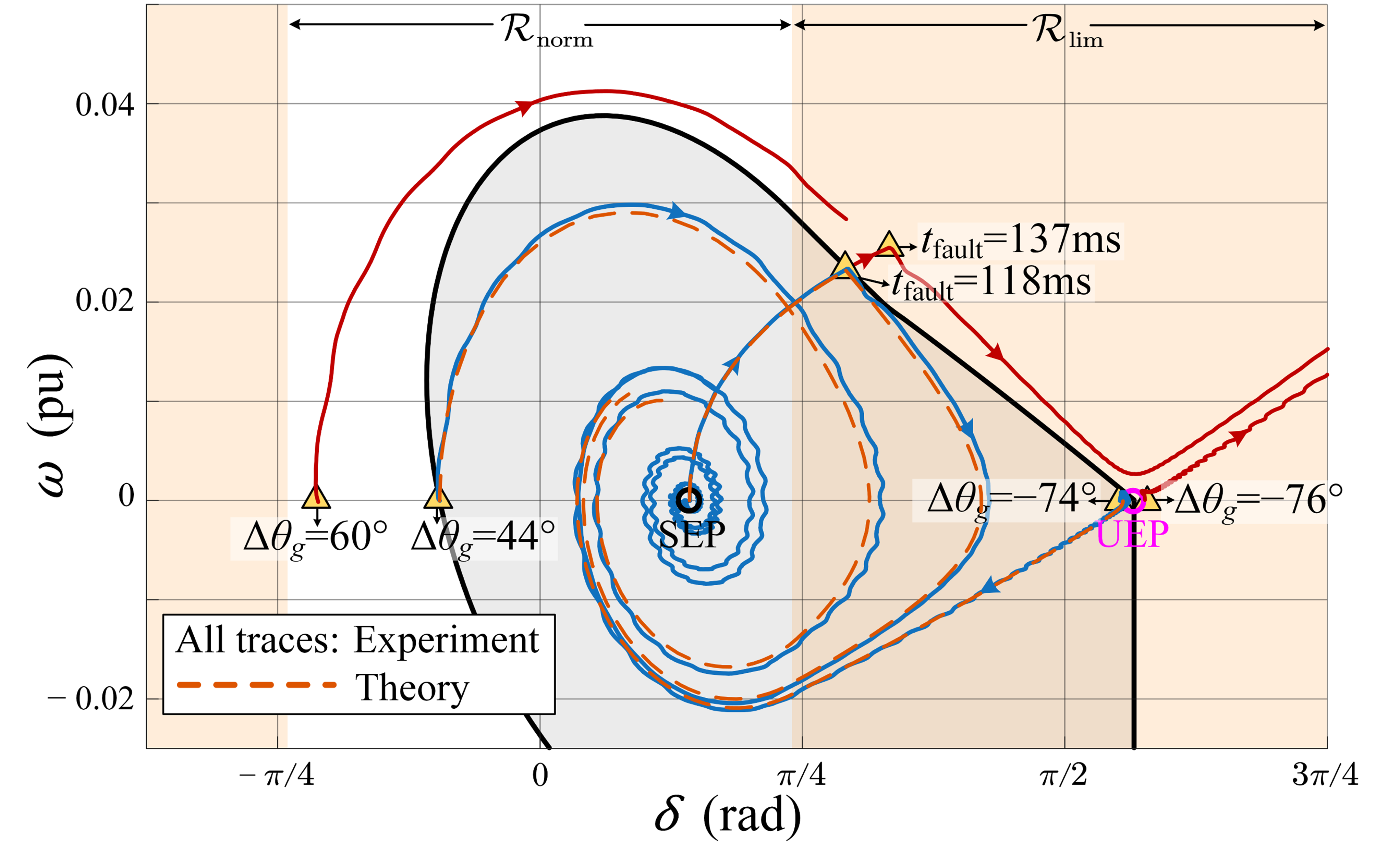}
    \vspace{-9pt}
    \caption{Cross-validation of experimental results for classical GFM inverter with theoretical analysis in the phase plane under different disturbances, using the EVI current-limiting strategy as an illustrative example. Blue/ red curves: stable/ unstable experimental trajectories;
    orange dashed curves: theoretical trajectories;
    gray shading: ROA by the proposed SdS energy function.}
    \label{GFMROAexp}	
    \vspace{-5pt}
\end{figure}

\subsubsection{Phase Jump}
\mbox{}\\ \hspace*{\parindent}
The same SdS energy criterion applies to grid phase
jumps by comparing the post-jump energy with
$\mathcal W_{\mathrm{cri}}$.
For illustration, \figref{GFMpj}(a) shows the stable
time-domain response under the EVI current-limiting
strategy at the predicted critical positive phase jump
of $\Delta\theta_g=44^\circ$.
Current limiting is activated at about
$0.16~\mathrm{s}$ and released at about $0.38~\mathrm{s}$.
Increasing the phase jump to $\Delta\theta_g=60^\circ$
causes loss of synchronism, as shown in \figref{GFMpj}(b).
In both cases, current limiting is initially inactive,
further demonstrating that the initial absence of
current limiting after a large disturbance neither
precludes its subsequent activation nor guarantees
transient stability.

In \figref{GFMpj}(c), the blue energy traces for
$44^\circ$ and $-74^\circ$ start near
$\mathcal W_{\mathrm{cri}}$ and converge to the origin.
The red traces for $60^\circ$ and $-76^\circ$
start above the critical energy and fail to converge.

Using EVI as an illustrative example,
\figref{GFMROAexp} presents the same voltage-sag
and phase-jump experimental data from
\figref{GFMsag2} and \figref{GFMpj} in the phase plane
to cross-validate the estimated ROA.
The stable experimental trajectories (blue) align with the theoretical trajectories (orange dashed)
and remain within the estimated ROA while crossing
between $\mathcal R_{\mathrm{norm}}$ and
$\mathcal R_{\lim}$.

\begin{figure}[t!]
\centering
        \vspace{0pt}
        \subfloat[Time-domain responses for $t_{\mathrm{fault}}=36~\mathrm{ms}$ (stable).]{%
            \includegraphics[width=0.33\textwidth]
            {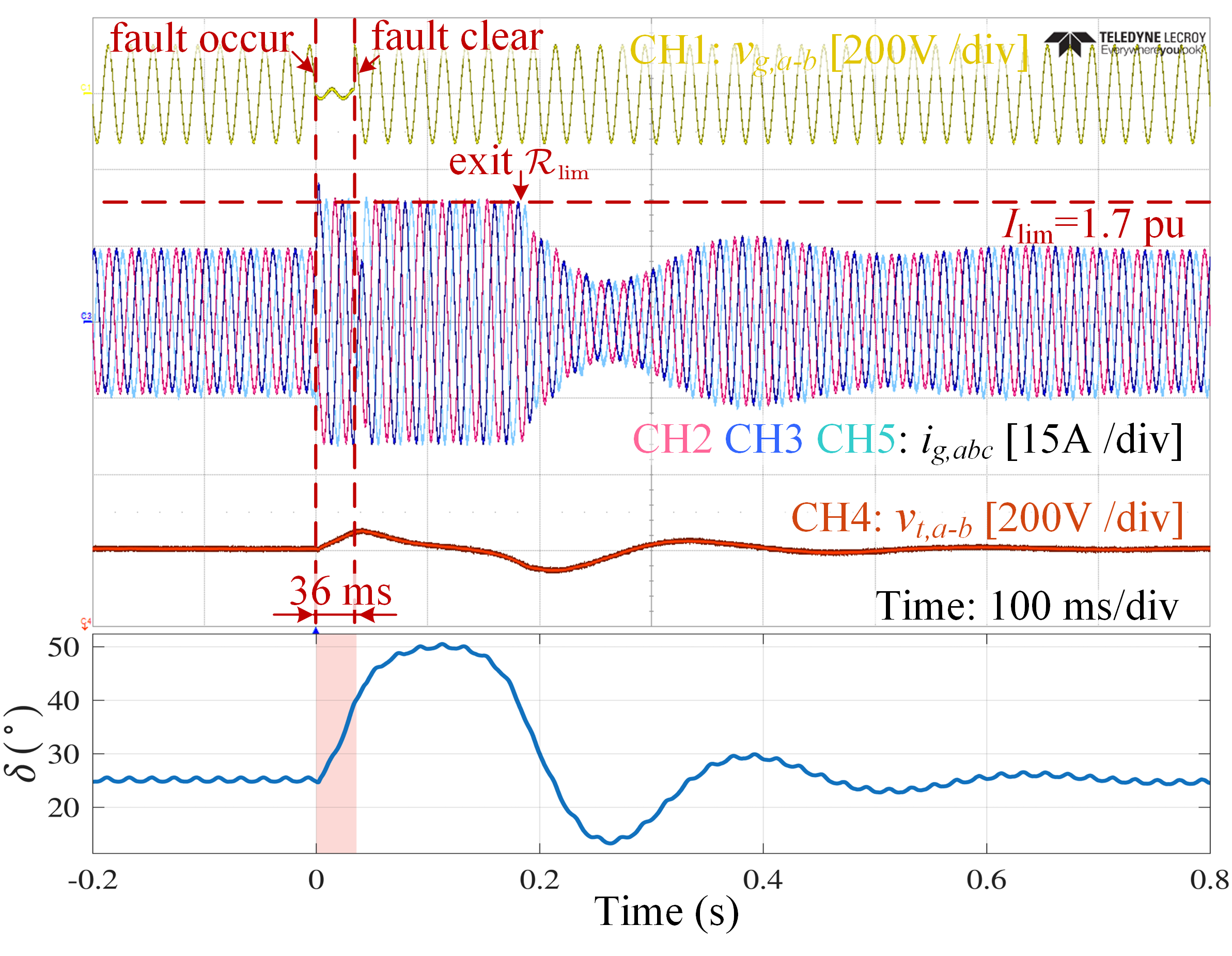}%
        }\\
        \vspace{-4pt}
        \subfloat[SdS energy function $\mathcal W$ for $t_{\mathrm{fault}}=36$ and $38~\mathrm{ms}$ (CCT $=\text{37}~\mathrm{ms}$).]{%
            \includegraphics[width=0.33\textwidth]
            {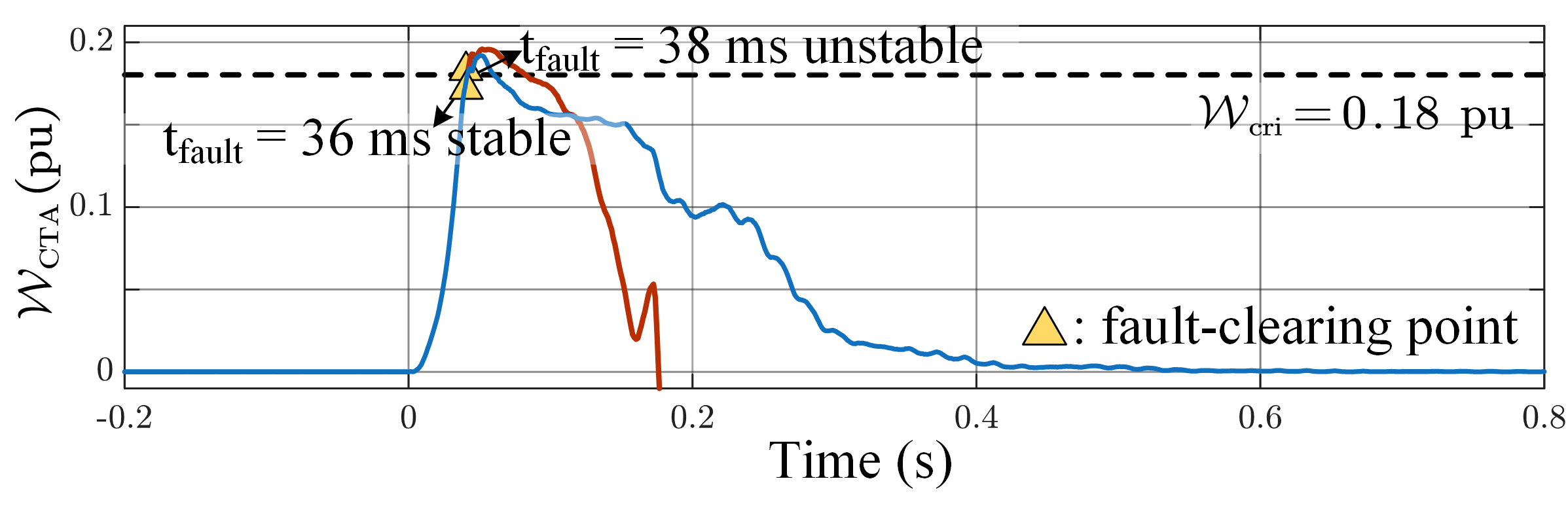}%
        }
        \vspace{-2pt}
        \caption{Experimental results of a DC-voltage-controlled GFM inverter with a circular limiter under a voltage sag fault.}
        \label{VFMsag1}
    \vspace{-8pt}
\end{figure}

\begin{figure}[t!]
\centering
        \vspace{0pt}
        \subfloat[Time-domain responses for $t_{\mathrm{fault}}=52~\mathrm{ms}$ (stable).]{%
            \includegraphics[width=0.35\textwidth]
            {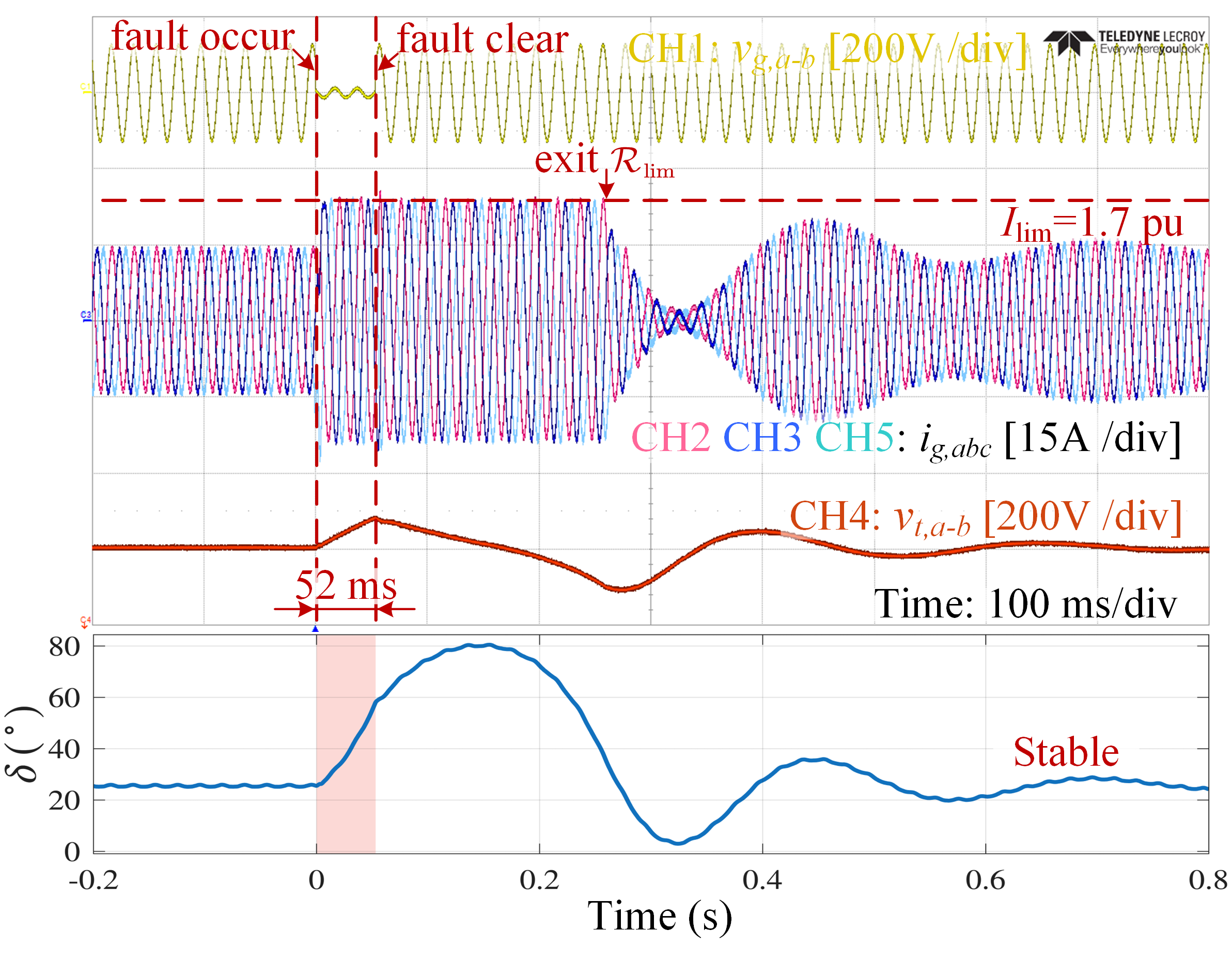}%
        }\\
        \vspace{-4pt}
        \subfloat[SdS energy function $\mathcal W$ for $t_{\mathrm{fault}}=52$ and $55~\mathrm{ms}$ (CCT $=\text{54}~\mathrm{ms}$).]{%
            \includegraphics[width=0.35\textwidth]
            {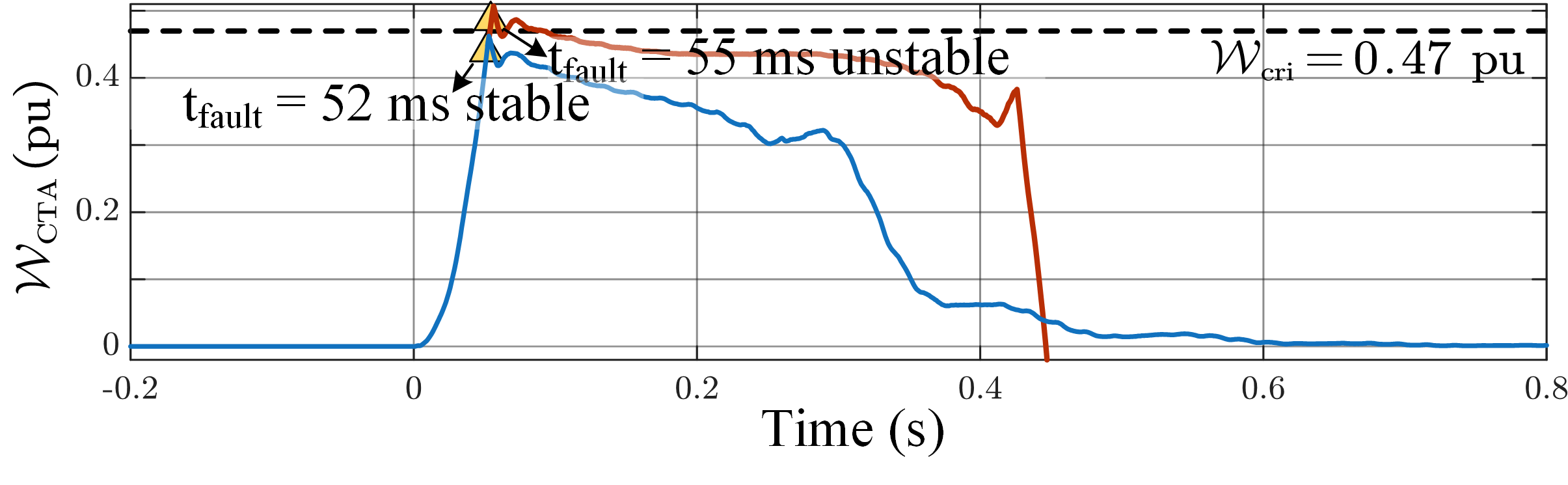}%
        }
        \vspace{-2pt}
        \caption{Experimental results of a DC-voltage-controlled GFM inverter with the EVI current-limiting strategy under a voltage sag fault.}
        \vspace{-4pt}
        \label{VFMsag2}
\end{figure}

\begin{figure}[t!]
    \centering
    \vspace{-3pt}
    \includegraphics[width=0.36\textwidth]{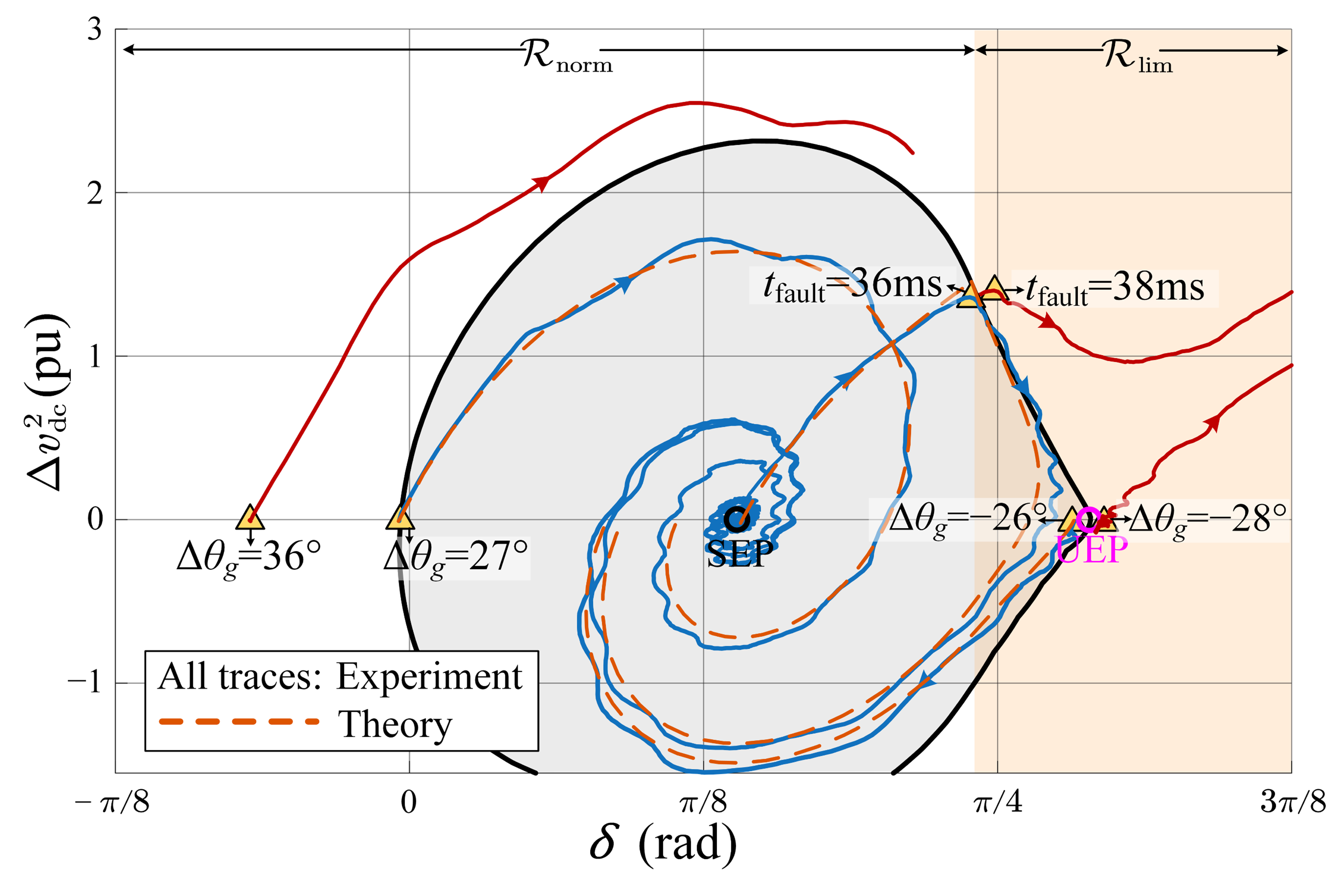}
    \vspace{-9pt}
    \caption{Cross-validation of experimental results for a DC-voltage-controlled GFM inverter with theoretical analysis in the phase plane under different disturbances, using the circular limiter as an illustrative example. The color scheme is the same as in \figref{GFMROAexp}.}
    \label{VFMROAexp}	
    \vspace{-5pt}
\end{figure}

\subsection{DC-Voltage-Controlled GFM Inverter}
\subsubsection{Voltage Sag Fault}
\mbox{}\\ \hspace*{\parindent}
The experiments in this subsection illustrate the proposed method
using the circular limiter and EVI as representative
hard current-limiting strategies for DC-voltage-controlled GFM inverter, consistent with
the emphasis of this paper.
For the circular current limiter,
\figref{VFMsag1}(a) shows the stable time-domain
response at fault duration $t_{\mathrm{fault}}=36~\mathrm{ms}$,
corresponding to the predicted CCT.
After fault clearance, current limiting continues
until about $0.18~\mathrm{s}$.
The corresponding blue energy trace in
\figref{VFMsag1}(b) lies near the predicted critical
level at fault clearance and subsequently converges
to the origin despite the continued current-limited
operation.
Increasing the fault duration to
$t_{\mathrm{fault}}=38~\mathrm{ms}$ causes loss of
synchronism, and the corresponding red energy trace
fails to converge to the origin.

For the EVI current-limiting strategy,
\figref{VFMsag2}(a) shows the stable time-domain
response at the predicted CCT of
$t_{\mathrm{fault}}=52~\mathrm{ms}$.
Current limiting is released at about
$0.26~\mathrm{s}$ after continuing beyond fault
clearance.
The energy at fault clearance in
\figref{VFMsag2}(b) does not exceed the indicated
critical level and subsequently converges to the
origin.
When $t_{\mathrm{fault}}$ increases to
$55~\mathrm{ms}$, the inverter loses synchronism.
The corresponding red energy trace exceeds the
indicated critical level at fault clearance and
fails to converge to the origin.

\subsubsection{Phase Jump}
\mbox{}\\ \hspace*{\parindent}
This subsection considers the circular limiter and
EVI as representative hard current-limiting strategies,
consistent with the emphasis of this paper.
For brevity, the experimental results are presented
directly in the phase plane together with the
estimated ROA in \figref{VFMROAexp}, where experimental voltage-sag trajectories are also
included.
The predicted critical phase jumps are
$\Delta\theta_g=27^\circ$ and $-26^\circ$, and the
corresponding experimental trajectories converge
to the original SEP.
For the larger phase jumps of
$\Delta\theta_g=36^\circ$ and $-28^\circ$, the
post-jump states lie outside the estimated ROA,
and instability is observed experimentally.

\section{Conclusions}
This paper proposes an \emph{SdS energy function}
for transient synchronization stability analysis of
GFM inverters considering various current limiting strategies.
The proposed methodology unifies normal and
current-limited operation within an energy-function
framework, enabling estimation of the overall ROA with mode switching taken into account. Compared with conventional single-mode energy functions,
the proposed method yields more accurate ROA estimates. The transient stability can be accurately predicted simply
by checking whether the fault-clearing state lies
within the estimated ROA, where its energy is below
the critical value.
This provides a unified transient-stability criterion
for different large disturbances, without prescribing
how current limiting subsequently activates and releases.
The effectiveness of the proposed approach and the
validity of the analysis have been confirmed by
experiments.
The methodology thus provides a quantitative assessment
tool on which subsequent controller design and
optimization can be based.
Future studies will extend the methodology to
multi-inverter systems.

\vspace{3pt}

\appendices
\setcounter{table}{0}
\setcounter{figure}{0}
\setcounter{equation}{0}
\renewcommand{\thetable}{A\arabic{table}}
\renewcommand\cellalign{tl}
\renewcommand{\theequation}{A\arabic{equation}}
\renewcommand{\thefigure}{A\arabic{figure}}
\section{Proof of Theorem~\ref{thm:sdsef_lasalle}}
\label{app:sdsef_lasalle}

For discontinuous switched systems, Lyapunov analysis can be
formulated using the Filippov solution concept and Clarke generalized
gradient \cite{shevitz2002lyapunov}. Let $K[f](\mathbf x)$ denote
the Filippov set-valued map and
$\partial_C\mathcal W(\mathbf x)$ the Clarke generalized gradient.
For the autonomous system considered here, the generalized derivative
of $\mathcal W$ is defined as
\begin{equation}
\widetilde{\dot{\mathcal W}}(\mathbf x)
:=
\bigcap_{\boldsymbol{\xi}\in\partial_C\mathcal W(\mathbf x)}
\boldsymbol{\xi}^{\rm T}K[f](\mathbf x),
\label{eq:app_generalized_derivative}
\end{equation}
and satisfies
\begin{equation}
\frac{{\rm d}}{{\rm d}t}\mathcal W(\mathbf x(t))
\in
\widetilde{\dot{\mathcal W}}(\mathbf x(t)),
\qquad
\text{a.e.}
\label{eq:app_chainrule}
\end{equation}
along the system trajectory \cite{shevitz2002lyapunov}.
Under the strict-crossing premise considered in this paper,
the above general result can be simplified as follows.

\begin{lemma}
\label{lem:boundary}
Under the strict-crossing premise, if
\begin{equation}
\nabla W_p(\mathbf x)^{\rm T}f_p(\mathbf x)\leq0,
\qquad
\mathbf x\in\mathcal R_p,
\label{eq:app_regional_nonpositive}
\end{equation}
for every operating region and the SdSEF is continuous across
switching interfaces, then
\begin{equation}
\widetilde{\dot{\mathcal W}}(\mathbf x)
\subseteq(-\infty,0]
\label{eq:app_boundary_nonpositive}
\end{equation}
also holds at every regular switching-interface point. Moreover,
for an interface $\Sigma_{pq}$,
\begin{equation}
0\in\widetilde{\dot{\mathcal W}}(\mathbf x)
\Longrightarrow
\dot W_p(\mathbf x)=0
\ \text{or}\
\dot W_q(\mathbf x)=0.
\label{eq:app_zero_derivative_implication}
\end{equation}
And no nonempty invariant subset can be entirely contained in a
switching interface.
\end{lemma}

\begin{proof}
Consider a switching interface $\Sigma_{pq} = \{\mathbf x:h_{pq}(\mathbf x)=0\}$, shared by two adjacent operating regions $\mathcal R_p$ and $\mathcal R_q$, with regional energy functions $W_p$ and $W_q$, respectively. 

Since the SdSEF is continuous on $\Sigma_{pq}$,
\[ W_p=W_q, \qquad \mathbf x\in\Sigma_{pq}, \] 
the two limiting gradients have identical tangential components. Therefore, there exists a scalar $\rho$ such that 
\[ \nabla W_q = \nabla W_p + \rho\nabla h_{pq}. \] 

Without loss of generality, assume the strict-crossing premise gives 
\[ \nabla h_{pq}^{\rm T}f_p>0, \qquad \nabla h_{pq}^{\rm T}f_q>0. \] 
If $\rho\ge0$, 
\[ \nabla W_p^{\rm T}f_q = \nabla W_q^{\rm T}f_q - \rho \nabla h_{pq}^{\rm T}f_q \le0, \] 
which implies \[ \nabla W_p^{\rm T}K[f] \subseteq (-\infty,0]. \] Since $\widetilde{\dot{\mathcal W}}$ is obtained by intersecting $\xi^{\rm T}K[f]$ over all Clarke generalized gradients, \[ \widetilde{\dot{\mathcal W}} \subseteq (-\infty,0]. \] The case $\rho\leq0$ follows analogously by interchanging $p$ and $q$.

Moreover, for $\rho\geq0$,
\[
0\in\widetilde{\dot{\mathcal W}}
\subseteq
\operatorname{co}
\left\{
\nabla W_p^{\rm T}f_p,\,
\nabla W_q^{\rm T}f_q-\rho\nabla h_{pq}^{\rm T}f_q
\right\}.
\]
Both endpoints of this interval are nonpositive. Hence, the
interval can contain zero only if either
$\nabla W_p^{\rm T}f_p=0$, or both
$\nabla W_q^{\rm T}f_q=0$ and $\rho=0$.
The case $\rho\leq0$ follows analogously. Therefore,
\begin{equation}
0\in
\widetilde{\dot{\mathcal W}}(\mathbf x)
\quad\Longrightarrow\quad
\dot W_p(\mathbf x)=0
\ \text{or}\
\dot W_q(\mathbf x)=0.
\label{eq:app_zero_derivative_implication}
\end{equation}

Finally, under the strict-crossing premise, every vector in
$K[f]$ has a nonzero normal component with respect to
$\Sigma_{pq}$. Hence, $K[f]$ contains no tangent direction,
and by Theorem~3.4 in \cite{shevitz2002lyapunov}, no nonempty
invariant subset can be entirely contained in $\Sigma_{pq}$.
\end{proof}

\begin{proof}[Proof of Theorem~\ref{thm:sdsef_lasalle}]
Within the compact set $\Omega_c$, the negative-semidefinite
regional derivatives in Definition~\ref{def:sdsef}, together with
Lemma~\ref{lem:boundary}, ensure that
\begin{equation}
\widetilde{\dot{\mathcal W}}(\mathbf x)
\subseteq
(-\infty,0],
\qquad
\forall\,\mathbf x\in\Omega_c.
\label{eq:app_global_nonpositive}
\end{equation}
Moreover, Lemma~\ref{lem:boundary} and the regional definition of
$\dot W_p$ give
\begin{equation}
\left\{
\mathbf x\in\Omega_c:
0\in\widetilde{\dot{\mathcal W}}(\mathbf x)
\right\}
\subseteq
\mathcal Z_c.
\label{eq:app_zero_set_inclusion}
\end{equation}

Therefore, all conditions of the nonsmooth LaSalle
invariance principle in \cite[Theorem~3.2]{shevitz2002lyapunov} are satisfied. Consequently, every trajectory in $\Omega_c$ approaches an invariant subset of $\mathcal Z_c$. Since
$\{\mathbf x_{\rm sep}\}$ is, by assumption, the largest invariant
subset of $\mathcal Z_c$, it follows that
\begin{equation}
\lim_{t\rightarrow\infty}\mathbf x(t)
=
\mathbf x_{\rm sep},
\qquad
\forall\,\mathbf x_0\in\Omega_c.
\end{equation}
\end{proof}

\setcounter{equation}{0}
\renewcommand{\theequation}{B\arabic{equation}}
\section{Proof of Lemma~\ref{lem:cta_curvature}}
\label{app:cta_curvature}
\setcounter{equation}{0}

\begin{proof}
Along the critical trajectory approaching the UEP,
write $y(\delta):=\Delta v_{\mathrm{dc}}^2$, $q(\delta):=\dot{\delta}$,
\eqref{VFMe} gives
\begin{equation}
\begin{aligned}
q(\delta)
&=k_p[P_{\mathrm{in}}-P_{e,\lim}(\delta)]
+k_i y(\delta),\\
y'(\delta)
&=\frac{2[P_{\mathrm{in}}-P_{e,\lim}(\delta)]}
{C_{\mathrm{dc}}q(\delta)}.
\end{aligned}
\label{eq:cta_slope}
\end{equation}
Differentiation yields $y''(\delta)=\mathcal{G}(\delta)/q^2(\delta)$,
where
\begingroup
\small
\begin{equation}
\begin{aligned}
\mathcal{G}(\delta)
={}&-\frac{2k_i}{C_{\mathrm{dc}}}
P_{e,\lim}'(\delta)y(\delta)-\frac{4k_i}{C_{\mathrm{dc}}^2}
\frac{[P_{\mathrm{in}}-P_{e,\lim}(\delta)]^2}
{q(\delta)}.
\end{aligned}
\label{eq:cta_G}
\end{equation}
\endgroup
Thus, the trajectory curvature is determined by the sign
of $\mathcal{G}$. Differentiating \eqref{eq:cta_G} gives
\begingroup
\small
\begin{equation}
\begin{aligned}
\mathcal{G}'(\delta)+\frac{2k_i[P_{\mathrm{in}}-P_{e,\lim}(\delta)]}
{C_{\mathrm{dc}}q^2(\delta)}\mathcal{G}(\delta)=-\frac{2k_i}{C_{\mathrm{dc}}}
P_{e,\lim}''(\delta)y(\delta).
\end{aligned}
\label{eq:cta_G_derivative}
\end{equation}
\endgroup
Introduce the positive integrating factor
\begin{equation}
\mu(\delta)=
\exp\!\left[
\int_{\delta_{\mathrm{sw}}}^{\delta}
\frac{2k_i[P_{\mathrm{in}}-P_{e,\lim}(\xi)]}
{C_{\mathrm{dc}}q^2(\xi)}
\,\mathrm{d}\xi
\right]>0
\end{equation}
and define $H(\delta):=\mu(\delta)\mathcal{G}(\delta)$.
Equation~\eqref{eq:cta_G_derivative} becomes
\begin{equation}
H'(\delta)=
-\frac{2k_i}{C_{\mathrm{dc}}}
\mu(\delta)P_{e,\lim}''(\delta)y(\delta).
\label{eq:cta_H_derivative}
\end{equation}
Along the approach to the UEP, $\mathcal{G}\to0$, and hence
$\lim_{\delta\to\delta_{\mathrm{UEP}}^-}H(\delta)=0$.

If $P_{e,\lim}''\ge0$ throughout the interval $\left[ \delta _{\mathrm{sw}},\delta _{\mathrm{UEP}} \right]$,
\eqref{eq:cta_H_derivative} gives $H'\le0$.
The terminal condition then implies $H\ge0$, and thus
$y''=H/(\mu q^2)\ge0$.
The inequalities reverse when $P_{e,\lim}''\le0$.
Therefore, the critical trajectory inherits the convexity
or concavity of $P_{e,\lim}$ over the interval.

Inspection of $P_{e,\lim}(\delta)$ for the three strategies considered here, as illustrated in \figref{Pe_curve}, shows that it is concaveunder EVI and
threshold VI strategy, but convex under the circular limiter.
\end{proof}

\section{Parameters} \label{parameter}
\renewcommand{\thetable}{C\arabic{table}}
If not specified, the detailed parameters and experiment setup throughout this paper are listed in \tabref{TableA}.

\begin{table}[!h]
\centering
\begin{threeparttable}
\vspace{-4pt}
    \caption{Detailed System Parameters and Experimental Setup}
    \label{TableA}
    \footnotesize
    \renewcommand{\arraystretch}{1.05}
    \setlength{\tabcolsep}{2pt}
    \begin{tabular}{
        >{\raggedright\arraybackslash}
        p{\dimexpr0.56\linewidth-2\tabcolsep-\arrayrulewidth\relax}|
        >{\raggedright\arraybackslash}
        p{\dimexpr0.22\linewidth-2\tabcolsep-\arrayrulewidth\relax}|
        >{\raggedright\arraybackslash}
        p{\dimexpr0.19\linewidth-2\tabcolsep\relax}
    }
    \hline\hline
    Parameters
        & Physical value
        & Per-unit (pu)\tnote{1} \\
    \hline

    Base capacity $S_b$
        & 1.5~kV$\cdot$A
        & 1 \\

    Base voltage $V_b$
        & 89.2~V
        & 1 \\

    Base frequency $\omega_0/2\pi$
        & 50~Hz
        & 1 \\

    Grid voltage $V_g$\tnote{2}
        & 89.2~V
        & 1 \\

    Reference /input power $P_{\mathrm{ref}}$, $P_{\mathrm{in}}$
        & 1.5~kW
        & 1 \\

    AC voltage reference $E$
        & 89.2~V
        & 1 \\

    DC voltage (reference)
    $V_{\mathrm{dc,ref}}$
        & 225~V
        & 2.5 \\

    Prescribed current limit $I_{\mathrm{lim}}$\tnote{3}
        & 23.5~A
        & 1.7 \\

    Total reactance to grid $X_{\Sigma,\mathrm{norm}}$\tnote{4}
        & $7~\mathrm{mH}\times\omega_0$
        & 0.43 \\

    Voltage sag depth
        & 8.9~V
        & 0.1 \\

    DC capacitance $C_{\mathrm{dc}}$
        & 6.1~mF
        & 10.3 \\

    LC filter inductance $L_f$
        & 1~mH
        & 0.059 \\

    LC filter capacitance $C_f$\tnote{5}
        & $5~\mu\mathrm{F}$
        & 0.008 \\

    Inner current control bandwidth
        & 1~kHz
        & -- \\

    Switching frequency
        & 10~kHz
        & -- \\

    \hline
    \multicolumn{3}{l}{
        Current-limiting strategy I -- circular limiter
    } \\
    \hline

    Grid impedance $Z_g$ 
        & $7~\mathrm{mH}$, $0.26~\Omega$
        & $j0.43+0.05$ \\

    Inner voltage control bandwidth
        & 200~Hz
        & -- \\

    \hline
    \multicolumn{3}{l}{
        Current-limiting strategy II -- EVI
    } \\
    \hline

    Grid impedance $Z_g$
        & $5~\mathrm{mH}$, $0.18~\Omega$ 
        & $j0.31+0.03$ \\

    Normal-mode virtual reactance $X_{v,\mathrm{norm}}$
        & $2~\mathrm{mH}\times\omega_0$
        & 0.12 \\

    Virtual-impedance $X/R$ ratio $\sigma$
        & 8
        & -- \\

    \hline
    \multicolumn{3}{l}{
        Current-limiting strategy III -- threshold VI
    } \\
    \hline

    Activation threshold $I_{\mathrm{th}}$\tnote{3}
        & 16.6~A
        & 1.2 \\

    Grid impedance $Z_g$
        & $7~\mathrm{mH}$, $0.26~\Omega$
        & $j0.43+0.05$ \\
        
    Virtual-impedance $X/R$ ratio $\sigma$
        & 8
        & -- \\

    virtual-impedance gain $\kappa_{\mathrm{VI}}$
        & --
        & 1.5 \\

    Inner voltage control bandwidth
        & 200~Hz
        & -- \\

    \hline
    \multicolumn{3}{l}{
        PSC I -- classical GFM
    } \\
    \hline

    Virtual inertia $J$
        & \multicolumn{2}{l}{$3$~s (default)} \\

    Damping coefficient $D$
        & \multicolumn{2}{l}{$15$~pu (default)} \\

    \hline
    \multicolumn{3}{l}{
        PSC II -- DC-voltage-controlled GFM
    } \\
    \hline

    PI controller $k_p$
        & \multicolumn{2}{l}{$1\times2\pi$~pu (default)} 
        \\

    PI controller $k_i$
        & \multicolumn{2}{l}{$5$~pu (default)} 
         \\

    \hline\hline
    \end{tabular}

    \begin{tablenotes}
        \scriptsize
        \item[1]
        The same per-unit values are used in the theoretical
        calculations, simulations, and experiments throughout
        this paper.
        \item[2]
        Line-to-line rms voltage.
        \item[3]
        Both $I_{\mathrm{lim}}$ and $I_{\mathrm{th}}$ are
        specified as peak current values.
        \item[4]
        All current-limiting strategies use the same total impedance in normal operation,
        $X_{\Sigma,\mathrm{norm}}=X_g+X_{v,\mathrm{norm}}$, for a fair comparison.
        \item[5]
        The filter capacitors are $\Delta$-connected.

    \end{tablenotes}
\end{threeparttable}
\end{table}


\ifCLASSOPTIONcaptionsoff
  \newpage
\fi
\bibliographystyle{IEEEtran}
\bibliography{References}

\end{document}